\documentclass{aa}

\usepackage{graphicx}
\usepackage{txfonts}
\usepackage{lipsum}
\usepackage{subcaption}         
                               
\usepackage{lscape}             
                            
\usepackage{placeins}           
                                
\usepackage{lineno}
\modulolinenumbers[1]
\usepackage{hyperref}
\usepackage{orcidlink}
\usepackage{amsmath}
\usepackage{siunitx}
\usepackage{amsmath}
\usepackage{booktabs}
\usepackage{multirow}  
\usepackage{graphicx} 
\usepackage{makecell}
\usepackage{tabularx}
\usepackage{CJK}
\usepackage{longtable}
\usepackage{caption}

\hypersetup{
    colorlinks=true,
    linkcolor=blue,
    filecolor=blue,      
    urlcolor=blue,
    citecolor=blue,
    pdftitle={Overleaf Example},
    pdfpagemode=FullScreen,
    }

\begin{document}

   \title{A new hot core in the outer Galaxy: Impact of metallicity on the formation of complex organic molecules}
   \titlerunning{Impact of metallicity on the formation of COMs}

   \author{Youxin Wang\inst{1}\fnmsep\thanks{Member of the International Max Planck Research School (IMPRS) for Astronomy $\&$ Astrophysics (\href{https://blog.mpifr-bonn.mpg.de/imprs/}{https://blog.mpifr-bonn.mpg.de/imprs/}) at the Universities of Bonn and Cologne.} \orcidlink{0000-0001-9155-0777}
        \and Arnaud Belloche\inst{1} \orcidlink{0000-0003-0046-6217} \and Robin T. Garrod\inst{2} \orcidlink{0000-0001-7723-8955}   
        }

   \institute{Max-Planck-Institut für Radioastronomie, Auf dem Hügel 69, 53121 Bonn, Germany\\
             \email{yxwang@mpifr-bonn.mpg.de}
            \and Departments of Chemistry and Astronomy, University of Virginia, Charlottesville, VA 22904, USA\\ }

   \date{Received September 30, 20XX}

  \abstract
   {Many complex organic molecules (COMs) in star-forming regions are believed to form on dust grains. We thus expect both the reduced metallicity and dust-to-gas ratio in the outer Galaxy to have an impact on the chemical composition of these regions.}
   {We investigate if certain COMs are more sensitive than others to metallicity by measuring the chemical composition of hot cores in the outer Galaxy.} 
   {We used the interferometer NOEMA to perform an imaging spectral line survey of G135.27+2.79, a hot core candidate located at a galactocentric distance of 13.1 kpc. We derived the rotational temperatures and column densities of the detected molecules while assuming local thermodynamic equilibrium and compared the chemical composition of G135.27+2.79 to other sources and to the predictions of the three-phase astrochemical code MAGICKAL.}
   {G135.27+2.79 hosts three continuum cores, labeled MM1, MM2, and MM3. Toward MM1, we detected 28 molecules, including 12 COMs, and several of their less abundant isotopologs. Most species trace a hot, compact region, confirming MM1 as a hot core (the third one identified in the outer Galaxy). MM1 drives a bipolar CO outflow. COMs show a velocity gradient along the outflow axis but opposite to that of CO, which may be related to a wide-angle disk wind. The chemical composition of MM1 correlates rather well with that of the inner and outer Galaxy hot cores G31.41+0.31 and WB89-789 SMM1, but its molecular abundances relative to methanol lie in between, which may reflect the influence of metallicity on COM formation. The model results agree reasonably well, though with a few notable exceptions, with the COM abundances of MM1 relative to methanol and with the abundance ratios between MM1 and G31.41+0.31. Sensitivity to the reduced metallicity and dust-to-gas ratio varies between molecules, with carbon chains and nitriles most negatively affected. The lower dust-to-gas ratio leads to slower adsorption under low-metallicity conditions so that more carbon is locked up into CO in the gas. Slow adsorption means that CO is hydrogenated more efficiently on grains, enhancing CO-related COM abundances above expectations.}
   {These results demonstrate that metallicity has a significant impact on the formation of COMs. A larger source sample is needed to investigate the robustness of the deviations noted between the model and the observations for a few species.}

   \keywords{Astrochemistry – stars: formation - line: identification – radio lines: ISM – ISM: molecules
               }

   \maketitle \nolinenumbers

\section{Introduction}
Complex organic molecules (COMs) are defined as carbon-bearing molecules containing six or more atoms \citep{2009ARA&A..47..427H}. These molecules are important tracers of the chemical complexity in the interstellar medium (ISM). COMs have been detected in a wide variety of astrophysical environments, ranging from warm inner envelopes surrounding protostars \cite[e.g.,][]{2016A&A...595A.117J,2020A&A...635A.198B,2023A&A...678A.137C} to protoplanetary disks \cite[e.g.,][]{2013prpl.conf2S032W,2019NatAs...3..314L,2025AJ....169..307F} and even to extragalactic environments, such as the center of the nearby starburst galaxy NGC 253 \citep{2021A&A...656A..46M,2025A&A...698A.261B,2025A&A...699A..70H}. How these species are formed has been the focus of much attention and debate \citep[e.g.,][]{2020ARA&A..58..727J}. COMs are generally thought to form and evolve in both the gas phase and the solid phase, the latter consisting of the surfaces and ice mantles of dust grains \citep[e.g.,][]{1992ApJS...82..167H,1998ARA&A..36..317V,2008ApJ...682..283G}. Observations over the past decade, particularly the detection of gas-phase COMs at low temperatures in prestellar cores and protostellar envelopes \citep[e.g.,][]{2012A&A...541L..12B,2014ApJ...795L...2V,2024A&A...686A.252B}, and more recently the identification of several COMs in interstellar ices using the James Webb Space Telescope \citep[JWST; e.g.,][]{2022A&A...665A..89R,2022ApJ...941L..13Y,2024A&A...683A.124R} suggest that the solid phase plays a major role in the production of COMs. 

Hot molecular cores associated with high-mass star-forming regions ($M$ >8~$M_\odot$; \citep{1995RMxAC...1..137W}) and their low-mass counterparts, known as hot corinos \citep{2004ASPC..323..195C}, serve as excellent laboratories for investigating the chemical complexity of the ISM. Hot cores are compact ($< 0.1$ pc), dense ($n > 10^6$ cm$^{-3}$) regions that form around nascent high-mass protostars. They represent a transient yet chemically rich phase during the early stages of high-mass star formation \citep[e.g.,][]{2000prpl.conf..299K,2005IAUS..227...59C}. As a protostar forms, it begins to heat its surrounding envelope. When the temperature rises above $\sim$100 K, thermal desorption of the ice mantles coating interstellar dust grains occurs, releasing a wide variety of COMs into the gas phase \citep{2020ARA&A..58..727J}. The advent of broadband instrumentation, especially in interferometers over the past decade, has significantly enhanced the feasibility of conducting unbiased spectral line surveys of hot cores and hot corinos, thereby advancing our understanding of COM formation. Notable examples include the Exploring molecular complexity with ALMA (EMoCA)/Re-exploring Molecular Complexity with ALMA (ReMoCA) project, which targeted hot cores in Sgr~B2(N) in the Galactic center region \citep[e.g.,][]{2016A&A...587A..91B,2019A&A...628A..10B,2025A&A...698A.143B,2017A&A...604A..60B,2024A&A...681A.104B}; the G31.41+0.31 Unbiased ALMA sPectral Observational Survey (GUAPOS) project, focused on the hot core G31.41+0.31 in the inner Galaxy \citep[e.g.,][]{2020A&A...644A..84M,2023A&A...677A..15M}; and the ALMA Protostellar Interferometric Line Survey (PILS) project, which studied the low-mass protostellar system IRAS 16293-2422 in the solar neighborhood \citep[e.g.,][]{2016A&A...595A.117J,2018A&A...616A..90C,2018A&A...620A.170J}. These surveys have enhanced our understanding of chemical complexity in the inner Galaxy and, surprisingly, revealed that the chemical compositions of Sgr B2(N2), G31.41+0.31, and IRAS 16293-2422B are remarkably similar when oxygen- and nitrogen-bearing COMs are normalized separately \citep{2020ARA&A..58..727J,2025A&A...698A.143B}. This raises the question whether this similarity still holds beyond the solar circle, where the metallicity is reduced. Our goal is to investigate if metallicity has an impact on the prevalence of COMs in star-forming regions of our Galaxy.

The outer Galaxy refers to regions beyond the solar circle, with Galactocentric distances, $D_{\rm gc}$, greater than 8.15 kpc \citep{2019ApJ...885..131R}. The disk of our Galaxy has a metallicity gradient between 5 and 17 kpc, as measured in various elements (e.g., C, N, O, Fe) that exhibit different radial trends \citep[e.g.,][]{2011AJ....142..136L}. For instance, at distances of 12-17 kpc from the center, the metallicity traced by Fe is about 1.7 to 3.4 times lower than at the solar circle. Several studies of star-forming regions in the outer Galaxy have investigated the impact of metallicity on molecular abundances \citep{2021ApJ...922..106B,2022A&A...660A..76F,2022A&A...664A.154F,2025ApJ...986..122K}. These studies generally report that the abundances of simple molecules and (cold) CH$_3$OH, when normalized to H$_2$, remain comparable to those in the inner Galaxy. However, these abundances depend critically on the adopted H$_2$ column density, which is typically inferred from dust continuum emission and therefore sensitive to assumptions about the gas-to-dust ratio (a value of 100 was used in these papers). However, the gas-to-dust ratio is expected to increase with galactocentric distance \citep{2017A&A...606L..12G}. Indeed, when accounting for this effect, 
\citet{2024A&A...691A.180F} and \citet{2025A&A...704A.171G} reported decreasing molecular abundances toward larger galactocentric radii. Reporting abundances relative to another molecule, such as CH$_3$OH, allows us to avoid the uncertainties related to the derivation of H$_2$ column densities.

Given that many COMs form on dust grains and contain multiple heavy atoms \citep{2006A&A...457..927G}, their formation is likely influenced by the metallicity and dust-to-gas ratio. Consequently, the chemical composition of the ice mantles of dust grains (and the gas phase, after desorption) in the outer Galaxy may differ from that in the inner Galaxy. In other words, the outer Galaxy serves as an ideal laboratory to study the impact of metallicity and the dust-to-gas ratio on the formation of COMs \citep{2024arXiv241104451S}. Moreover, studies in the outer Galaxy are critical for bridging our understanding between the normal-metallicity environment in the inner Galaxy and the low-metallicity environment observed in nearby dwarf galaxies such as the Large Magellanic Cloud (LMC) and Small Magellanic Cloud (SMC). Recently, six hot cores have been identified in the LMC \citep{2016ApJ...827...72S,2020ApJ...891..164S,2018ApJ...853L..19S,2022ApJ...931..102S,2024A&A...688A...3H} and two in the SMC \citep{2023ApJ...946L..41S}. However, only two hot cores, WB89-789 SMM1 and Sh 2-283-1a SMM1, have been detected in the outer Galaxy so far \citep{2021ApJ...922..206S,2026ApJ...999..194I}. To better understand the impact of metallicity on the formation of COMs, two actions are essential: building a larger sample of hot cores in the outer Galaxy observed through unbiased spectral line surveys and comparing their chemical composition with the predictions of astrochemical models. 

This study focuses on the search for hot cores embedded in the outer Galaxy, and we investigate the impact of metallicity on the formation of COMs. To this end, we first conducted single-dish telescope observations to identify promising hot core candidates. These candidates were then followed up at high angular resolution with an unbiased spectral line imaging survey using the Northern Extended Millimeter Array (NOEMA) with the goal of characterizing their physical structure and chemical composition. In this work, we present and analyze NOEMA observations of G135.27+2.79, the hot core candidate of our sample that was observed first. The source selection, the observational setup, and the data reduction are described in Sect.~\ref{Observations}. Section~\ref{Results} presents an analysis of the observations and a comparison to other sources. Section~\ref{modeling} compares the chemical composition derived for G135.27+2.79 to the predictions of the chemical model MAGICKAL. The results are discussed in Sect.~\ref{discussion}, and Sect.~\ref{Conclusions} presents our conclusions.

\section{Observations and data reduction} \label{Observations}
\subsection{Source selection} \label{Sourceselection}
Our sample was constructed by identifying candidate high-mass star-forming regions in the outer Galaxy based on the presence of water or Class II methanol masers, which are indicators of young stellar objects (YSOs; \citealt{2000prpl.conf..299K}). Class II methanol masers, in particular, are known to trace exclusively high-mass YSOs \citep[e.g.,][]{2003A&A...403.1095M,2006ApJ...638..241E}. We focused on sources located in the second and third Galactic quadrants with heliocentric distances of less than 10 kpc, enabling comparison with well-studied regions in the inner Galaxy. We selected all sources fulfilling these criteria beyond a galactic radius, $D_{\rm gc}$, of 12 kpc from the Effelsberg water maser surveys of IRAS sources \citep{1988A&A...191..323W,1993A&AS...98..589W}, the Parkes/ATCA/MERLIN methanol multi-beam survey \citep{2017MNRAS.469.1383G}, the 6.7 GHz methanol maser catalog \citep{2005A&A...432..737P}, and the Bar and Spiral Structure Legacy  (BeSSeL) Very Long Baseline Interferometry survey \citep[][and reference therein]{2019ApJ...885..131R}. Moreover, we kept only sources with a maser position that was known sufficiently accurately compared to the beams of the IRAM 30m telescope at 3~mm~/1.3~mm and the Atacama Pathfinder Experiment (APEX) telescope at 1.3 mm, either from interferometric observations or from cross-scans in the case of Effelsberg. We also included an IRAS source at $D_{\rm{gc}} = 15.8$ kpc with no maser detection but detected in H$_2$CO \citep{2008AsBio...8...59B}. Our final sample comprises 19 sources spanning $D_{\rm gc}$ from 12 to 17 kpc. The seven targets in the second quadrant were observed with the IRAM 30m telescope and the twelve sources in the third quadrant with the APEX telescope. These observations achieved deep integrations, with rms sensitivities at a spectral resolution of $\sim$1 km~s$^{-1}$ of about 5 mK for the APEX observations, and ranging from 5.9 to 7.1 mK and from 8.1 to 18.2 mK for the IRAM 30 m observations at 3 mm and 1.3 mm, respectively. This enabled a sensitive search for high-temperature rotational emission of COMs. Cold methanol emission was detected in all sources, but hints of the presence of a hot core were found in only four targets (weak emission of CH$_{3}$CN or torsionally excited
CH$_{3}$OH).

The first of our hot core candidates that we subsequently observed at higher angular resolution, G135.27+2.79 (also known as WB89-437, IRAS 02395+6244), is a luminous far-infrared source detected with IRAS \citep{1989A&AS...80..149W}. The distance to G135.27+2.79 was determined through trigonometric parallax measurements of its associated H$_2$O maser, yielding a heliocentric distance of $6.0\pm0.2$ kpc \citep{2009ApJ...696.1981H}, and a Galactocentric distance of 13.1 kpc \citep[assuming a Galactocentric distance of the Sun of 8.15 kpc,][]{2019ApJ...885..131R}. The trigonometric distance is shorter than the kinematic distance assumed earlier by \cite{1989A&AS...80..149W}. Adjusted to the new distance, the IRAS luminosity of 7.1$\times$10$^4$ L$_\odot$ reported by \cite{1989A&AS...80..149W} becomes 3.1$\times$10$^4$ L$_\odot$, which corresponds to a B0 star if it is a single object \citep{1973AJ.....78..929P}. At this distance, the expected metallicity for G135.27+2.79 is $[\mathrm{Fe/H}] \approx -0.3$~dex, based on the Galactic radial abundance gradient $\frac{d[\mathrm{Fe/H}]}{dD_\mathrm{gc}} = -0.061~\mathrm{dex~kpc^{-1}}$ \citep{2011AJ....142..136L}. This corresponds to a metallicity $Z\sim0.5\,Z_\odot$ (see Sect. \ref{modeling} for the corresponding elemental abundances). The source is associated with several H$_2$O maser spots \citep{1993A&AS...98..589W,2009ApJ...696.1981H}, strong CO wing emission tracing an outflow \citep{1996AAS..119..439W}, but no radio continuum emission was detected, suggesting that it is extremely young or that it consists of a cluster of lower mass stars \citep{1996ApJ...458..653R}. Additionally, several simple C-bearing species (e.g., HCN, C$_4$H, and c-C$_3$H$_2$) and the cold ($T_{\rm ex}$ < 20 K) emission of two COMs (CH$_3$CCH and CH$_3$OH) were detected toward this source \citep{2021ApJ...922..106B,2022A&A...660A..76F,2022A&A...664A.154F,2025A&A...704A.171G,2025ApJ...986..122K}.

\subsection{NOEMA observations} \label{NOEMA_obs}

We conducted a spectral line survey toward G135.27+2.79 using the NOEMA interferometer\footnote{\url{https://iram-institute.org/science-portal/noema/}} in the frequency range 215.1$-$246.4 GHz. The observations were carried out in the C configuration with nine antennas on 2023 October 5, December 3, and December 5. The phase center was ($\alpha$,$\delta$)$_{(J2000)}$ = 02$^h$43$^m$28$^{s}\!\!{.}$57, +62$^\circ$57$^\prime$08$^{\prime\prime}$\!\!\!.4, which is the water maser position from the BeSSeL survey \citep{2009ApJ...696.1981H,2019ApJ...885..131R}. The half-power beam width (HPBW) of the primary beam of the 15 m antennas is about 23" at 1.3 mm. 

The spectral line survey was divided into two spectral setups, S1 and S2. Each setup was observed in two polarizations and delivered four spectral windows of bandwidth 3.9 GHz each, two per sideband (LO, LI, UI, UO). The on-source integration time was 6 h for S1 and 5.2 h for S2. The baselines of the array ranged from 19.5 m to 328.0 m for S1 and from 17.7 m to 368.0 m for S2. The observations were performed with a channel spacing of 250 kHz using the PolyFiX correlator. The spectra were smoothed to a resolution of 500 kHz (0.6$-$0.7 km~s$^{-1}$). Bandpass calibration was performed using observations of 3C84, and flux calibration was performed using observations of LKHA101 and MWC349. The maximum recoverable scale was about 8", which was calculated using Eq. (3.28) in \cite{cortes2025alma}. Phase and amplitude calibration was performed using J0244+6228 and 0224+671. Details about the observational setups are given in Table~\ref{setup}. Following \citet{2019A&A...621A..76M}, we assume an absolute flux calibration uncertainty of 15$\%$.

\subsection{Data reduction}
The data were calibrated with GILDAS/CLIC\footnote{\url{https://www.iram.fr/IRAMFR/GILDAS/}} and then imaged with IMAGER version 4.4.\footnote{\url{https://imager.oasu.u-bordeaux.fr/}} We ran the IMAGER pipeline step by step to perform self-calibration, imaging, and splitting of the continuum and line emission. To improve the data quality, the self-calibration was conducted in SURVEY mode without a line catalog. Within spectral windows S1-UI and S2–LI, the CO line is very strong, dominates the integrated flux, and is spatially extended, making the SURVEY mode inappropriate for self-calibration. Therefore, we split each of these windows into two sub-windows and applied self-calibration separately: in SPLIT mode for the sub-window containing the CO line and in SURVEY mode for the other one. The splitting of the line and continuum emission was performed with the C-SCM method of IMAGER. The size (HPBW) of the synthesized beam, the corresponding linear size, and the rms noise level in the final cubes are given in Table~\ref{noise}. The noise level of each spectral window corresponds to the median of the noise levels measured in all channel maps using the procedure $go$ $noise$ in GREG/GILDAS$^{2}$.

\section{Results and analysis} \label{Results}
\subsection{Continuum emission} \label{continuum}

\begin{figure*}[!htbp]
    \centering
    \includegraphics[width=0.39\linewidth]{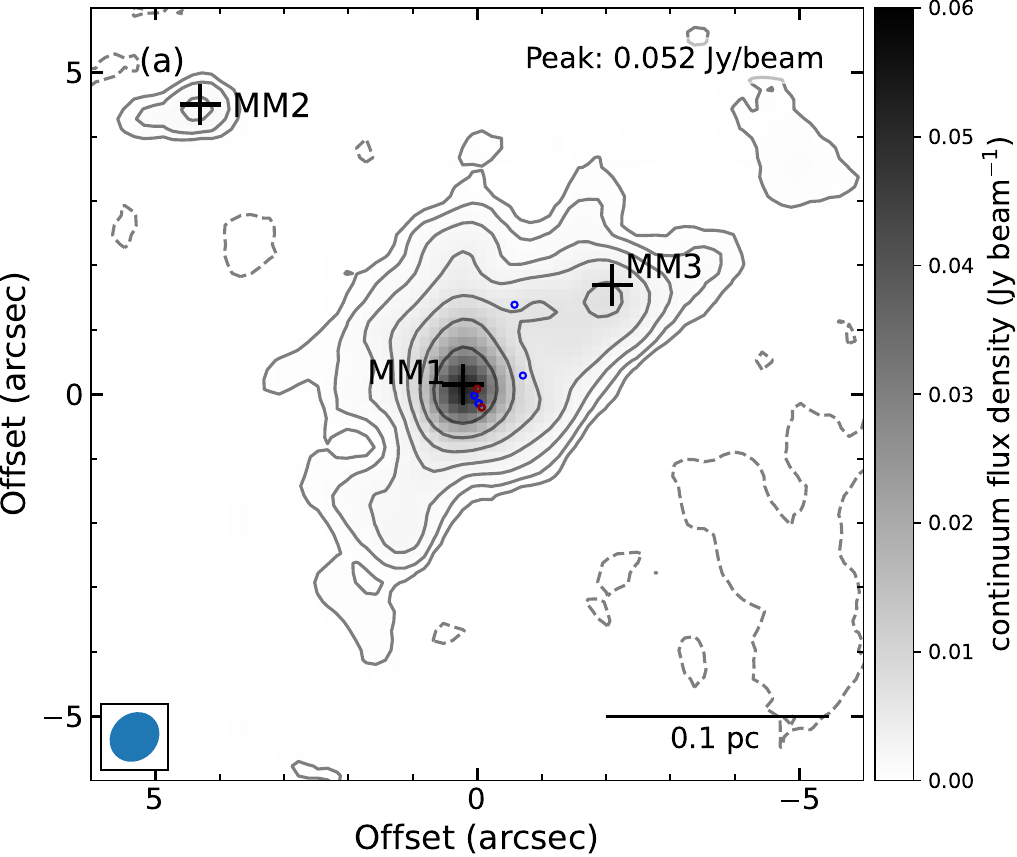}
    \includegraphics[width=0.39\linewidth]{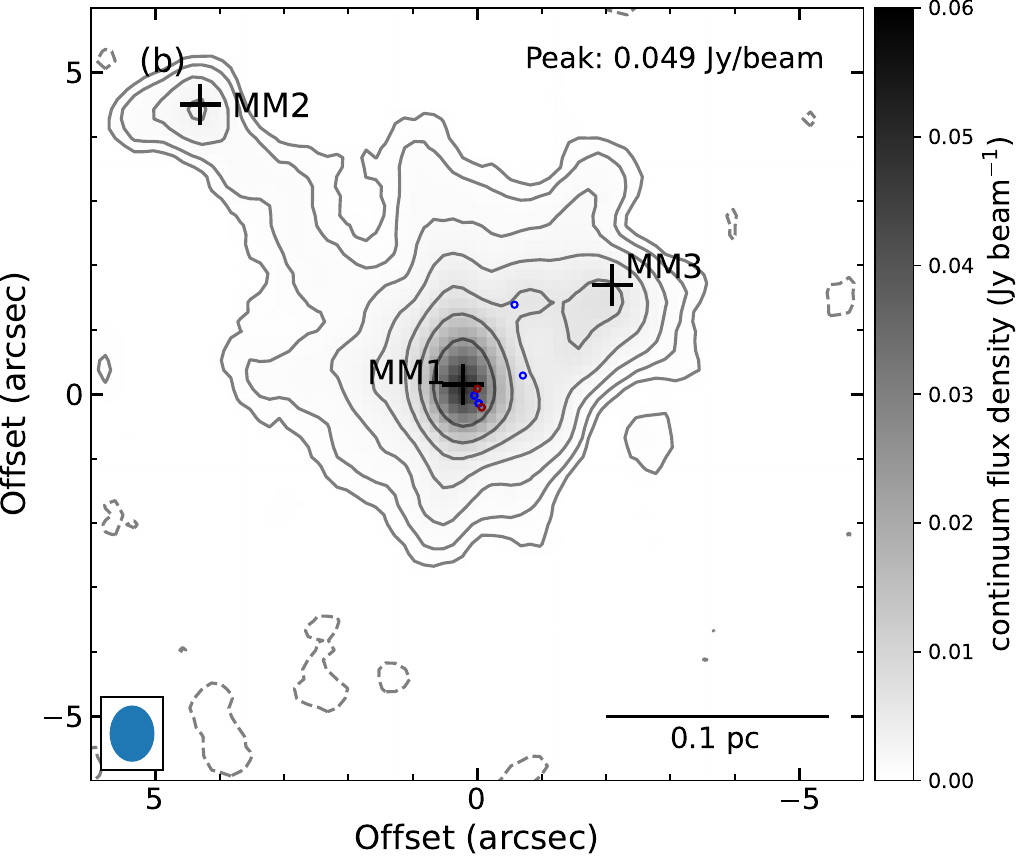}
    \caption{Continuum maps of G135.27 at 233 GHz (a) and 229 GHz  (b) in gray scale, extracted from setups S1-UI-CO and S2-LI-CO, respectively. Contour levels start at 3$\sigma$ and increase by a factor of 2, where $\sigma = 0.13$ and $0.12$ mJy~beam$^{-1}$, respectively. Negative contours at $-3\sigma$ are shown as dashed lines. The black crosses mark three continuum peaks of G135.27, labeled MM1, MM2, and MM3 (see Sect.~\ref{continuum}), with the peak flux density of MM1 indicated in the upper right corner. The empty circles denote the positions of H$_2$O masers \citep{2009ApJ...696.1981H}, with blue and red indicating maser spots blue- and redshifted with respect to the source systemic velocity of $-$71.7 km~s$^{-1}$, respectively \citep{2008AsBio...8...59B}. The beam size (HPBW) is displayed as an ellipse in the bottom left corner. The offsets are defined with respect to the phase center (see Sect.~\ref{NOEMA_obs}). The maps are not corrected for primary-beam attenuation.}
    \label{fig_dust}
\end{figure*}

Although the uv coverage and array configurations differ between the two NOEMA setups, the resulting continuum emission maps are largely consistent in structure. Figure~\ref{fig_dust} presents the continuum maps of spectral windows S1-UI-CO and S2-LI-CO, which
have a similar angular resolution (HPBW $\sim$0.8"). While slight variations in morphology are evident, the overall structure remains comparable: each map reveals three continuum peaks, which we named MM1, MM2, and MM3. MM1 and MM2 are located at equatorial offsets (0.23", 0.15") and (4.3", 4.5") from the phase center, respectively. These positions were determined by averaging the results from all sub-bands using 2D Gaussian fits. Determining the peak position of MM3 with a Gaussian fit is challenging due to its proximity to MM1. Instead, we constructed the continuum profile along the direction going through the peak positions of redshifted and blueshifted SiO components detected around MM3 (see Sect.~\ref{SiO} for details) and identified the corresponding peak continuum position of MM3 as ($-$2.1", 1.7").

To investigate the nature of the detected emission in MM1, the in-band spectral index, $\alpha$, was analyzed across the eight sub-bands. $\alpha$ was determined by fitting the flux densities to the formula $F_\nu$ $ \propto$ $\nu^{\alpha}$ using the scipy function curve$\_$fit. The fluxes $F_\nu$ in all sub-bands were extracted using an aperture with a diameter of 1.8", which corresponds to approximately twice the beam size of the lowest spatial resolution (S1-LO). Figure~\href{https://zenodo.org/records/17779973}{E.1} shows the result of the in-band spectral index fitting. We find $\alpha$ = 3.3$\pm$1.4, indicating that the continuum emission is dominated by dust emission at a scale of 1.8". We then obtain $\beta = \alpha - 2 = 1.3$, which is similar to what \cite{2017A&A...604A..60B} derived for the hot core Sgr~B2(N2). Such a flat index suggests a dust opacity spectrum intermediate between models of dust grains of \cite{1994A&A...291..943O} without ice mantles and with thin ice mantles.

The optical depth $\tau$ at the peak position of MM1 was estimated using the following equation \citep{2017A&A...604A..60B}: 

\begin{equation}
    \tau = -{\rm ln}(1-\frac{F_{\nu}^{\rm beam}}{\Omega_{\rm beam}B_{\nu}(T_{\rm d})}),
\end{equation}
with $\Omega_{\rm beam}$ = $\frac{\pi}{4\rm ln2}$ $\times$ $HPBW_{\rm maj}$ $\times$ $HPBW_{\rm min}$ as the solid angle of the synthesized beam, $T_{\rm d}$ as the dust temperature, and $B_{\rm \nu}$$(T_{\rm d}$) as the Planck function. The term $F_{\nu}^{\rm beam}$ is the peak continuum flux density in Jansky per beam, corrected for primary beam attenuation. Assuming that dust and gas are well coupled, we adopted a dust temperature ($T_{\rm d}$) of 75 K. This value corresponds to the rotational temperature ($T_{\rm rot}$) of CH$_3$CCH toward the continuum peak position, determined from our population diagram analysis (see Sect.~\ref{PD} and Fig.~\href{https://zenodo.org/records/17779973}{H.2}). We used the temperature derived from CH$_3$CCH because, as with the dust continuum emission, this molecule traces more extended structures than other COMs. The opacity calculations for all sub-bands yield values lower than 0.04, confirming that the dust continuum emission at the scale traced with NOEMA in this source is optically thin.

\subsection{H$_2$ column density and total gas mass of MM1} \label{H2}

The H$_2$ column density of MM1, $N_{\rm H_2}$, was estimated from the millimeter dust continuum emission. Since this emission is optically thin at the observed frequencies (Sect.~\ref{continuum}), $N_{\rm H_2}$ was calculated using the following equation:
\begin{equation}
    N_{\rm H_2} = \frac{F_{\nu} \xi_{\rm d}}{\Omega \kappa_{\nu}B_{\nu}(T_{\rm d}) \mu_{\rm H_{2}} m_{\rm H}},
\end{equation}
where $F_{\nu}$ is the peak continuum flux density, $\Omega$ is the beam solid angle, $\kappa$$_{\nu}$ is the dust mass opacity (in square centimeter per gram of dust) at frequency $\nu$, $\xi_{\rm d}$ is the gas-to-dust ratio, $\mu_{\rm H_2}$ is the mean molecular weight per H$_2$ \citep[2.8,][]{2008A&A...487..993K}, and $m_{\rm H}$ is the hydrogen mass. We assumed $\kappa_{\nu}$ = $\kappa_0$($\nu/\nu_{0}$)$^{\beta}$, with $\kappa_0$ = 1.6 cm$^2$ g$^{-1}$ at 1.3 mm as \cite{2017A&A...604A..60B}, since we have a similar value of $\beta$ (see Sect.~\ref{continuum}).

To estimate the gas-to-dust ratio, $\xi_{\rm d}$, we used the C$^{18}$O and dust emission integrated over an aperture of 1.8" in diameter. Here again, we used CH$_3$CCH to estimate the temperature at a diameter of 1.8" (which is similar to the C$^{18}$O emission size; see Fig. \ref{fig_all_size}), assuming that the rotational temperatures of CH$_3$CCH and C$^{18}$O, as well as the dust temperature, are the same under local thermodynamic equilibrium (LTE) conditions. Employing the CH$_3$CCH spectrum averaged over a ring of inner and outer diameters of 0.9" and 2.7", respectively, we derived a temperature of 63 K from its population diagram (see Sect.~\ref{PD} and Fig.~\href{https://zenodo.org/records/17779973}{H.2}). The gas mass surface density was derived from the C$^{18}$O column density, since the C$^{18}$O line is optical thin, with an opacity of $\sim$0.2 according to our Weeds model (see Sect.~\ref{weeds}). The total column density $N_{\rm tot}$ of C$^{18}$O was calculated using the following equation:
\begin{equation}
    N_{\rm tot} = \frac{8\pi k_{\rm B}\nu^2W}{hc^3A_{ul}bg_u}Q\exp({\frac{E_u}{k\rm _BT_{\rm rot}}}),
    \label{eq_column}
\end{equation}
where $k_{\rm B}$ is the Boltzmann constant, $W$ is the integrated intensity over the full line profile after correction for primary beam attenuation, $h$ is the Planck constant, c is the speed of light, $A_{ul}$ is the Einstein coefficient for spontaneous emission, $b$ is the beam filling factor with a value of one for C$^{18}$O, $g_{u}$ is the degeneracy of the upper level, $Q$ is the partition function, and $E_u$ is the upper-level energy. This yielded a C$^{18}$O column density of 3.3$\times$10$^{16}$~cm$^{-2}$, with the integrated intensity measured from the spectrum averaged over an aperture of diameter 1.8". We applied Eq. (1) of \cite{2017A&A...606L..12G}, which assumes a CO abundance of 9.5$\times$10$^{-5}$ relative to H$_2$ in the solar neighborhood, to derive the total gas mass surface density from C$^{18}$O. Following their approach, we adopted a C/H gradient of $-0.08$ dex kpc$^{-1}$ from \cite{2011AJ....142..136L} and used the $^{16}$O/$^{18}$O isotopic ratio relation, $^{16}$O/$^{18}$O = (58.8$\pm$11.8)$D_{\rm gc}$ + (37.1$\pm$82.6), from \cite{1994ARA&A..32..191W}. The dust mass surface density was calculated as $\Sigma_{\rm dust}$ = $\frac{S_{\nu}/\Omega}{\kappa{_{\nu}}B_{\nu}(T_{\rm d})}$, with $S_{\rm \nu}$ the flux density integrated over an aperture of diameter 1.8" and $\Omega$ the solid angle of this aperture. This yields a gas-to-dust mass ratio of $303\pm66$.\footnote{Using a smaller aperture equal to the beam size gives a gas-to-dust ratio of 335$\pm$73, which is consistent with this result within the uncertainties.} The Galactocentric gradient of gas-to-dust ratio determined by \cite{2017A&A...606L..12G} yields a ratio of $\sim$380 at a Galactocentric distance of 13.1 kpc, but the dispersion is large. Our result is consistent with this estimate within the uncertainties, and we thus adopted a value $\xi_{\rm d}$ of 303 for the rest of the analysis. This yields a peak H$_2$ column density of (1.1$\pm$0.1)$\times$10$^{24}$ cm$^{-2}$, without accounting for the calibration uncertainty. Assuming that the source diameter is equal to the beam size, we obtained a mean gas number density of $n_{\rm H_2}$ = 1.5$\times10^{7}$ cm$^{-3}$. Assuming a typical density profile with a slope of -1.5, the gas number density at 0.3", corresponding to the typical size of the COM emission (see Sect. \ref{size}), is $\sim8\times10^{7}$ cm$^{-3}$.

Furthermore, the gas mass of the envelope within a diameter of 3.0" can be estimated using the following equation:
\begin{equation}
    M_{\rm gas} = \frac{S_{\nu}D^2 \xi_{\rm d}}{\kappa{_{\nu}}B{_\nu}(T{\rm _d})},
\end{equation}
where $S_{\nu}$ is the flux density integrated over the aperture and $D$ is the heliocentric distance to G135.27. Assuming a radial power-law profile for the temperature, we obtained an exponent of $-$0.25 using the two CH$_3$CCH temperatures mentioned above (75 K for 0.9" and 63 K for 1.8"), yielding a temperature of 55 K at a diameter of 3.0" by extrapolation. Assuming the same temperature for the dust, we obtained a total gas mass of 49 M$_\odot$ within a radius of 9000 au.

\subsection{Kinematics} \label{outflow}
\subsubsection{Outflow morphology} \label{morphology}

\citet{1996AAS..119..439W} reported strong line wings in CO 1$-$0 and 2$-$1 toward G135.27 based on IRAM 30m observations, suggesting the presence of an outflow. Follow-up millimeter interferometric observations with BIMA, using a synthesized beam of $21.8'' \times 14.9''$, confirmed this outflow, revealing blue- and redshifted lobes oriented toward the northeast and southwest, respectively, as well as a region of overlap between the lobes near the dust continuum peak \citep{2001A&A...377..644B}. Our NOEMA data allowed us to explore the outflow morphology at a higher angular resolution.

\begin{figure}
    \centering
    \includegraphics[width=0.95\linewidth]{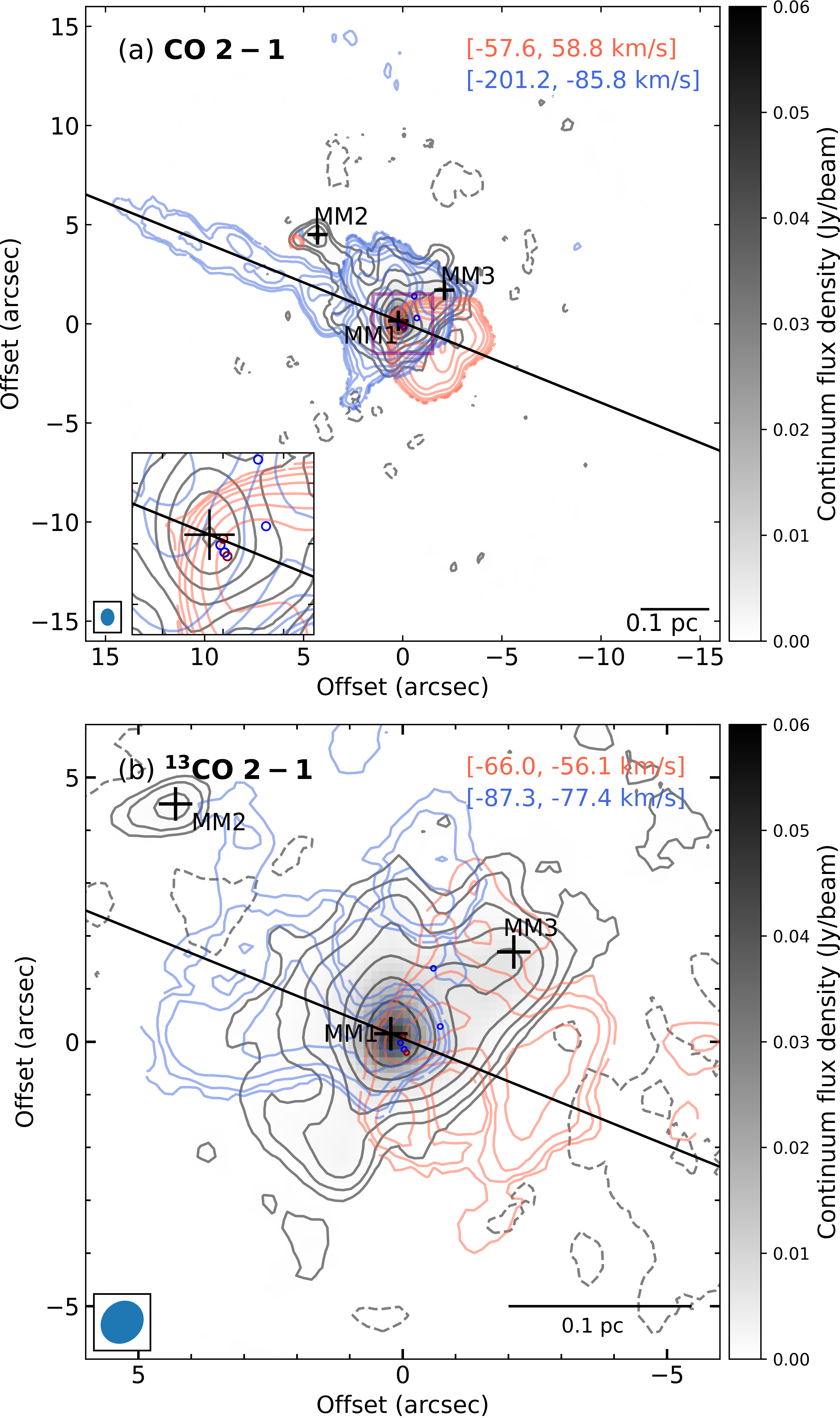}
    \caption{Morphology of the outflow traced by CO (a) and $^{13}$CO (b). In each panel, the blue and red contours represent the blue- and redshifted emission, respectively. The integration ranges indicated in the top-right corner. The blue and red contours start at 3$\sigma$ and then increase by a factor of two, where $\sigma$ = 34.1 (CO, red), 34.0 (CO, blue) in panel a, and 10.2 mJy~beam$^{-1}$~km~s$^{-1}$ for both red and blue contours in panel b. The solid line represents the outflow axis. The black contours show the 1.3 mm continuum emission as in Fig.~\ref{fig_dust}, and the symbols are the same as in that figure. In panel a, a zoomed-in view of size 1"$\times$1" around MM1 (within the purple square) is shown in the bottom-left corner to better display the positions of the H$_2$O masers. The maps are not corrected for primary-beam attenuation.}
    \label{fig_outflow}
\end{figure}

Figure~\ref{fig_outflow} presents the redshifted and blueshifted intensity maps of CO and $^{13}$CO integrated over the ranges listed in Table~\href{https://zenodo.org/records/17779973}{F.1} (see Appendix~\href{https://zenodo.org/records/17779973}{F} for details), overlaid on the 1.3 mm continuum emission. Blueshifted emission and redshifted emission are predominantly observed in the north-east and south-west, respectively. The outflow axis is oriented at a position angle of approximately 68$^{\circ}$ east from north, which we defined by the longest extent of the blueshifted emission and passing through the continuum peak position of MM1. Notably, there is a region of overlap between the blue- and redshifted emissions near the center of MM1, possibly suggesting a wide opening angle of the southeastern lobe along the line-of-sight, maybe similar to the large opening angle of the redshifted lobe seen in the plane of the sky. The outflow direction is consistent with that reported in \cite{2001A&A...377..644B}. Multiple H$_2$O maser features were reported around MM1 at angular separations of approximately 0.1–0.2", 0.8", and 1.5" from MM1, spanning a velocity range from $-$76 to $-$62 km~s$^{-1}$ \citep[see Fig. 2 and Table 3 in][]{2009ApJ...696.1981H}. Their proximity to MM1 and location within the overlap region of the blue- and redshifted emission suggest that they are associated with the outflow driven by MM1.

\begin{figure}[!htbp]
    \centering
    \includegraphics[width=0.42\textwidth]{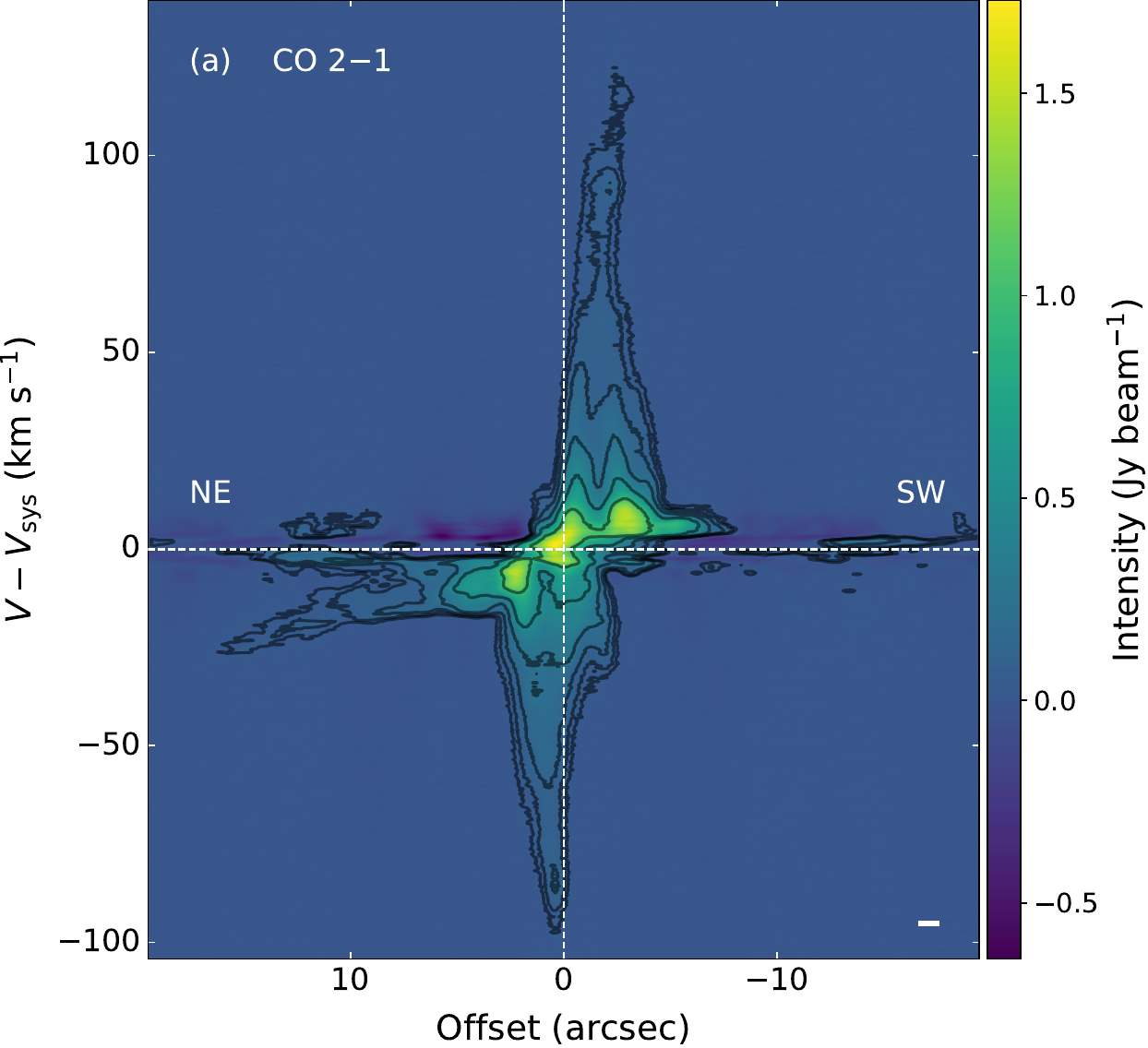}
    \includegraphics[width=0.9\linewidth]{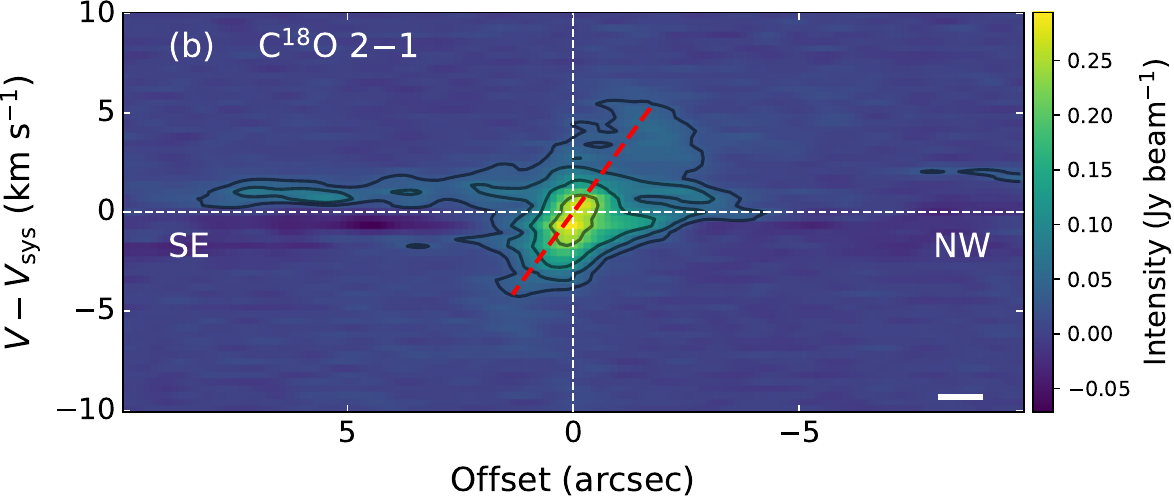}
    \caption{Position–velocity diagrams of CO (panel a) and C$^{18}$O (panel b) extracted along the outflow axis (northeast to southwest) defined in Fig.~\ref{fig_outflow} and perpendicularly to it (southeast to northwest), respectively. The y axis shows the velocity with respect to the systemic velocity of the source, displayed by the horizontal dashed line. The vertical white line represents the continuum peak position of G135.27 MM1. The horizontal bar in the bottom-left corner of each panel displays the beam size along the direction of the cut. The dashed red line in panel b highlights the velocity gradient traced by C$^{18}$O.}
    \label{fig_PV}
\end{figure}

Figure~\ref{fig_PV}a displays the position-velocity (PV) diagram of CO 2$-$1 taken along the outflow axis defined in Fig.~\ref{fig_outflow}. It reveals two components. The blueshifted component that extends up to about 16" from MM1 in Fig.~\ref{fig_PV} has projected velocities that reach $\sim$27 km~s$^{-1}$ at that distance. The second component is compact and has much higher red- and blueshifted projected velocities (up to about 100 km~s$^{-1}$). It is detected up to $\sim-$2.5" and $\sim$0.3" toward the southwest and northeast of MM1, respectively. We estimated the dynamical timescales of these two velocity components using the relation
\begin{equation}
t_{\mathrm{dyn}} = \frac{R_{\mathrm{proj}}}{V_{\mathrm{proj}}} \times \frac{\cos(i)}{\sin(i)}, 
\end{equation}
where \(R_{\mathrm{proj}}\) and \(V_{\mathrm{proj}}\) are the projected length and line of sight gas velocity, respectively, and \(i\) is the inclination angle of the outflow with respect to the line of sight. The blue- and redshifted lobes exhibit a distinct bipolar outflow structure with a small overlap near the center (see Fig.~\ref{fig_outflow}), indicating that the outflow axis is inclined at an intermediate angle to the line of sight. We thus adopted an inclination angle of 45$^{\circ}$. The dynamical timescales for the compact and extended velocity components are 88$-$580 yr and 17000$-$46000 yr, respectively (see Table~\ref{tab_dynamical_times}). These distinct timescales likely reflect different episodes of mass ejection: the compact component traces a recent, high-velocity outflow event occurring close to the protostar, whereas the extended component corresponds to slower, more dispersed material ejected during an earlier, sustained phase. The dynamical timescale of the extended component is comparable to those measured in Galactic high-mass protostellar objects \citep{2002A&A...383..892B, 2015MNRAS.453..645M}, as well as in the outer Galaxy source Sh2$-$283$-$1a SMM1 \citep[$D_{gc} = 15.7$ kpc,][]{2025ApJ...988..111I}.

\begin{table}[htbp]
\centering
\caption{Dynamical timescales of the outflow components assuming an inclination angle of 45$^\circ$.}
\label{tab_dynamical_times}
\begin{tabular}{cccc}
\hline\hline
Component & $R_{\mathrm{proj}}$ & $V_{\mathrm{proj}}$ & $t_{\mathrm{dyn}}$  \\
 & (") & (km~s$^{-1}$) & (yr) \\
\hline
Compact-red & 2.5 & 122 & \(5.8 \times 10^2\) \\
Compact-blue & 0.3 & 97 & 88 \\
Extended-red & 8 & 5  & \(4.6 \times 10^4\) \\
Extended-blue & 16 & 27  & \(1.7 \times 10^4\) \\
\hline
\end{tabular}
\end{table}

Compared to CO, C$^{18}$O usually traces circumstellar envelopes \citep[see, e.g.,][]{2025ApJ...989...56H}. Its PV diagram extracted perpendicularly to the CO outflow axis shows a clear velocity gradient from the bottom left to the top right, with redshifted and blueshifted emission on opposite sides of the protostar (Fig.~\ref{fig_PV}b). This indicates systematic motion of the gas, possibly due to rotation within the circumstellar envelope \citep[e.g.,][]{2012ApJ...748...16T,2025ApJ...989...56H}.

\subsubsection{Kinematics traced by CH$_3$OH and CH$_3$OCHO} \label{gradient}

\begin{figure}[!htbp]
    \centering
    \includegraphics[width=0.49\linewidth]{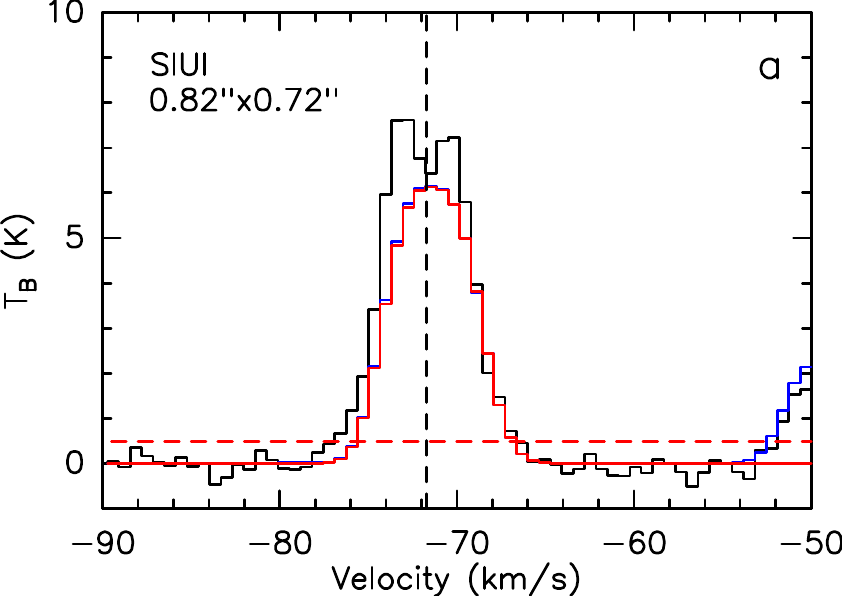}
    \includegraphics[width=0.49\linewidth]{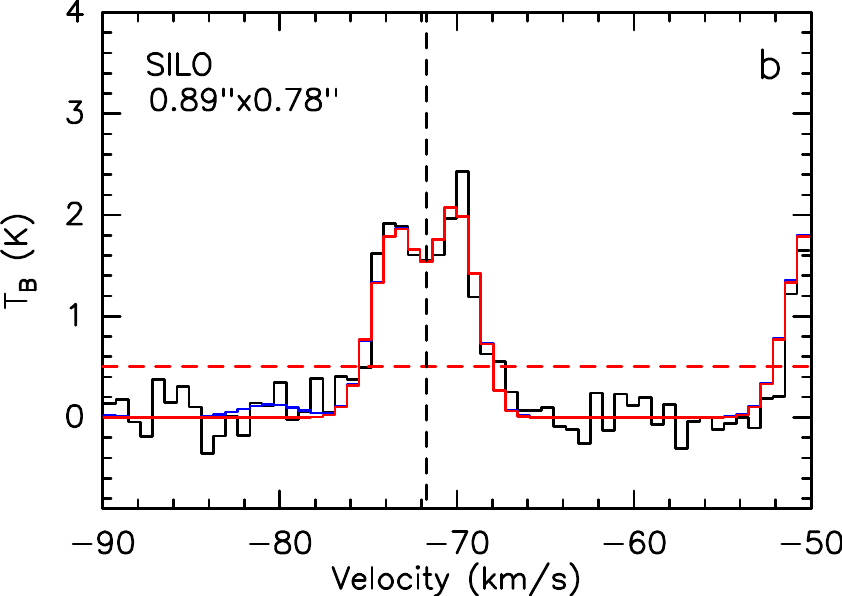}
    \caption{Spectra of a CH$_3$OH~$v_t=0$~($\nu_{\rm rest}$=232.783 GHz, $E_{\rm u}$=447 K) and b CH$_3$OCHO $v_t=0$ ($\nu_{\rm rest}$=218.281 GHz, $E_{\rm u}$=100 K) toward the G135.27 MM1 continuum peak position. In each panel, the velocity axis refers to the rest frequency of the transition. The vertical and horizontal dashed lines indicate the systemic velocity and the 3$\sigma$ noise level, respectively. The blue spectrum shows the LTE model incorporating all identified molecular species, while the red spectrum displays the LTE model including only CH$_3$OH (panel a) and CH$_3$OCHO (panel b). The observational setup and spectral window are labeled in the top-left corner of each panel, along with the corresponding HPBW. }
    \label{fig_spec_two_velocity}
\end{figure}

The spectra of CH$_3$OH and CH$_3$OCHO toward MM1 reveal two distinct velocity components, as illustrated in Fig.~\ref{fig_spec_two_velocity}, which shows one rotational transition of each molecule, selected for the absence of contamination from other species. The spectra of most other COMs can also be fitted with two velocity components (see details in Sect.~\ref{Method} and Table~\ref{tab_best_fit}). Small differences in the peak velocities of the two components reported in Table~\ref{tab_best_fit} among different molecules are smaller than the channel width and likely arise from the limited spectral resolution. In contrast, most simple molecules exhibit only one velocity component, indicating a complex kinematic structure within MM1. 

\begin{figure}[!htbp]
    \centering
    \includegraphics[width=\linewidth]{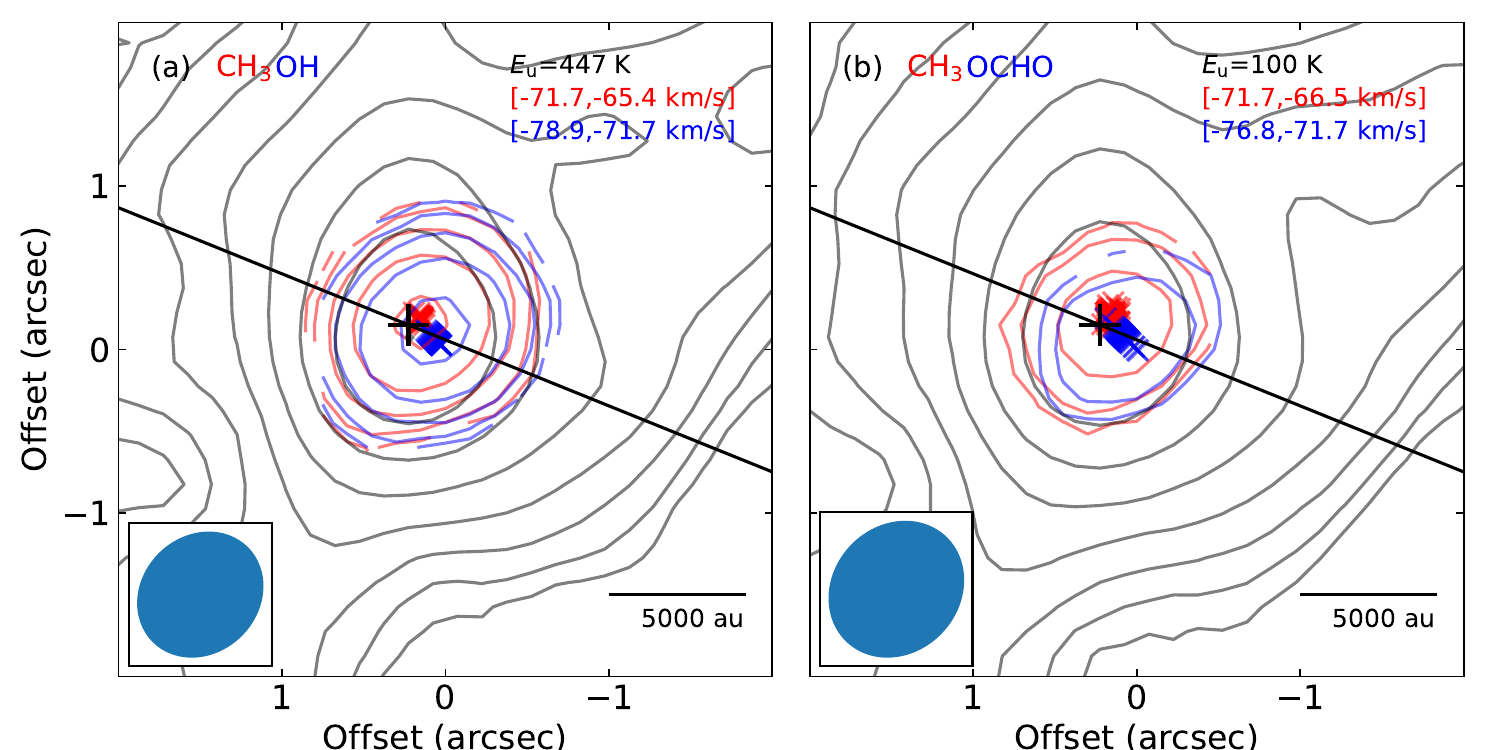}
    \caption{Blueshifted (blue contours) and redshifted (red contours) emission of the  transitions shown in Fig.~\ref{fig_spec_two_velocity}. Contours start from 3$\sigma$ and increase by a factor of two, where $\sigma$ = 10.9 (CH$_3$OH, red) and 11.6 (CH$_3$OH, blue) in panel a, and 9.9 (CH$_3$OCHO, red) and 9.8 mJy~beam$^{-1}$~km~s$^{-1}$ (CH$_3$OCHO, blue) in panel b. The redshifted and blueshifted emission peaks of all selected transitions of CH$_3$OH and CH$_3$OCHO are marked with red and blue crosses, respectively. The integration ranges are indicated in the top-right corner of each panel. The continuum beam size is indicated as an ellipse in the bottom-left corner of each panel. The black line represents the outflow axis traced by the large-scale CO emission. The black cross indicates the continuum peak position, while the black contours represent the continuum emission starting at the 3$\sigma$ level and increasing by factors of two.} 
    \label{fig_inverse}
\end{figure}

Figure~\ref{fig_inverse} presents maps of the red- and blueshifted emission of the transitions shown in Fig.~\ref{fig_spec_two_velocity}. Additionally, the redshifted emission peaks of all CH$_3$OH and CH$_3$OCHO transitions that have a peak signal-to-noise ratio higher than three and are not contaminated by other species are marked with red crosses, while their blueshifted emission peaks are indicated by blue crosses. The redshifted and blueshifted components of CH$_3$OH and CH$_3$OCHO do not peak at the same position. They are shifted along a direction roughly parallel to the outflow axis, with the redshifted component closer to MM1 and the blueshifted component further away toward the southwest ($\sim$ 0.1" or 600 au from MM1). This velocity gradient is also visible in the moment 1 maps shown in Fig.~\href{https://zenodo.org/records/17779973}{G.1}. Its direction is surprisingly opposite to the orientation of the outflow. This anti-alignment implies that the kinematics traced by CH$_3$OH and CH$_3$OCHO are not dominated by the outflow. A similar anti-alignment has also been detected on small scales ($\sim$ 300 au) in the NGC 1333 IRAS 4A2 protostar \citep{2024A&A...686L..13D}. We discuss the possible origin of this opposite velocity gradient in Sect.~\ref{opposite_grad}.

\subsubsection{A second outflow traced by SiO} \label{SiO}

\begin{figure}[!htbp]
    \centering
    \includegraphics[width=0.4\textwidth]{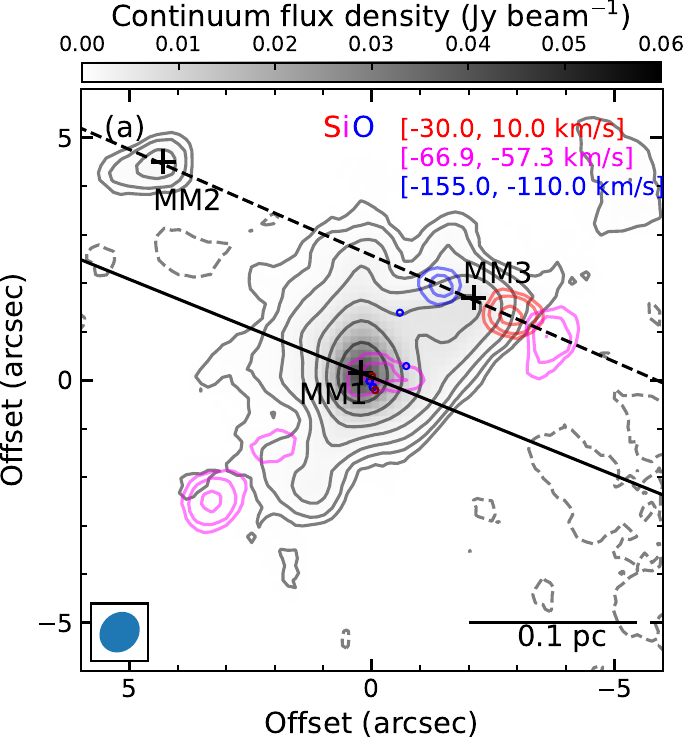}
    \includegraphics[width=0.45\textwidth]{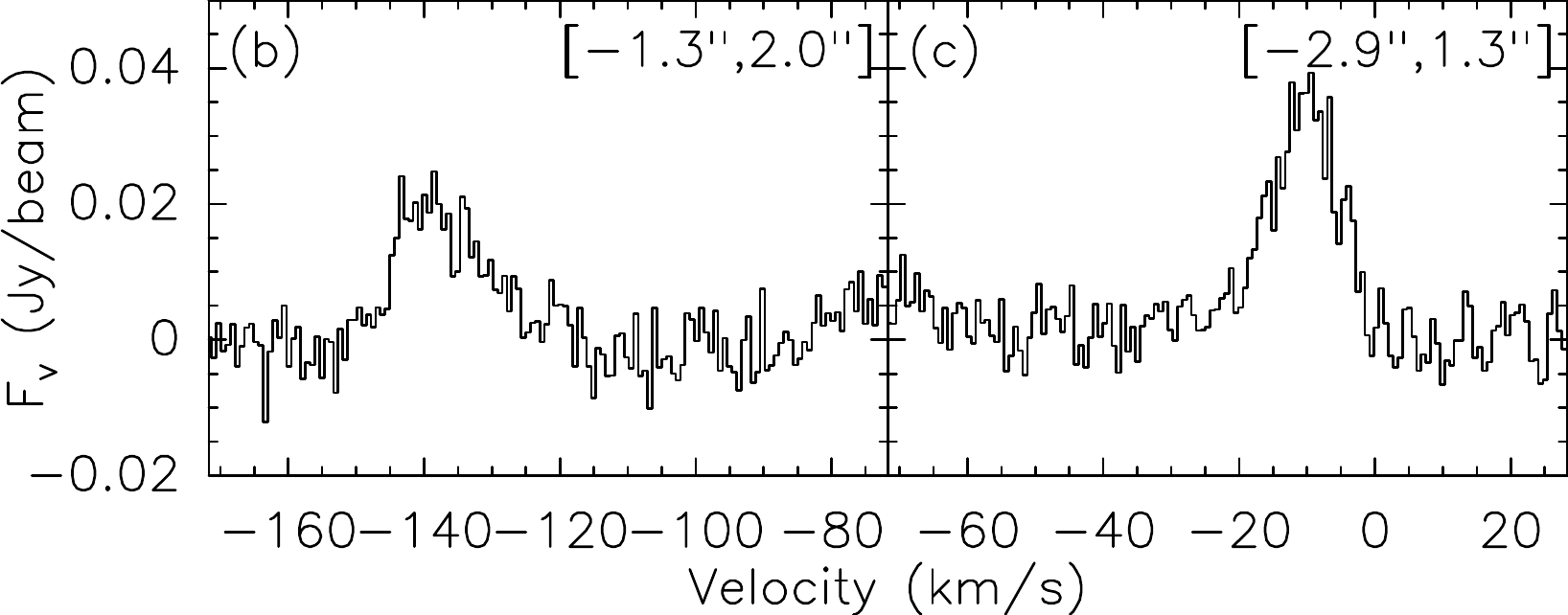} 
    \caption{ Panel a: Extremely high velocity blueshifted (blue contours) and redshifted (red contours) lobes of the outflow associated with MM3 traced by SiO 5$-$4. The lower-velocity redshifted emission is shown in magenta. The integration range for each component is indicated in the top-right corner. The contours start at 3$\sigma$ and increase by a factor of two, where $\sigma$ = 27.5 (red), 13.5 (magenta), and 29.1 (blue) mJy~beam$^{-1}$~km~s$^{-1}$. The black contour levels of the continuum emission and the symbols are the same as in Fig.~\ref{fig_dust}. The solid and dashed lines mark the axes of the outflow associated with MM1 and MM3, respectively. The SiO spectra extracted from the peak positions of the blue- and redshifted emission are shown in panels b and c respectively. The positional offsets relative to the phase center are indicated in the top-right corner of each panel. }
    \label{fig_SiO}
\end{figure}

Notably, SiO serves as another powerful tracer of protostellar outflows or jets because its abundance is significantly enhanced in shocked gas due to the sputtering of silicon from dust grains \citep[e.g.,][]{1997ApJ...487L..93B,2016ARA&A..54..491B,2025ApJ...979...17L}. We present the red- and blueshifted emission of SiO in Fig.~\ref{fig_SiO}a. Notably, an additional bipolar outflow centered on MM3 was identified through high-velocity SiO emission relative to the systemic velocity of MM3 ($-$70.7 km s$^{-1}$), which was determined from the CN and HC$_3$N spectra toward the MM3 position. Both molecules exhibit a local maximum in their emission at this position. The position angle of this outflow is about \(66^{\circ}\) east from north, which is similar to that of MM1.

Figures~\ref{fig_SiO}b-c display SiO (J = 5$-$4) spectra, extracted at the peaks of the blueshifted and redshifted emission, respectively. The SiO line shows a symmetric shift in the blue and red directions relative to the systemic velocity of MM3, with peak emission occurring at |$V_{\rm peak}$ - $V_{\rm sys}$| $\sim$ 64 km~s$^{-1}$. The spectral characteristics of the high-velocity SiO emission closely resemble those of extremely high velocity (EHV) outflows, which are typically associated with the earliest stages of protostellar evolution and exhibit velocities exceeding 30 km~s$^{-1}$ \citep[e.g.,][]{1996ARA&A..34..111B,2010A&A...522A..91T,2019A&A...632A.101T}. This strongly suggests that MM3 is a YSO driving an outflow. Part of the blueshifted and redshifted CO emission seen around MM3 in Fig.~\ref{fig_outflow} may result from this second outflow, which would in part explain the complex morphology of the CO emission.

At lower velocity, only redshifted SiO emission is detected, with no corresponding blueshifted emission (see magenta contours in Fig.~\ref{fig_SiO}a). Its distribution shows three distinct features. The central feature near MM1 and the top-right feature near MM3 are both elongated along their respective outflow directions, suggesting that they originate from these outflows. Surprisingly, the third feature consists of two spots located in the southeast of MM1, at a position angle of about 130 deg east from north and apparently not related to MM1's outflow. Spectra from the farthest spot (3.3",$-$2.5") reveal lines from other molecules, including CO, CS, SO, SO$_2$, and tentative detections of H$_2$CO and CH$_3$OH. The presence of sulfur-bearing species (SO and SO$_2$), as well as SiO, is commonly linked to shock activity \citep[e.g.,][]{2008A&A...482..549J,2021A&A...655A..65T,2022A&A...667A..20A}. CH$_3$OH is also efficiently released into the gas phase via sputtering and desorption during shock events \citep{2014MNRAS.440.1844S}. Therefore, the chemical composition of the stronger spot suggests that it results from some shock activity, the origin of which is still unknown.

\subsection{Spectral line modeling with Weeds} \label{weeds}

\subsubsection{Method of modeling}  \label{Method}

To identify the molecular lines in the observed spectra and derive column densities and excitation temperatures, we used the software Weeds \citep{2011A&A...526A..47M}. Weeds\footnote{\url{https://www.iram.fr/IRAMFR/GILDAS/doc/html/weedshtml/weeds.html}}, a part of the GILDAS/CLASS software, enables the generation of synthetic spectra under the LTE assumption. The calculation of synthetic spectra relies on spectroscopic data obtained from a local database. In this work, we used the same database as \citet{2025A&A...698A.143B}, which includes spectroscopic data from the Cologne Database for Molecular Spectroscopy \citep[CDMS\footnote{\url{https://cdms.astro.uni-koeln.de/}},][]{2001A&A...370L..49M,2005JMoSt.742..215M,2016JMoSp.327...95E}, the Jet Propulsion Laboratory (JPL\footnote{\url{https://spec.jpl.nasa.gov/}}) spectroscopy database \citep{1998JQSRT..60..883P}, and the Lille Spectroscopic Database \citep[LSD\footnote{\url{https://lsd.univ-lille.fr/}},][]{2025A&A...699A.348M}, as well as additional predictions contributed by spectroscopists. For CH$_2$DOH, we used a spectroscopic entry provided by L. Coudert that is based on \citet{2014JChPh.140f4307C} rather than the JPL entry used by \citet{2025A&A...698A.143B}, which has numerous issues at higher frequencies. For CO, H\textsubscript{2}S, HDO, DCN, and \textsuperscript{33}SO\textsubscript{2}, which were not modeled in \cite{2025A&A...698A.143B}, we used entries from the CDMS database, or from JPL in the case of HDO. 

Practically, the modeling in Weeds requires five input parameters for each species: line width, velocity offset relative to the systemic velocity, $V_{\rm sys}$, of the source, temperature, column density, and emission size. The line width and velocity offset relative to $V_{\rm sys}$ are determined by performing one-dimensional Gaussian fits to optically thin and unblended transitions of each molecule. The same line width and velocity offset are then adopted for all transitions of that molecule. Almost all simple species show a single velocity component. CH$_3$CCH and CH$_2$DOH clearly show only one velocity component, while most COMs exhibit two components toward the continuum peak position of MM1. For COMs with weak lines, it is difficult to determine how many velocity components they trace due to low signal-to-noise ratios. Therefore, we adopted two velocity components for most COMs but a single velocity component for most simple species, CH$_3$CCH, CH$_2$DOH, and COMs with weak lines. The emission sizes were derived as described in Sect.~\ref{size}. Rotational temperatures and column densities were initially derived using population diagrams (see Sect.~\ref{PD}) and subsequently refined through visual inspection.

We modeled each species separately and then added up the contributions of all detected species to create the final synthetic spectrum. This approach works well for optically thin lines that overlap in frequency or for lines (optically thick or thin) that come from different regions within the beam. However, it is not accurate for optically thick lines that overlap in frequency and originate from the same region or along the same line of sight. In such cases, the synthetic spectrum can overestimate the actual line flux, particularly for CH$_3$OH spectral lines with low upper-level energies in this work. The emission of vibrationally excited states of a given molecule were modeled independently of the vibrational ground state and the emission of isotopologs of a given molecule were also modeled separately.

\subsubsection{Sizes of the COM emission} \label{size}

\begin{figure*}[!htbp]
    \centering
    \includegraphics[width=0.24\textwidth]{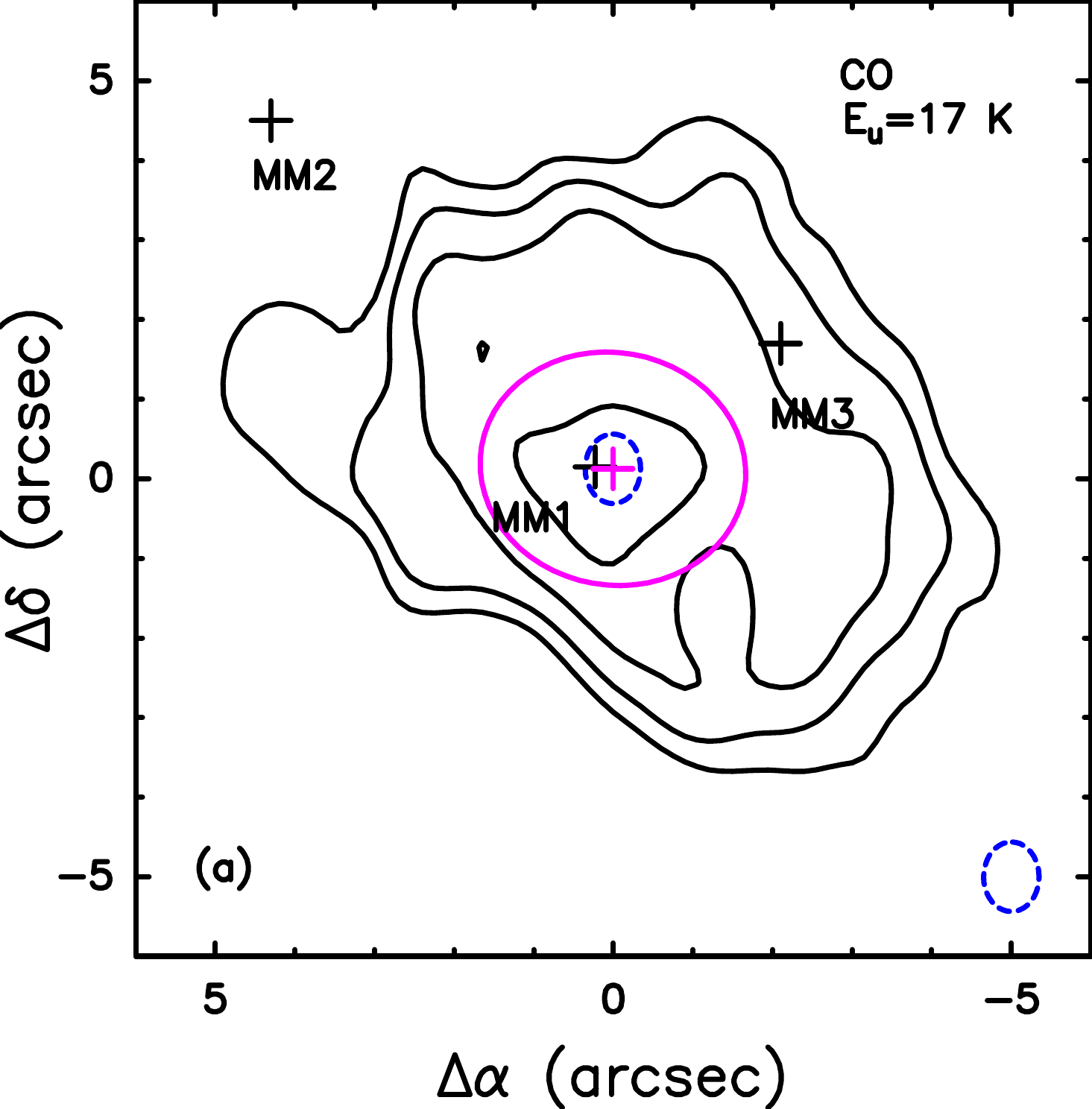}
    \includegraphics[width=0.24\textwidth]{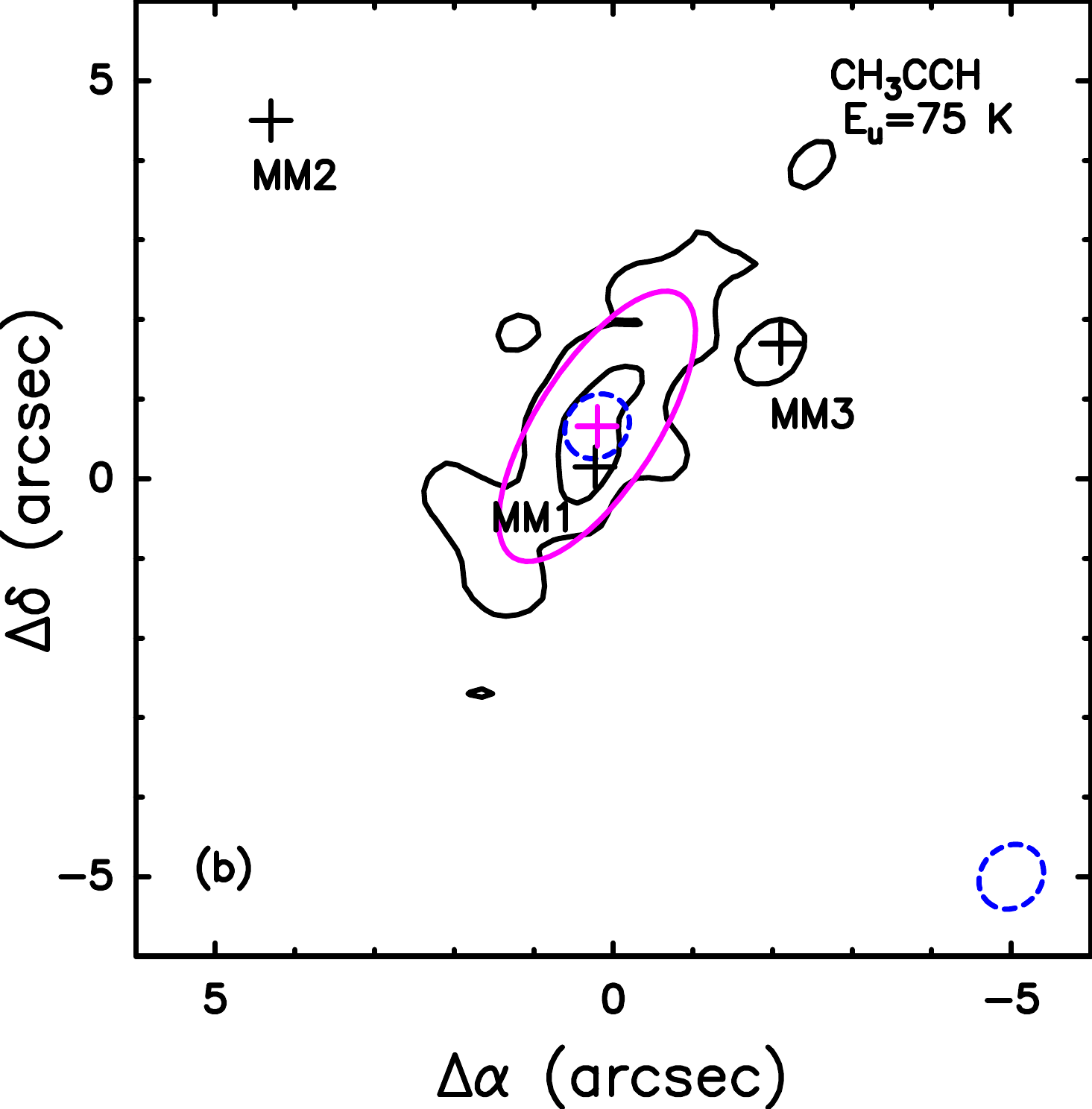}
    \includegraphics[width=0.24\textwidth]{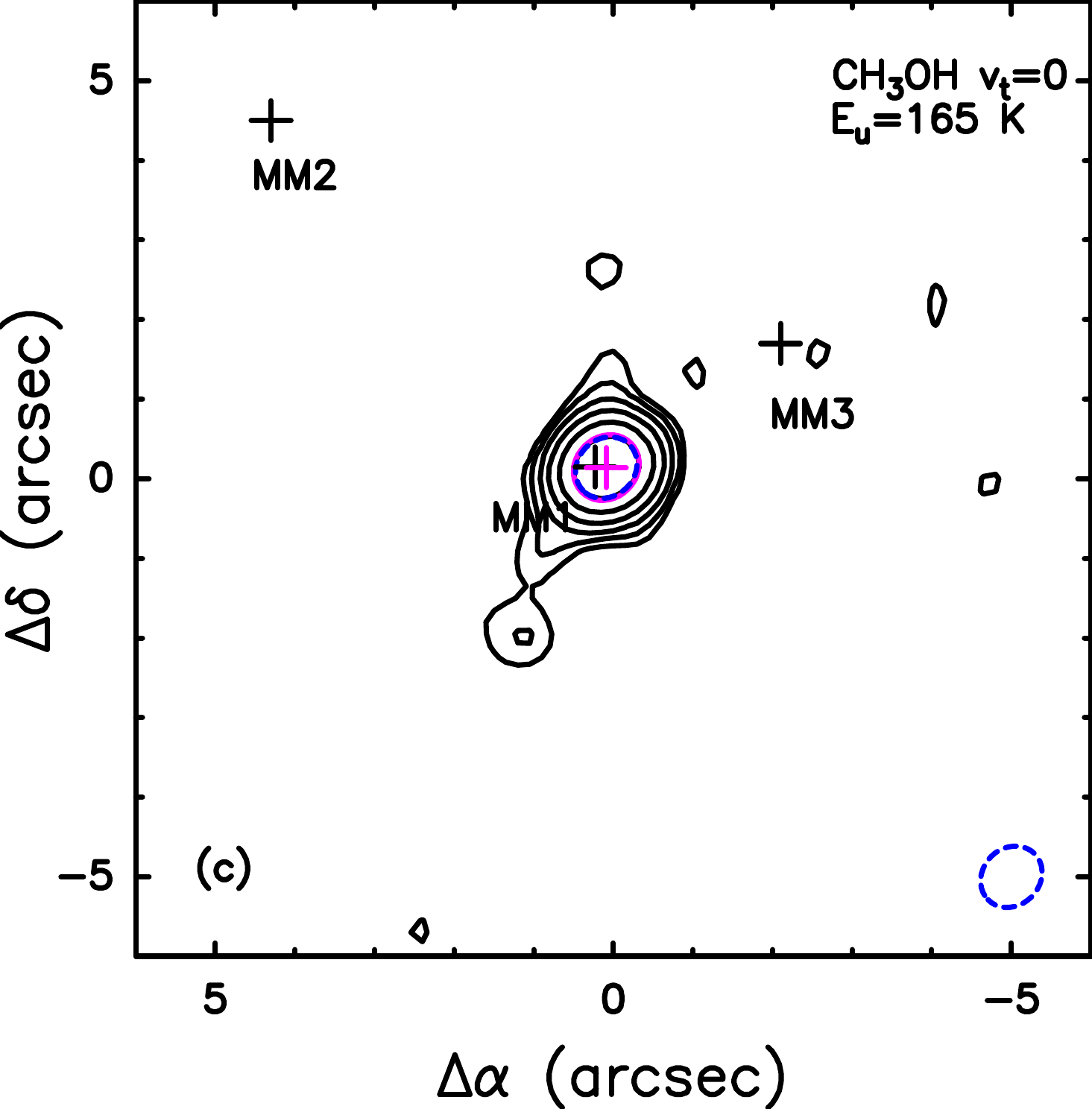}
    \includegraphics[width=0.24\textwidth]{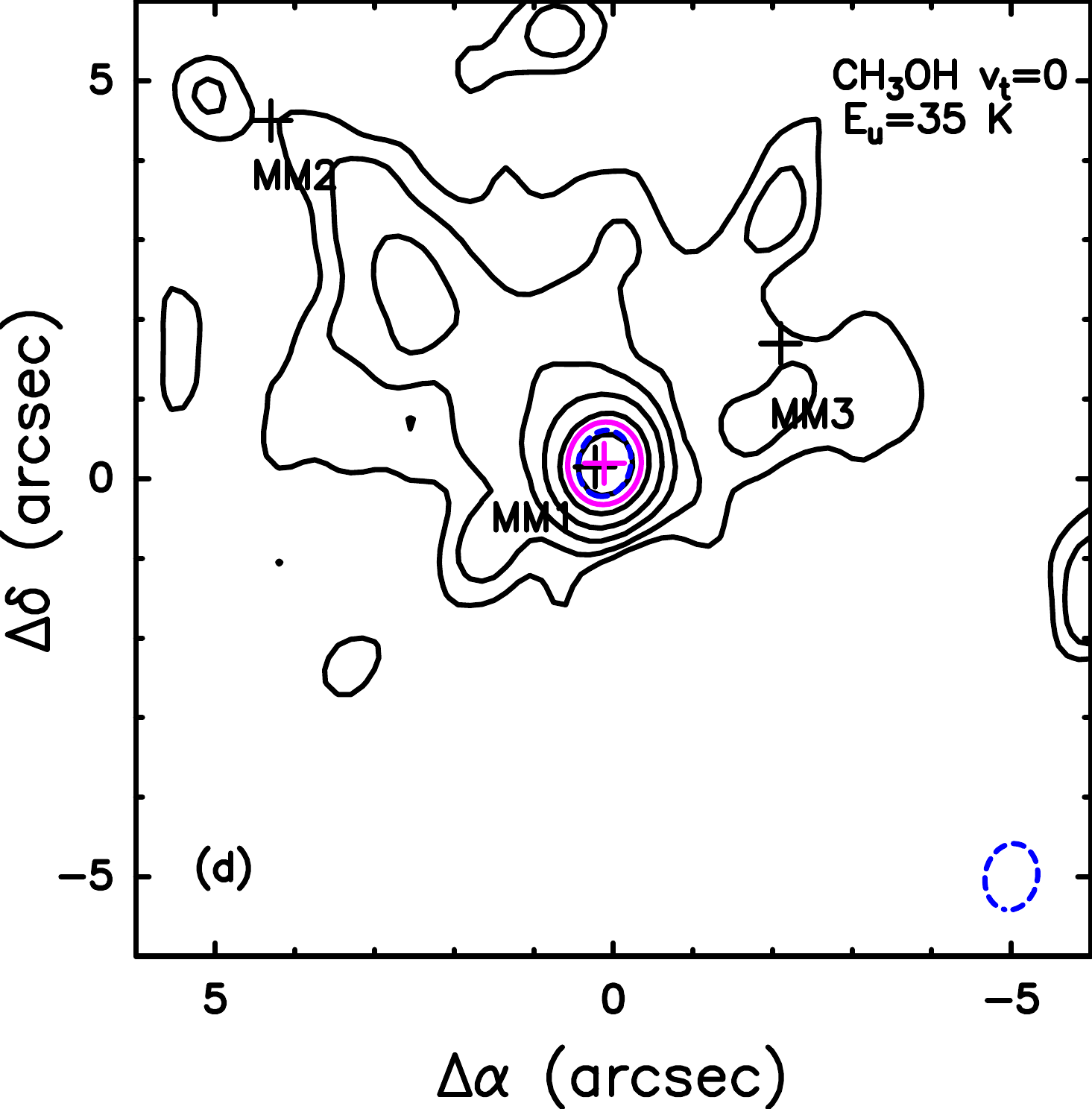}
    \caption{Integrated intensity maps of CO 2-1 (a), CH$_3$CCH 13$_0$-12$_0$ (b), CH$_3$OH $10_{2,9}-9_{3,6}$ (c), and CH$_3$OH $5_{0,5}-4_{0,4}$ (d). The values of the noise level, $\sigma$, are 80.7, 11.6, 14.3, and 16.5 mJy~beam$^{-1}$~km~s$^{-1}$, respectively. The contours start at 3$\sigma$ and increase by a factor of two at each step. In each panel, the molecule name and the upper-level energy of the line are shown in the upper right corner, the black crosses indicate the continuum peak positions, the magenta cross and ellipse represent the peak position and FWHM of the 2D Gaussian fit, and the blue dashed ellipses (at the continuum peak and in the bottom right corner) represent the beam size.}
    \label{fig_gauss}
\end{figure*}

To determine the emission size of each spectral line, we fit its integrated intensity map with a 2D Gaussian function. We selected only the spectral lines that have a peak temperature above 3$\sigma$ and are not contaminated by other species. Figures~\ref{fig_gauss}a, \ref{fig_gauss}b, and \ref{fig_gauss}c present examples of CO ($E_{\rm u} = 17$ K), CH$_3$CCH ($E_{\rm u}=75$ K), and CH$_3$OH ($E_{\rm u}=165$ K), respectively, representing three spatial scale groups across all molecules. The fitted emission sizes of CO and CH$_3$CCH are larger than the beam, indicating well-resolved emission, while CH$_3$OH shows an emission size only slightly exceeding the beam, suggesting marginally resolved emission. From the fitted sizes, we calculated the deconvolved size of the emission region for all molecules. For CH$_3$OH $v_{\rm t}=0$, the median deconvolved size is about 0.3", but the size decreases with increasing upper level energy, from $\sim$0.6" at low energy to $\sim$0.1" at $E_u$ $\sim$800 K (top panels of Fig.~\ref{fig_size}). However, the deconvolved sizes also depend on the line opacity, as illustrated in the bottom panels of Fig.~\ref{fig_size}. All mean sizes larger than 0.4" in Fig.~\ref{fig_size}c correspond to lines with an opacity higher than 1 (see  Fig.~\ref{fig_size}f). The Gaussian fitting may be biased by the line saturation, suggesting that the large sizes of the low-energy transitions are most likely overestimated.

Figure~\ref{fig_all_size} shows the deconvolved size for all detected molecules. In this work, we adopt a fixed deconvolved size of 0.3" for all COMs (except CH$_3$CCH and CH$_3$CHO) in our Weeds modeling, as most COMs are not fully resolved and exhibit similar morphology. CH$_3$CCH and CH$_3$CHO trace more extended structures, as was reported in other sources \citep[e.g.,][]{2015A&A...576A..45F,2022A&A...665A..96B}, and we used their median deconvolved sizes for the Weeds modeling (1.9" and 0.8", respectively).

\subsubsection{Population diagrams} \label{PD}

Population diagrams are constructed to determine the rotational temperature, $T_{\rm rot}$, and total column density, $N$, for each molecule under the LTE assumption. We used the following equation \citep{1999ApJ...517..209G}:

\begin{equation}
    ln\frac{N_u}{g_u} = ln\frac{8\pi k_{\rm B}\nu^2W}{hc^3A_{ul}bg_u} = ln\frac{N_{\rm tot}}{Q} - \frac{E_u}{k\rm_BT_{\rm rot}},
    \label{eq_PD}
\end{equation}
where $N_{\rm u}$ is the column density of the upper level and $b$~=~$\frac{\theta_{\rm s}^2}{\theta_{\rm s}^2 + HPBW^2}$ is the beam filling factor, with $\theta_{\rm s}$ as the emission size.

\begin{figure}[!htbp]
    \centering
    \includegraphics[width=0.5\textwidth]{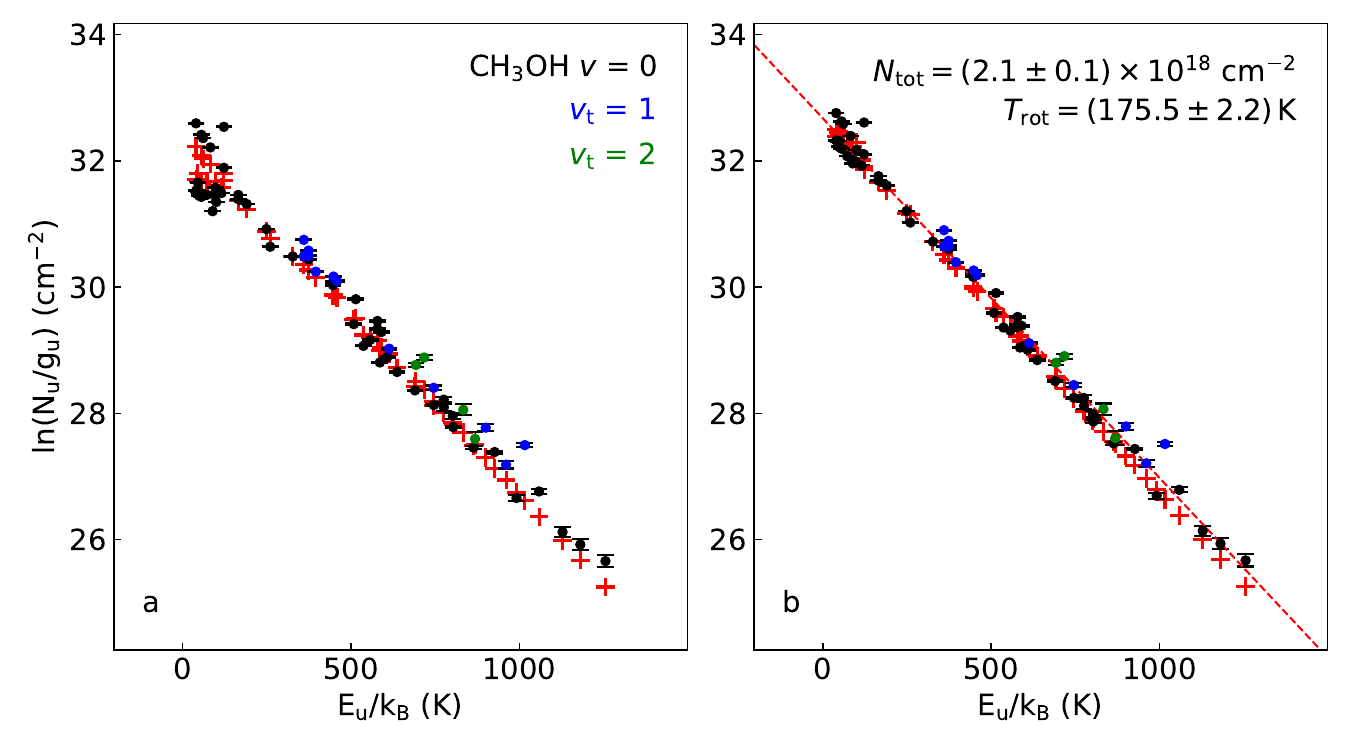}
    \caption{Panel a shows the observed data points (see the legend in the upper-right corner) and synthetic data points (in red). In panel b, the opacity correction was applied to both the observed and synthetic populations and the observed populations were corrected for contamination by other molecules. The red line represents a linear fit to the observed data points in linear-logarithmic space. The fit results are indicated in panel b.}
    \label{fig_PD}
\end{figure}

In our population diagram analysis, we selected transitions that were minimally contaminated by lines from other species and had a peak signal-to-noise ratio greater than 3$\sigma$. In cases where a transition was partially contaminated, the contribution from the contaminating species was subtracted from the measured integrated intensities, based on the full LTE model. Each population diagram was also corrected for optical depth following the method described in \cite{1999ApJ...517..209G}. The opacities are obtained from our best-fit Weeds model. However, the optical depth correction may not be sufficient to mitigate the effects for transitions with extremely high opacities, because gradients of excitation temperature along the line of sight are not accounted for in the population diagram method. Therefore, transitions with opacity higher than 2 were excluded from the population diagram analysis. Figure~\ref{fig_PD} illustrates the population diagram for CH$_3$OH, showing both the observed and modeled data points. In the right panel, deviations between the observed and synthesized data points are generally small after applying the two corrections, indicating that the chosen input parameters for the Weeds modeling are reliable and the LTE assumption is valid. The population diagrams of other molecules can be found in Fig.~\href{https://zenodo.org/records/17779973}{H.1}. Derived column densities and rotational temperatures from the population diagrams are summarized in Table~\ref{tab_best_fit}.

This approach is applicable to all detected COMs except CH$_3$COOH, C$_2$H$_5$CN, and a-(CH$_2$OH)$_2$, as the transitions of these three molecules are weak, with only a few exhibiting a signal-to-noise ratio of approximately three, and confined to a narrow range of upper level energies. This limitation makes it challenging to derive reliable rotational temperatures and column densities using the population diagram method for these three molecules. Here, for CH$_3$COOH and a-(CH$_2$OH)$_2$, we adopted the same temperature as derived for CH$_3$OH (175 K), while a temperature of 190 K from CH$_3$CN was used for C$_2$H$_5$CN. The population diagram method was also applied to simple molecules that exhibit multiple transitions spanning a wide range of upper level energies, including HDO, OCS, SO$_2$ and $^{34}$SO$_2$, H$_2$CS, H$_2$CO, HNCO, HC$_3$N, CH$_2$CO, and \textit{t}-HCOOH. For other simple molecules for which the population diagram method could not be used, we adopted the rotational temperature of CH$_3$CCH because CH$_3$CCH and these simple molecules trace similarly extended structures. For several isotopologs, such as H$_2^{13}$CO, we assumed the same temperature as derived for their main isotopolog.

\subsubsection{Weeds modeling result} \label{weeds_result}

Based on our Weeds modeling, we systematically identified the molecular emission lines detected toward the continuum peak position of MM1. Figure~\ref{fig_all_spec} presents the observed spectrum in black, zoomed-in along the y axis in order to highlight the weak lines, overlaid with the Weeds synthetic spectrum of all identified molecules in dark red. The unzoomed spectrum is shown in Fig.~\ref{fig_all_spec_unzoom}. In total, we identified 485 spectral lines with signal-to-noise ratios exceeding 3, from 28 molecules, several of their less abundant $^{13}$C, $^{18}$O, $^{17}$O, $^{34}$S, and $^{33}$S isotopologs, as well as one deuterated COM (CH$_2$DOH). The detected molecules span a broad range of chemical complexity, from simple diatomic species to COMs (12 in total) containing up to ten atoms. Additionally, three molecules ($^{13}$C$^{34}$S, OC$^{33}$S, and CH$_2$(OH)CHO) are considered tentatively detected due to their low signal-to-noise ratios or partial line blending. The list of detected molecules is presented in Table~\ref{molecules_MM1}. The input parameters adopted for the Weeds model of each species as well as the number of detected lines are detailed in Table~\ref{tab_best_fit}. Numerous lines with a peak signal-to-noise ratio higher than 3 remain unidentified in the spectrum of MM1 (see Table~\href{https://zenodo.org/records/17779973}{J.1}). We consider it likely that most of these unidentified lines correspond to rotational transitions in vibrationally excited states of molecules whose ground state and some lower vibrational states are included in our model, but for which spectroscopic predictions of higher states are still missing in the spectroscopic databases.

\begin{table*}[]
    \centering
    \caption{Molecules identified toward G135.27 MM1.}
    \setlength{\tabcolsep}{3pt}
    \begin{tabular}{l l l l l l l l l l l l l}
    \hline \hline
    \noalign{\vskip 2pt}
    \multicolumn{3}{c}{2 atoms} &  \multicolumn{2}{c}{3 atoms} & 4 atoms & 5 atoms & 6 atoms & 7 atoms & 8 atoms & 9 atoms & 10 atoms  \\
    \hline
    \noalign{\vskip 4pt}
    CN & SO & CS & H$_2$S & OC$^{34}$S & H$_2$CS & CH$_2$CO & CH$_3$OH & CH$_3$CHO & CH$_3$COOH & CH$_3$OCH$_3$ & CH$_3$C(O)CH$_3$ \\
    CO & $^{34}$SO & $^{13}$CS & HDO & OC$^{33}$S$^{(a)}$ & H$_2$CO & HC$_3$N & $^{13}$CH$_3$OH & CH$_3$CCH & CH$_3$OCHO & C$_2$H$_5$OH & a-(CH$_2$OH)$_2$ \\
    $^{13}$CO & $^{33}$SO & C$^{34}$S & DCN & SO$_2$ & H$_2$$^{13}$CO & t-HCOOH & CH$_2$DOH &  & CH$_2$(OH)CHO$^{(a)}$  & C$_2$H$_5$CN  &  \\ 
    C$^{18}$O & S$^{18}$O & C$^{33}$S & OCS & $^{34}$SO$_2$ & HNCO &  & CH$_3$CN &  &   &  &  \\
    C$^{17}$O & SiO & $^{13}$C$^{34}$S$^{(a)}$ & O$^{13}$CS & $^{33}$SO$_2$ & & & NH$_2$CHO &  &  &  & \\
    \hline
    \end{tabular}
    \setlength{\tabcolsep}{6pt} 
    \vspace{-8pt}
    \tablefoot{\textsuperscript{(a)} tentative detection.}
    \label{molecules_MM1}
\end{table*}

We applied the same Weeds modeling procedure to identify the spectral lines detected toward the continuum peak positions of MM2 and MM3. Toward MM2, only eight species were detected, while 20 species were identified toward MM3. All identified molecules are simple species, except for CH$_3$OH detected toward both MM2 and MM3 and CH$_3$CCH toward MM3 (see Tables~\href{https://zenodo.org/records/17779973}{I.1}~and~\href{https://zenodo.org/records/17779973}{I.2}). The input parameters of the Weeds models of MM2 and MM3 are detailed in Tables~\ref{tab_best_fit_MM2}~and~\ref{tab_best_fit_MM3}, respectively. Although CH$_3$OH is detected toward MM2 and MM3, its emission does not peak at their continuum peak positions, as can be seen in Fig.~\ref{fig_gauss}d. The emission of low-energy transitions of methanol traces extended structures across the cores, suggesting that it arises from larger scales rather than being directly associated with MM2 and MM3 themselves. In contrast, the CH$_3$CCH emission has a local peak close to the continuum position of MM3 (see Fig.~\ref{fig_gauss}b), indicating that it traces the dense core material directly associated with MM3. 

We note that uncertainties in the derived $N_{\rm tot}$ arise from the adopted parameters in the Weeds modeling. They were estimated by varying one parameter at a time while keeping the other four input parameters fixed. Under this assumption, uncertainties are below 10$\%$ for both $N_{\rm tot}$ and the line width, and about 20$\%$ for $T_{\rm rot}$. The emission sizes of individual molecules range from 0.2" to 0.33", corresponding to an uncertainty of about 40\% for each species (see Fig.~\ref{fig_all_size}). Adopting a fixed value of 0.3" may introduce an uncertainty of up to a factor of two in $N_{\rm tot}$, which dominates the total error budget. Since both the COMs and CH$_3$OH were analyzed using the same method and beam size and exhibit similar spatial distributions, systematic uncertainties related to the line width and flux calibration should cancel out in the abundance ratios. For species with comparable emission sizes, beam-dilution effects are also reduced, whereas smaller sizes may still introduce uncertainties of up to a factor of two. Consequently, the column density ratios are less affected by systematic uncertainties than the column densities themselves.

\subsection{Comparison to galactic hot cores} \label{MW}

\begin{figure}[!htbp]
    \centering
    \includegraphics[width=0.43\textwidth]{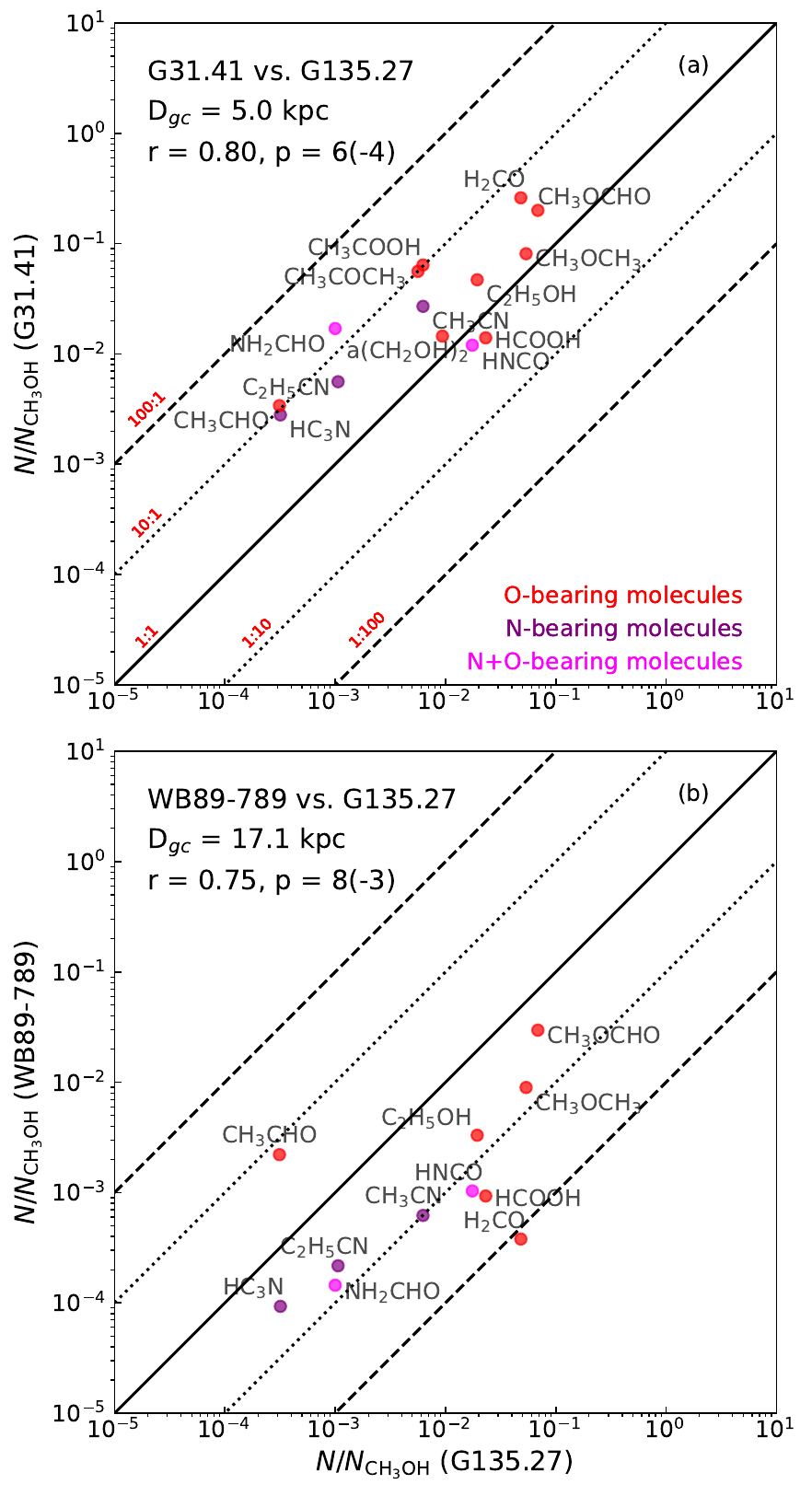}
    \caption{Correlation plots of column densities normalized to methanol in G135.27 MM1 compared to G31.41 (panel a) and WB89-789 (panel b). Oxygen-, nitrogen-, and nitrogen+oxygen-bearing molecules are marked in red, purple, and magenta, respectively. Dashed and dotted lines indicate deviations by factors of 100 and 10, respectively. The Pearson correlation coefficient ($r$) and $P$-value for each comparison are shown in the top-left corner of each panel.}
    \label{fig_com_MW}
\end{figure}

Figure~\ref{fig_com_MW} presents a comparison of the abundances of COMs relative to CH$_3$OH between G135.27 MM1 and other galactic hot cores. The reference species selected for this comparison is CH$_3$OH, as it is considered a parent molecule in the synthesis of more COMs \citep[e.g.,][]{2009ARA&A..47..427H}. We have selected two hot cores the chemical composition of which has been studied in detail in the past: G31.41+0.31 (hereafter G31.41) and WB89-789 SMM1 (hereafter WB89-789). G31.41 is located in the inner Galaxy at a Galactocentric distance of 5.0 kpc, as determined via trigonometric parallax measurements from the BeSSeL Survey \citep{2019A&A...632A.123I}. Its chemical composition was derived in the frame of the GUAPOS project \citep{2020A&A...644A..84M,2023A&A...677A..15M,2021A&A...653A.129C,2022A&A...658A.150G,2024MNRAS.529.3244L}. We adopted a CH$_3$OH column density of $1.0 \times 10^{19}$ cm$^{-2}$ for G31.41, derived from the best-fit model of high-energy ($\sim$300$-$800 K) CH$_3$OH $v_{\rm t}=1$ transitions \citep[see Table 2 in][]{2023A&A...677A..15M}. The chemical composition of WB89-789, the first hot core identified in the outer Galaxy, was published in \cite{2021ApJ...922..206S}. We adopted a Galactocentric distance of 17.1 kpc, based on the BeSSeL Distance Calculator V2\footnote{\url{http://bessel.vlbi-astrometry.org/node/378}}, rather than the 19.0 kpc used in \cite{2021ApJ...922..206S}, which was derived from optical spectroscopy of a K3 III star \citep{2007A&A...464..909B}. Based on the Galactic gradient reported by \citet{2023A&A...670A..98Y}, we assumed a $^{12}$C/$^{13}$C isotopic ratio of 100 at 17.1 kpc to estimate the CH$_3$OH column density from that of its isotopolog $^{13}$CH$_3$OH, rather than the 
value of 150 at 19 kpc adopted by \citet{2021ApJ...922..206S}. We obtained a CH$_3$OH column density of $2.9 \times 10^{17}$ cm$^{-2}$ for WB89-789. 

Overall, the COM abundances (relative to methanol) in G135.27 MM1 show a good correlation with those measured in G31.41 and WB89-789, although a noticeable scatter is present. To quantify this correlation, we calculated the Pearson correlation coefficients between the relative abundances of COMs in G135.27 MM1 and those in the two sources, obtaining $r = 0.80$ and $r = 0.75$ for G31.41 and WB89-789, respectively, both with $p$-values less than 0.01. Considering that the typical uncertainty in the abundance ratios is on the order of 20--30\% (corresponding to $\sim$0.1 dex in logarithmic scale, and at most a factor of two in some cases), the correlations remain statistically significant within the uncertainties. Moreover, our analysis reveals no systematic difference between N-bearing and O-bearing COMs among these three hot cores.

The abundances relative to methanol of most COMs in G135.27 MM1 are lower than those in G31.41 but higher than those in WB89-789, suggesting that G135.27 MM1 occupies an intermediate position in terms of chemical richness. Given that G135.27 MM1 is located between G31.41 in the inner Galaxy and WB89-789 in the outer Galaxy, its metallicity is likely intermediate as well. This trend may reflect the influence of metallicity on the formation and abundance of COMs. 

One exception to this trend is CH$_3$CHO, which displays a different pattern: its abundance relative to methanol is similar in G31.41 and WB89-789, both higher than in G135.27 MM1. This is likely because CH$_3$CHO forms significantly ($\sim$ 40$\%$) via gas-phase reactions ($\mathrm{C_2H_5} + \mathrm{O} \rightarrow \mathrm{CH_3CHO} + \mathrm{H}$) during the warm-up stage \citep{2022ApJS..259....1G}. This gas-phase formation pathway makes the abundance of CH$_3$CHO less directly dependent on the initial ice composition and metallicity, and more sensitive to the evolving physical conditions and chemical kinetics of the protostellar environment.

\subsection{Comparison to extragalactic hot cores}

Apart from those discovered in the Milky Way, hot cores have also been identified in nearby galaxies, namely the LMC and SMC. The metallicity of the LMC is reported to be lower than the solar value by a factor of two to three \citep[e.g.,][]{2016MNRAS.455.1855C,2021MNRAS.507.4752C}. \citet{2021ApJ...922..206S} reported the first extragalactic hot core in the LMC, ST11, identified through the detection of a hot SO$_2$ component (190 K), but no COMs were detected, not even CH$_3$OH. Subsequent studies have revealed additional hot cores in the LMC, including N113-A1 and N113-B3 \citep{2018ApJ...853L..19S}, ST16 \citep{2020ApJ...891..164S}, N105-2A and N105-2B \citep{2022ApJ...931..102S}, and N132-14A \citep{2024A&A...688A...3H}. In the SMC, which exhibits an even lower metallicity than the LMC, with values estimated to be 4 to 10 times lower than the solar metallicity \citep[e.g.,][]{2018MNRAS.475.4279C}, \citet{2023ApJ...946L..41S} reported the detection of two hot cores, S07 and S09. Similar to ST11 in the LMC, the two hot cores in the SMC are identified through the presence of hot SO$_2$ emission (171 K for S07 and 146 K for S09). In these sources, however, CH$_3$OH primarily traces colder gas at temperatures of about 40 K, distributed over extended regions of size 0.2-0.3 pc, rather than the compact hot core material. 

\begin{figure}[!htbp]
    \centering
    \includegraphics[width=0.5\textwidth]{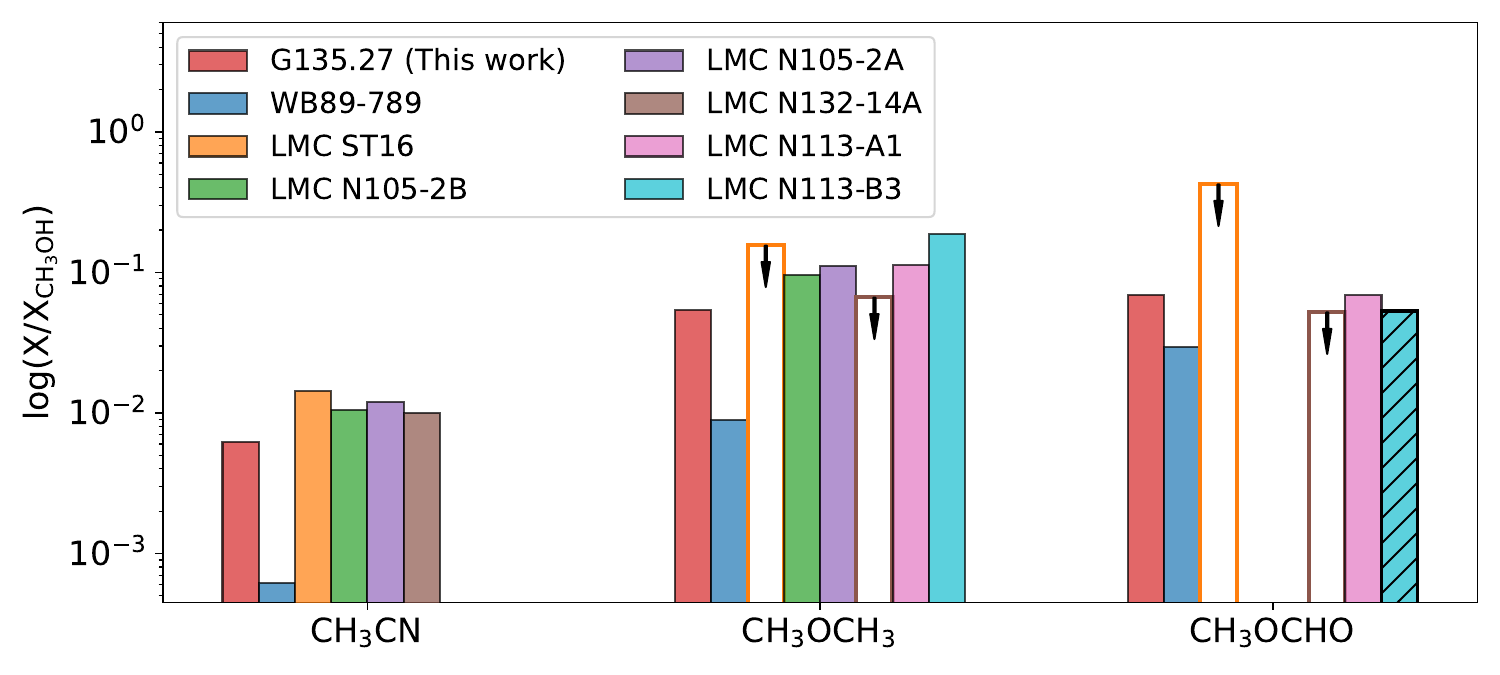}
    \caption{ Comparison of the abundances of COMs relative to CH$_3$OH between G135.27 MM1, WB89-789, and hot cores in the LMC. The LMC hot cores are ST16 \citep{2020ApJ...891..164S}, N113-A1 and N113-B3 \citep{2018ApJ...853L..19S}, N105-2A and N105-2B \citep{2022ApJ...931..102S}, and N132-14A \citep{2024A&A...688A...3H}. Unfilled bars with an arrow indicate upper limits, while hatched bars denote tentative detections. }
    \label{fig_comparison_LMC}
\end{figure}

Figure~\ref{fig_comparison_LMC} presents a comparison of the abundances of COMs (CH$_3$CN, CH$_3$OCH$_3$, and CH$_3$OCHO) relative to CH$_3$OH between G135.27 MM1, WB89-789, and the LMC hot cores where hot methanol has been detected. Overall, the abundances of CH$_3$CN, CH$_3$OCH$_3$, and CH$_3$OCHO relative to CH$_3$OH show no clear trends among hot cores in these low-metallicity environments, with the exception of WB89-789. The values in G135.27 MM1 are comparable to those observed in the LMC hot cores. This similarity is reasonable given that the expected metallicity of G135.27 ($\sim$0.5 $Z_\odot$) is close to that of the LMC \citep[$\sim$0.4 $Z_\odot$,][]{2021MNRAS.507.4752C}. For WB89-789, the abundances of all three COMs relative to CH$_3$OH are markedly lower than those observed in other sources. The chemical composition of WB89-789 was derived by \citet{2021ApJ...922..206S} assuming a beam fill factor of 1. However, the emission of the COMs identified by \citet{2021ApJ...922..206S} is barely resolved, with fitted sizes (not deconvolved) that are close to the beam size. This suggests that much smaller sizes should be assumed to model the COM emission and derive their proper column density in WB89-789. Reducing the emission size results in higher column densities, but also higher opacities, which may have an impact on the balance between optically thin and thick lines specific to each molecule. A reanalysis of the ALMA data using smaller sizes is needed before we can draw any firm conclusion concerning WB89-789.

Naturally, we needed to consider whether the difference in spatial scales could lead to variations in the derived abundances. The COM emission extends over $\sim$5000 au in G135.27 MM1, and $\sim$0.1 pc in the LMC hot cores. \citet{2021ApJ...922..206S} determined COM abundances in WB89-789 at both 5000 au and 0.1 pc scales and found that, while absolute abundances vary somewhat with scale, the ratios of COMs to CH$_3$OH for these three species remain similar within a factor of two. This suggests that the difference in spatial scales between the observations of G135.27 and the LMC hot cores does not significantly affect the comparison shown in Fig.~\ref{fig_comparison_LMC}, and that the relative abundances can be meaningfully compared despite the different scales. Nevertheless, variations in environmental parameters, such as cosmic-ray ionization rates and UV radiation fields, may also influence the chemistry and contribute to differences between the outer Galaxy and the LMC.

\section{Astrochemical modeling} \label{modeling}

To interpret the COM abundances detected toward the hot core in G135.27 MM1, we simulate hot-core chemistry under both local- and outer-Galaxy conditions using the astrochemical code MAGICKAL \citep[Model for Astrophysical Gas and Ice Chemical Kinetics and Layering,][]{2022ApJS..259....1G}. MAGICKAL is a three-phase (gas phase, grain/ice-surface and bulk-ice mantle) chemical kinetics code with an extensive chemical network intended for the study of COM chemistry. Along with standard gas-phase reactions and diffusive grain-surface chemistry, the model includes the nondiffusive surface and ice-mantle chemical treatment introduced by \cite{2020ApJS..249...26J} and further developed by \cite{2022ApJS..259....1G}. Here, MAGICKAL is applied in single-point mode, using a two-stage physical treatment that has been used in various past studies \citep[e.g.,][]{2008ApJ...682..283G,2019A&A...628A..10B,2022ApJS..259....1G,2024ApJ...972..146S,2025A&A...698A.143B}. Stage 1 consists of an isothermal collapse ($T_{\mathrm{gas}} = 10$~K) from initial gas density $n_{\mathrm{H}} = 3 \times 10^3$~cm$^{-3}$, reaching a final value of $2 \times 10^8$~cm$^{-3}$ in $\sim$1~Myr. During the collapse, the visual extinction evolves in proportion with $n_{\mathrm{H}}^{2/3}$, starting at a value $A_{\mathrm{V,init}} = 2$ or 3 mag and taking a maximum value of 500 mag. While the gas temperature is held steady, the dust temperature evolves with visual extinction according to the expression of \cite{2017A&A...604A..58H}, resulting in an initial value $T_{\mathrm{dust}} \simeq 12.7$~K or $11.8$~K, and final value of $5.5$~K. In stage 2, the gas density is held steady, while the gas and dust temperatures rise in concert to a final value of 400~K. As in past work, three timescales for the warm up stage are considered in this model: {\em fast} ($5 \times 10^{4}$~yr), {\em medium} ($2 \times 10^{5}$~yr) and {\em slow} ($1 \times 10^{6}$~yr). The same canonical interstellar cosmic-ray ionization rate of $\zeta_0 = 1.3 \times 10^{-17}$~s$^{-1}$ is used in all models, throughout stages 1 and 2.

The adoption of two $A_{\mathrm{V,init}}$ values in stage 1 is intended to allow for a modest exploration of the effects of external UV while the core is still relatively diffuse; as noted by \cite{2022ApJS..259....1G}, certain COMs are more strongly affected by UV-driven processing of the early ices formed on grains, making the choice of initial visual extinction potentially important for those species. Furthermore, it is unclear how the initial visual extinction profile in a nascent hot core may vary between local- and outer-Galaxy conditions; however, the higher gas-to-dust ratio in the outer Galaxy (see below) would suggest lower extinction values than within the solar circle, all else being equal.

Using the same physical setups in each, the models are run under chemical conditions appropriate to the local galactocentric radius (8.2 kpc) and that of G135.27 (13 kpc). In practice, this means the adoption of different initial elemental abundances of C, N, and O, and the gas-to-dust ratio by mass. Local-Galaxy initial abundances are $n(\rm C^{+}) = 1.40 \times 10^{-4}$ $n_{\mathrm{H}}$, $n($N$) = 2.14 \times 10^{-5} n_{\mathrm{H}}$, and $n(\rm O) = 3.20 \times 10^{-4} n_{\mathrm{H}}$ \citep{2022ApJ...935..133J}. These are scaled according to \cite{2011AJ....142..136L}, giving values at 13~kpc of $n($C$^{+}) = 5.47 \times 10^{-5} n_{\mathrm{H}}$, $n($N$) = 1.20 \times 10^{-5} n_{\mathrm{H}}$, and $n(\rm O) = 1.66 \times 10^{-4} n_{\mathrm{H}}$. These correspond to the initial forms and values used for these elements at the beginning of stage 1 in the chemical models in the low-metallicity case. We note that C, N, and O abundances are also available from the studies by \cite{2020MNRAS.496.1051A,2022MNRAS.510.4436M}, who derive total elemental abundances in H\,\textsc{ii} regions from ionic abundances, using ionization correction factors. \cite{2011AJ....142..136L}, on the other hand, determine elemental abundances from photospheric absorption lines of Cepheids, providing direct total abundances without ionization corrections. This, combined with the larger sample size provided by Luck and Lambert ($\sim$400 versus fewer than 50 in the other studies), made that study a more favorable choice for our modeling purposes.

Under local conditions, a gas-to-dust ratio by mass of 100 is adopted, which is scaled to 303 at 13~kpc for G135.27 (see Sect.~\ref{H2}). Making the usual assumption of a representative grain radius of 0.1~$\mu$m, this leads to a gas-to-dust ratio by number of $7.57 \times 10^{11}$ locally, or $2.88 \times 10^{12}$ at 13~kpc.

The chemical network used here is an updated version of that used by \cite{2022ApJS..259....1G}, and includes a more extensive gas-phase radical chemistry, as well as the adoption of more accurate quantum tunneling treatments for certain surface reactions. The network is described in more detail by Bonfand et al. (in prep.).

\subsection{Chemical model predictions for low- versus normal-metallicity} \label{Model_prediction}

\begin{figure*}[!htbp]
    \centering
    \includegraphics[width=0.94\textwidth]{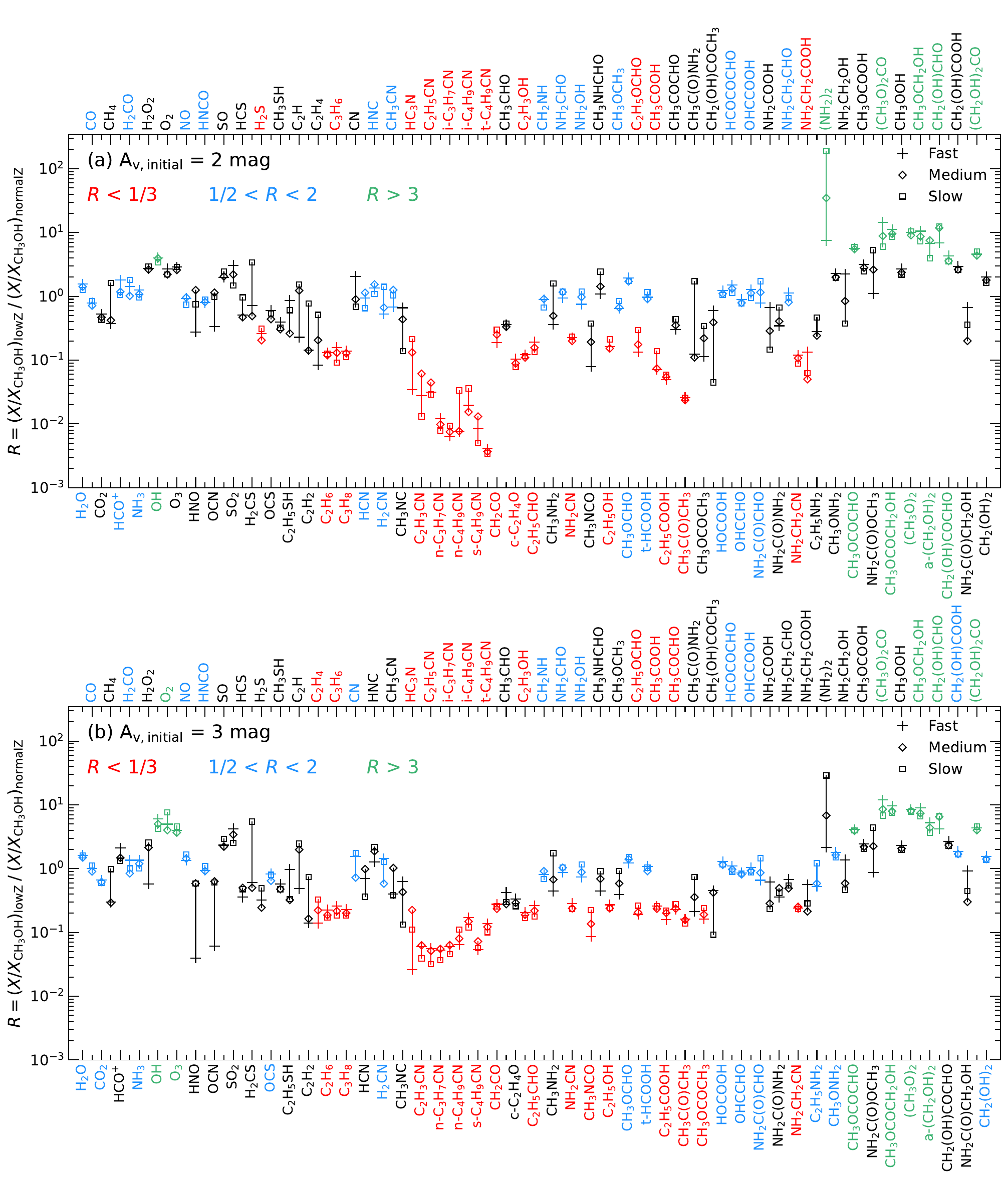}
    \caption{Ratio $R$ of molecular abundances (relative to methanol) between low- and normal-metallicity hot-core models for a sample of 96 molecules, shown for three warm-up timescales: 5 $\times$ 10$^4$ yr (fast), 2 $\times$ 10$^5$ yr (medium), and 1 $\times$ 10$^6$ yr (slow). Panels a and b show models with an initial visual extinction of 2 and 3 mag, respectively. Molecules with $R$ lower than one-third at all three timescales are highlighted in red, while those with $R$ higher than three at all three timescales are highlighted in green, and molecules with $R$ within a factor of two of unity (1/2 < $R$ < 2) at all three timescales are highlighted in blue. The other molecules are displayed in black.}
    \label{fig_Rob_mod}
\end{figure*}

Figure~\ref{fig_Rob_mod} illustrates the ratio \( R \) of molecular abundances (relative to methanol) of low- and normal-metallicity hot-core models for a set of 96 molecules. The results are shown for the initial visual extinction values and warm-up timescales introduced in the previous paragraphs. This comparison reveals how molecular abundances respond to changes in metallicity across the Milky Way. For each panel, molecules that have a significantly lower abundance relative to methanol ($R$ < 1/3) in the low-metallicity models whatever the warm-up timescale are highlighted in red (e.g., C$_2$H$_5$CN, C$_2$H$_5$OH, CH$_3$C(O)CH$_3$, CH$_3$COOH), indicating strong sensitivity to metallicity. In contrast, molecules with $R$ values within a factor of two of unity across all timescales (1/2 < $R$ < 2) are shown in blue, suggesting that their formation pathways are not more sensitive to metallicity variations than methanol (e.g., CH$_3$OCHO, NH$_2$CHO). 
Molecules with abundances relative to methanol significantly higher ($R$ > 3) in the low-metallicity models whatever the warm-up timescale are highlighted in green (e.g., a-(CH$_2$OH)$_2$, CH$_2$(OH)CHO). Since methanol itself shows a moderate decrease of its abundance relative to H$_2$ (less than a factor of three) in the low-metallicity models, $R$ > 3 indicates that the abundances of these molecules relative to H$_2$ are even enhanced at low metallicity. The other molecules that do not fit into these three categories are displayed in black.

Most of the simpler species shown in Fig.~\ref{fig_Rob_mod}, such as water, CO, CO$_2$, and NH$_3$ (which are major ice components) are fairly stable in their peak warm-up stage gas-phase abundances with respect to methanol, at low-Z versus normal-Z values. These species also show minimal variation as a function of warm-up timescale or with different initial visual extinctions. Since each of these species is mostly preserved in the solid phase until temperatures are high enough for them to desorb substantially, it is instructive to consider their abundances in the ices at the end of the cold collapse stage. In the models, much of the production of COMs occurs during this earlier period. 

In absolute terms, water ice, for example, falls by a factor $\sim$1.4 (1.6) for $A_{\rm {V,init}}= 2$ mag ($A_{\rm {V,init}}= 3$ mag) in the 13 kpc model versus the local-Galaxy model, which is a little less than the drop-off in the total oxygen budget of 1.93$\times$. This means that some other major O-bearing species must have fallen by a factor greater than this value. Indeed, CO ice at the end of the cold collapse stage is a factor $\sim$4.1 (3.5) lower in the 13 kpc model, while CO$_2$ ice falls by a factor $\sim$4.6 (4.0). These decreases are greater than the individual drop-off factors in either the oxygen (1.93$\times$) or carbon budget (2.559$\times$). For ammonia, NH$_3$, the main solid-phase carrier of nitrogen in the models, the absolute ice abundance falls by a factor $\sim$1.8 (1.9) for $A_{\rm{V,init}}= 2$ mag ($A_{\rm{V,init}}= 3$ mag) in the 13 kpc model versus the local-Galaxy model, which is much more consistent with the drop in nitrogen budget of 1.778$\times$.

On a simple analysis, one might expect that the decline in CO abundance would follow the reduction in budget of the least abundant of either the elemental C or O; this is because CO is efficiently formed in the gas phase and is quite chemically stable, allowing it to incorporate a large fraction of the overall budget of each element (any grain-surface conversion to CO$_2$ or other species notwithstanding). Regarding oxygen, the slight overabundance of water ice may be explained by the lower CO and CO$_2$ abundances themselves; the fact that there is a reduction in the carbon budget reduces the possible abundances of these species, allowing water to incorporate a larger proportion of the reduced overall oxygen budget. The fact that the final CO (and CO$_2$) ice abundance is depressed beyond the factor associated with the lower carbon budget indicates that there is another factor at work.

Methane is typically the other major carbon-carrying species in the ice mantles \citep[e.g.,][]{2023NatAs...7..431M}, and its absolute abundance falls by a factor 7.5 (9.5) in the 13 kpc models versus the local-Galaxy models. It is thus disproportionately reduced (by $\sim$2.9--3.7$\times$) beyond the expected reduction associated with lower metallicity. This much lower solid-phase CH$_4$ abundance, and the associated lower abundances of other hydrocarbons, including radicals (such as CH$_3$) and larger, stable species (such as C$_2$H$_6$) leads to lower abundances of certain more complex organics that happen to be directly dependent on the addition of methyl or larger radical groups to other radicals, or on the hydrogenation of unsaturated hydrocarbon species (originating in the gas phase -- see below) whose abundances are also depressed. This effect manifests in the gas-phase behaviors shown in Fig.~\ref{fig_Rob_mod}, wherein ethane (C$_2$H$_6$), propane (C$_3$H$_8$), and several nitriles (e.g., ethyl cyanide, C$_2$H$_5$CN) show declines of around an order of magnitude or more in their ratios with methanol, between the low-Z and normal-Z models.

Methanol itself, with which all other species are compared in Fig.~\ref{fig_Rob_mod}, is less affected in its collapse-stage solid-phase abundance than CO, falling by a factor $\sim$2.3 (2.5). The peak gas-phase methanol abundances during the warm-up stage are similarly affected, although there is a degree of variation between values from the different combinations of warm-up timescale and $A_{\rm {V,init}}$ that does not exhibit a clear pattern. The fact that the production of both solid- and gas-phase methanol is closely related to the conversion of grain-surface CO means that the latter appears very stable in Fig.~\ref{fig_Rob_mod}.

In fact, while the final CO ice abundance in the collapse stage is only lower by a factor of a few in the low-Z case, during much of this stage the surface coverage of CO on the grains is five to ten times lower (only eventually rising further when gas densities become much greater). This lower coverage is caused by the $\sim$3$\times$ lower rate of adsorption onto the dust grains associated with the larger gas-to-dust ratio. This has the effect that there is less reactive material on the surface at any one time in the low-Z model with which H may react; most of the other material on the surface of the grain/ice is water or CO$_2$, which are essentially inert to reactions with H atoms. Although the adsorption of gas-phase atomic H is also slowed down on a per-atom basis in the low-Z case, the gas-phase abundance of H undergoes a commensurate rise, so that its total rate of adsorption is largely unaffected during the period when freeze-out is strongest. This drives up the surface population of atomic hydrogen, such that the CO that is yet present on the surface tends to be driven to a further degree of hydrogenation. This favors the production of species such as glycolaldehyde, CH$_2$(OH)CHO, and ethylene glycol, a-(CH$_2$OH)$_2$, along with other species whose production is driven by nondiffusive reactions between CO-product radicals such as HCO, CH$_3$O, and CH$_2$OH. Such species, clustered toward the right of Fig.~\ref{fig_Rob_mod}, show substantial increases in their peak gas-phase abundances relative to methanol during the warm-up stage. The abundance gains across these various species add up so as to correspond to the excessive fall in CO and CO$_2$ ice abundances.

Curiously, the peak gas-phase abundance of methyl formate, CH$_3$OCHO, is not so strongly enhanced in the low-Z case. This likely reflects a relative scarcity of H-addition/abstraction reactions in the network for that species, as compared with glycolaldehyde. Our network includes reactions and rates taken from \citet{2018MNRAS.479.2007A}, who calculated mechanisms for the addition or abstraction of atomic H with species including methyl formate and glycolaldehyde, including the interconversion of the latter with ethylene glycol, via one of two intermediate radicals. However, their calculations did not go as far in the consideration of the methyl formate conversion, and indeed our network does not include such interconversion steps. Thus, while CH$_3$OCH$_2$OH is formed on the grains in stage 1, similarly to a-(CH$_2$OH)$_2$, it cannot be interconverted to CH$_3$OCHO. Glycolaldehyde abundance in stage 2 benefits from the latter back conversion from ethylene glycol, serving to increase its abundance substantially. Therefore, the apparent lack of enhancement in methyl formate in the low-Z models may be an artifact of the network under the extreme conditions sampled in these models.

While the grain surfaces during the collapse stage host the production of various molecules, including methane (via hydrogenation of atomic C) and various COMs, the low-Z conditions also produce important changes in the behavior of the gas-phase chemistry that is relevant to those species. Crucially, the much lower rate of freeze-out of gas-phase species onto the grains in the low-Z case provides more time for atomic carbon to be locked up into, primarily, gas-phase CO (although, as described above, it is also more rapidly converted into larger species once it reaches the grains). As a result, less atomic C is available to be directly accreted onto the dust grains, resulting in the lower methane ice abundances. The production of solid-phase methane is indeed seen more generally in models to be formed early in the evolution of the dust-grain ices, while the availability of free carbon in the gas phase is high, prior to the strong growth of CO \citep[see][]{2011ApJ...735...15G}.

Exacerbating the effects of slower freeze-out is the lower C:O ratio in the low-Z models, which may be just as important to the ultimate abundances of various organic molecules as the actual decline in either C or O elemental abundances would be. In the normal-Z case, the initial C:O ratio is 0.44, versus 0.33 at low-Z. In the models, this leaves much less carbon available for incorporation into pure hydrocarbon forms due to incorporation into CO.

Differences between the $A_{\rm{V,init}}= 2$ mag and $A_{\rm{V,init}}= 3$ mag cases are most clearly manifest in the abundances of the nitrile species, with lower values with respect to methanol seen in the $A_{\rm{V,init}}= 2$ mag case. It should be noted that the choice of comparison with respect to methanol, which is less affected by lower Z due to the robustness of the CO behavior, tends to overemphasize the effects of lower metallicity on the nitriles. However, the ratios for those species are in some cases around an order of magnitude lower for $A_{\rm{V,init}}= 2$ mag. This is mainly related to the higher degree of conversion of neutral atomic C to C$^+$ at this extinction, leaving less neutral carbon that can be converted into neutral carbon chains, such as C$_3$, that contribute directly to, for example, ethyl cyanide (C$_2$H$_5$CN) on the dust grains. Again, this carbon is preferentially converted into CO in the gas.

Within each panel in Fig.~\ref{fig_Rob_mod}, there exists a degree of variation for individual species as a function of warm-up timescale. This indicates that those species that do not vary strongly in this way owe much of their production to the earlier, cold stage. In the most extreme case, that of hydrazine, (NH$_2$)$_2$, cosmic-ray induced UV photolysis of solid-phase ammonia is enhanced by longer timescales, prior to desorption of the ice mantles.

\subsection{Comparison of model predictions to G135.27 MM1} \label{Comp_model_obs}

\begin{figure*}[!h]
    \centering
    \includegraphics[width=0.92\textwidth]{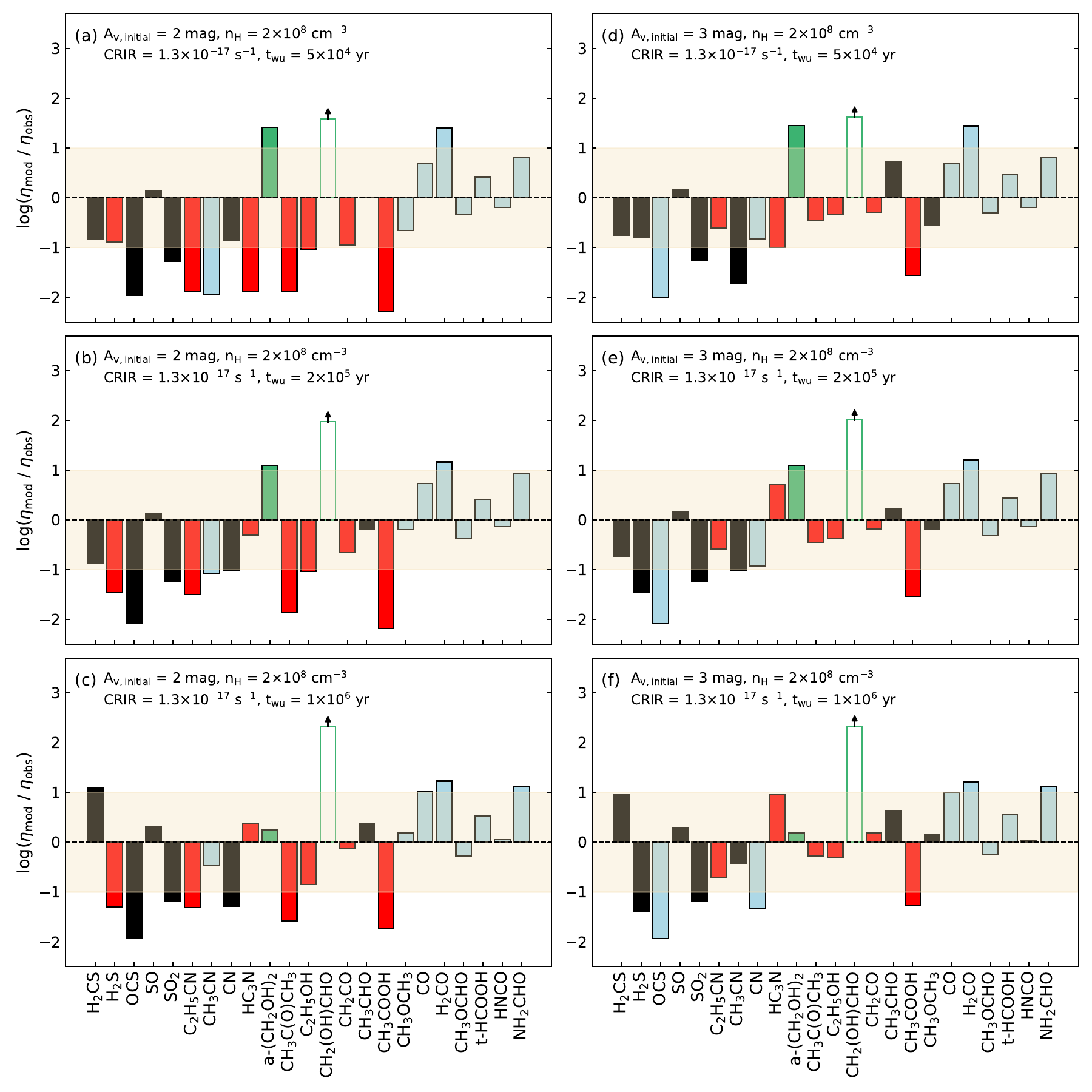}
    \caption{Comparison of the model abundances to observational results for different initial visual extinctions, \( A_{\mathrm{V,\,init}} = 2\,\mathrm{mag} \) (left panels) and \( 3\,\mathrm{mag} \) (rights panels), at varying warm-up timescales: fast (top panels), medium (middle panels), and slow (bottom panels). Bars indicate the number of orders of magnitude by which each modeled abundance ratio exceeds (positive bars) or falls below (negative bars) the corresponding observed ratio. The shaded region denotes agreement within one order of magnitude. Unfilled bars with an arrow represent cases where the comparison is based on an observational upper limit. The color coding is the same as in Fig.~\ref{fig_Rob_mod}.}
    \label{fig_comp_obs_mod}
\end{figure*}

Figure~\ref{fig_comp_obs_mod} compares the measured abundances (relative to methanol) obtained for G135.27 MM1 in Sect.~\ref{weeds} to the model predictions for different warm-up timescales. Unfilled bars with arrows represent observational upper limits, while red, blue, green, and black bars correspond to species exhibiting different abundance patterns with respect to metallicity variations in the models, as illustrated in Fig.~\ref{fig_Rob_mod}. With the cosmic-ray ionization rate and final gas density held fixed, the key parameters distinguishing the models are the warm-up timescale and the initial visual extinction. To quantify the overall agreement with the observations, we compute a matching parameter (the root mean square of the ratios between the model predictions and the observed abundances) for each of the six models. The resulting matching parameters are shown in Fig.~\href{https://zenodo.org/records/17779973}{K.1}. Some of the detected species are not included in our matching calculations, as they are not present in the chemical network used in the model grid. CH$_2$(OH)CHO is excluded because it is only tentatively detected and we treat it as an upper limit, and all models appear to overestimate it. CN and CO are excluded, as they primarily trace more extended cold gas rather than the hot core environment. We also exclude the sulfur-bearing species, since the dominant form of sulfur in dense clouds is uncertain, making the chemistry poorly constrained. In total, 16 species have been included in the matching routine, comprising 11 of the 13 detected COMs (excluding CH$_3$CCH and CH$_2$(OH)CHO) and five simple molecules (H$_2$CO, HC$_3$N, HNCO, t-HCOOH, and CH$_2$CO). As illustrated in Fig.~\href{https://zenodo.org/records/17779973}{K.1}, compared to the models with \( A_{V,\,\mathrm{init}} = 2~\mathrm{mag} \), those with \( A_{V,\,\mathrm{init}} = 3~\mathrm{mag} \) show significantly better consistency with the observed abundances, whatever the warm-up timescale. The slow warm-up timescale produces the smallest dispersion compared to the other timescales, regardless of whether $A_{V,\,\mathrm{init}} = 2~\mathrm{mag}$ or $3~\mathrm{mag}$ is adopted. Therefore, the best-fit model for G135.27 MM1 is provided by the one adopting a slow warm-up timescale of \( t_{\mathrm{wu}} = 1 \times 10^6~\mathrm{yr} \), and an initial visual extinction of \( A_{V,\,\mathrm{init}} = 3~\mathrm{mag} \), which is shown in Fig.~\ref{fig_comp_obs_mod}f.

In general, most common hot-core species are reasonably well reproduced by the best-fit model, i.e., are within one order of magnitude of the observed values (as marked by the shaded region). COMs that are more sensitive to metallicity than methanol and are marked as red bars (C$_2$H$_5$CN, C$_2$H$_5$OH, CH$_3$C(O)CH$_3$, and CH$_3$COOH) are all somewhat underproduced in the model, although their abundances are mostly within one order of magnitude of the observed values. CH$_3$OCHO and NH$_2$CHO, which are both similarly sensitive to metallicity as methanol (blue bars), show opposite trends in the best-fit model: CH$_3$OCHO is slightly underproduced, while NH$_2$CHO is overproduced by about one order of magnitude. a-(CH$_2$OH)$_2$, whose predicted abundance is enhanced by the effect of metallicity, is consistent with the observations.  Additionally, COMs shown in black bars are consistent with the model. On the other hand, simple molecules show larger discrepancies, with some exhibiting significant deviation relative to the observations. For example, three of the five S-bearing species are underproduced by more than one order of magnitude. However, the initial S budget is uncertain, and the initial value used in the models (8 $\times$ 10$^{-8}$ $n_{\rm H}$) may be a substantial underestimate.

\section{Discussion} \label{discussion}
\subsection{Nature of G135.27's continuum sources} \label{nature}

We used the LTE spectral modeling tool Weeds to analyze the continuum-subtracted spectra of the three continuum sources detected toward G135.27, and derive their chemical compositions. These three dense cores, which we called MM1, MM2, and MM3, exhibit distinct chemical and physical properties that reflect their differing evolutionary stages. Toward MM1, we identified 13 COMs, 12 firmly and one tentatively (CH$_2$(OH)CHO). They range from relatively simple species such as CH$_3$OH (six atoms) to larger molecules such as CH$_3$COCH$_3$ (ten atoms). Hundreds of spectral lines associated with these COMs were detected. The detection of such a wide variety of COMs, along with their numerous transitions, reflects a chemically rich environment in MM1 (see Sect.~\ref{weeds_result} and Fig.~\ref{fig_all_spec}). The derived temperatures span a range from 65 to 190 K, with a mean value of approximately 124 K and a median value of 120 K, based on a population diagram analysis and synthetic spectrum fitting with Weeds (see Table~\ref{tab_best_fit}). Both the mean and median rotational temperatures exceed the typical thermal desorption thresholds of COMs from dust grain surfaces \citep[e.g.,][]{2006A&A...457..927G,2020ARA&A..58..727J}. Additionally, the emission sizes are generally on the order of 0.3", corresponding to a radius of approximately 900 au at the distance of G135.27 (see Sect.~\ref{size}). The combination of compact emission, high rotational temperatures, and the detection of a rich inventory of COMs indicate that MM1 is a hot core, making it the third such object discovered in the outer Galaxy after WB89-789 identified by \cite{2021ApJ...922..206S} and Sh 2-283-1a recently identified by \cite{2026ApJ...999..194I}. Additionally, MM1 drives a bipolar outflow traced by CO and $^{13}$CO. The central overlap of the lobes and their extent in the plane of the sky suggests a wide opening angle along the line of sight (Fig.~\ref{fig_outflow}). High-velocity components reveal both a compact, recent ejection close to the protostar and an extended, slower component from an earlier phase, with nearby H$_2$O masers further supporting ongoing outflow activity. However, only one SiO feature is detected near MM1 (Fig.~\ref{fig_SiO}): it is redshifted with low velocities and elongated along the outflow direction. WB89-789, where CO was not observed, shows a symmetric SiO distribution with secondary peaks tracing shocked gas from bipolar outflows.

In contrast, the continuum sources MM2 and MM3 exhibit much simpler chemical signatures compared to MM1. Only the COM CH$_3$OH is detected toward both cores, with CH$_3$CCH additionally identified in MM3. Assuming that the COM emission scales with the continuum peak flux density yields COM line intensities in MM2 and MM3 that would have been detected in our dataset. Therefore, the non-detection of COMs in MM2 and MM3 is unlikely to be due to a lack of sensitivity, unless they harbor COM emission over a much
smaller size than MM1. As demonstrated in Sect.~\ref{weeds_result}, CH$_3$OH appears to originate from larger-scale gas rather than being directly associated with these cores. Although the CH$_3$CCH emission in MM3 peaks near the continuum position, indicating that it traces the dense core material directly associated with MM3, CH$_3$CCH commonly traces less warm and more extended gas than other COMs \citep[e.g.,][]{2025A&A...696A.198C}. Additionally, rotational temperatures are below 100 K in both cores, with H$_2$CO yielding 41 K for MM2 and 89 K for MM3. These simpler chemical signatures and lower temperatures suggest that MM2 and MM3 are at an earlier evolutionary stage, prior to the onset of the hot core chemistry characteristic of MM1. This interpretation is further supported by the narrower line widths measured toward MM2 and MM3 ($\sim$ 1.5 and 2~km~s$^{-1}$, respectively) compared to MM1 (4~km~s$^{-1}$), which reflects a lower level of turbulence or dynamical activity within these cores. Moreover, the CO and $^{13}$CO spectra toward MM2 exhibit redshifted absorption against the continuum, that is an inverse P Cygni profile, providing evidence of infall motions \citep[e.g.,][]{2001ApJ...562..770D,2013A&A...558A.126M,2015ApJ...814...22E}. The presence of such an infall feature suggests that MM2 is a dynamically young core likely undergoing gravitational collapse. On the other hand, no outflow activity is detected toward MM2 in any of the observed molecular tracers, including CO and SiO. This absence of outflow signatures indicates that MM2 may be at the prestellar stage. For MM3, the detection of EHV SiO emission reveals the presence of a powerful outflow. Such EHV SiO emission is typically associated with the early protostellar stage of star formation (see Sect.~\ref{SiO}). These characteristics, including cold gas, limited chemical complexity, and active outflow processes, indicate that MM3 is a young, embedded protostellar object.

\subsection{Origin of the velocity gradient traced by COMs} \label{opposite_grad}

As mentioned in Sect.~\ref{gradient}, COMs (CH$_3$OH and CH$_3$OCHO) trace a velocity gradient that is aligned with the outflow axis but opposite in direction to the large-scale CO outflow. To understand the physical processes responsible for the observed opposite velocity gradient, it is essential to consider the formation and release mechanisms of CH$_3$OH and CH$_3$OCHO. These two species are closely linked, both theoretically and observationally \citep{2020A&A...634A..52S,2022MNRAS.511.3463Q}. CH$_3$OH forms on icy dust grain surfaces through successive hydrogenation of CO \citep[e.g.,][]{2002ApJ...571L.173W}, while CH$_3$OCHO is primarily produced via surface reactions between HCO and CH$_3$O \citep{2020A&A...634A..52S}. Both molecules are generally believed to form on grain surfaces at relatively low temperatures \citep[e.g.,][]{2008ApJ...682..283G,2020ApJS..249...26J}, and are subsequently released into the gas phase as the grains are heated by the central protostar. In addition to thermal desorption, shock-related heating and sputtering may also play a significant role in some environments, such as the quiescent Galactic Center cloud G+0.693$-$0.027 \citep[e.g.,][]{2008ApJ...672..352R,2018MNRAS.478.2962Z} and the shocked region located in the outflow of G31.41 \citep{2025A&A...704A.288L}. Grain processing caused by the passage of a shock driven by an outflow can lead to nonthermal desorption, providing an alternative pathway for injecting CH$_3$OH and CH$_3$OCHO into the gas phase \citep[e.g.,][]{2010MNRAS.409...29F,2024A&A...681A.104B}. 

A similar opposite velocity gradient has been observed in the low-mass protostar NGC 1333 IRAS 4A2 where CH$_3$OH and SiO show a velocity gradient on a scale of ~150 au along the outflow axis with a direction opposite to the kinematics of the outflow on larger scales. To interpret this phenomenon, \cite{2024A&A...686L..13D} proposed three possible explanations: (1) rotating disk or inner envelope, (2) jet or outflow precessing or with an axis in the plane of the sky, and (3) wide-angle disk wind. They favored the last scenario.

In the case of G135.27, the C$^{18}$O PV diagram extracted perpendicular to the outflow axis shows a velocity gradient consistent with rotation in the protostellar envelope (Fig.~\ref{fig_PV}b). In contrast, the gradients traced by CH$_3$OH and CH$_3$OCHO are aligned with the outflow axis and therefore cannot be explained by rotation. Their orientation opposite to the large-scale CO outflow also rules out the possibility that they trace the outflow itself. Instead, these features could be explained by interaction with a wide-angle wind, where shocks driven into the envelope release COMs from dust grains and generate the observed inverse gradients, consistent with the interpretation proposed for NGC 1333 IRAS 4A2. The shock-sensitive nature of CH$_3$OH and CH$_3$OCHO further supports this scenario.

However, the spatial scales traced by the COM velocity gradient differ between G135.27 and NGC 1333 IRAS 4A2. In G135.27, the offset between blue- and redshifted COM emission indicates that the opposite velocity gradient extends over 0.1" (600 au), a size significantly larger than the extent of the gradient reported in NGC 1333 IRAS 4A2 ($\sim$150 au). This difference in scale may result from the mass difference between these two star-forming regions: NGC1333 IRAS 4A2 is a low-mass protostar with an envelope mass of 2.3 M$_{\odot}$ and a luminosity of 4.7 L$_{\odot}$ \citep{2020A&A...635A.198B}, while G135.27 is a protostar with an envelope mass of 49 M$_{\odot}$ (see Sect.~\ref{H2}) and a luminosity of 3.1$\times$10$^{4}$ L$_{\odot}$ (see Sect.~\ref{Sourceselection}), which qualify it as a high-mass star-forming region. The typical disk radii in low-mass star-forming regions are smaller than 50 au \citep{2019ApJ...882...49L,2019A&A...621A..76M}, while disk radii in high-mass star-forming regions can reach up to 1000 au, such as the 858 au disk around IRAS 20126+4104 \citep{2016ApJ...823..125C}. This comparison supports the idea that a wide-angle disk wind in G135.27 can extend to $\sim$600 au, further reinforcing the interpretation that the opposite velocity gradient traced by CH$_3$OH and CH$_3$OCHO is associated with such a wind.

\subsection{Metallicity-dependent behavior of classes of COMs} \label{model_beha}

\begin{figure}[!htbp]
    \centering
    \includegraphics[width=0.45\textwidth]{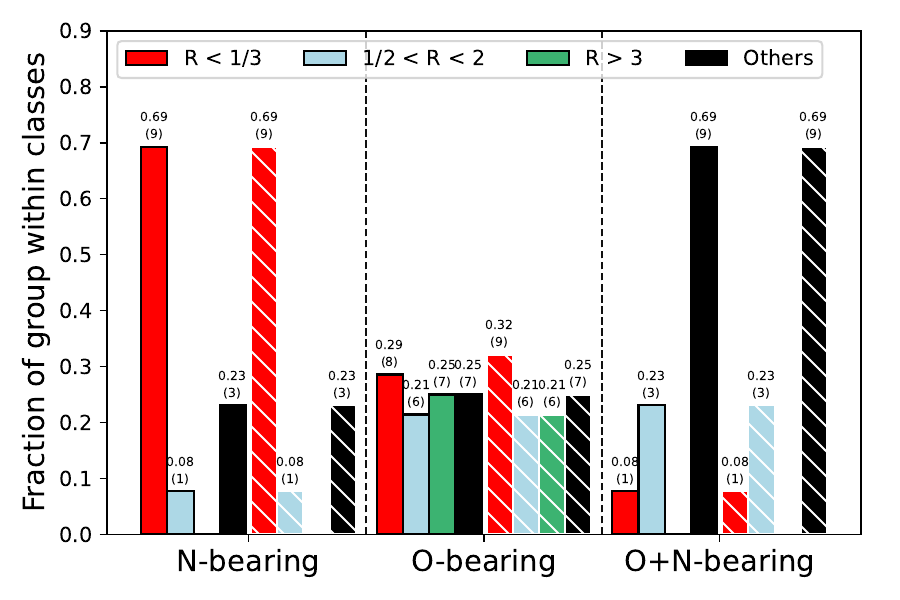}
    \caption{Fraction and number of molecules exhibiting different abundance patterns with respect to metallicity variations for N-, O-, and O+N-bearing COMs, based on chemical modeling results. Their values are shown at the top of each bar. The color coding is the same as in Fig.~\ref{fig_Rob_mod}. Solid bars and hatched bars denote the fractions for initial visual extinctions of 2 and 3 mag, respectively.}
    \label{fig_model_behavior}
\end{figure}

Figure~\ref{fig_model_behavior} shows the fraction and number of COMs exhibiting different abundance patterns in response to metallicity variations for N-, O-, and O+N-bearing COMs, using our chemical models. These statistics build directly on the molecular-level behavior illustrated in Fig.~\ref{fig_Rob_mod}. The distribution of molecules with respect to metallicity sensitivity varies markedly among classes of COMs. Across both $A_{\rm{V,init}}= 2$ mag and 3 mag, N-bearing COMs exhibit the highest fraction of species whose abundances are more sensitive to metallicity than methanol ($R$ < 1/3), with roughly 70$\%$ of all N-bearing COMs affected. In contrast, O-bearing COMs show a smaller fraction (about one third) of species more sensitive to metallicity than methanol. Only $8\%$ O+N-bearing COMs are more sensitive to metallicity than methanol. For molecules with metallicity sensitivity comparable to CH$_{3}$OH (1/2 < $R$ < 2), the trend is reversed: the fraction increases from N-bearing to O-bearing, and then to O+N-bearing COMs, also supporting that N-bearing COMs are the most sensitive to metallicity variations. Additionally, only O-bearing COMs show enhancement at low metallicity.

Stronger variation among the N-bearing species is associated in the models with molecules within that group whose formation has a direct dependence on radicals related to methane and other hydrocarbons. These species' abundances are diminished as the result of the more complete conversion of gas-phase carbon into CO, associated with the slower freeze-out caused by a greater gas-to-dust ratio.

The formation mechanisms of O-bearing COM species are typically more dependent on CO-related chemistry, involving radicals such as HCO and CH$_{3}$O/CH$_{2}$OH that are closely tied to CO hydrogenation. These radicals are strongly enhanced, beyond any enhancement in the abundance of CO itself, due to elevated surface hydrogenation, which allows their product COMs to maintain or exceed the ratios (versus methanol) achieved under low-Z conditions. However, there remain some O-bearing species, such as ethanol (C$_{2}$H$_{5}$OH), that have influential formation mechanisms not directly related to CO chemistry and are more reliant on the addition of methyl (CH$_{3}$) or larger alkyl radicals to other species. These species tend to be somewhat diminished in the same way as various N-bearing COMs.

Those molecules bearing both O and N tend to be among that group of species in the models that are more reliant on CO chemistry, so that, in spite of the lower abundance of elemental nitrogen, almost all of them are able to avoid drastic reductions in their ratios with methanol, compared with the normal-Z case that is they avoid falling into the R$<$ 1/3 category.

\subsection{Impact of metallicity on the COM chemical composition} \label{impactofZ}

\begin{figure}[!htbp]
    \centering
    \includegraphics[width=0.5\textwidth]{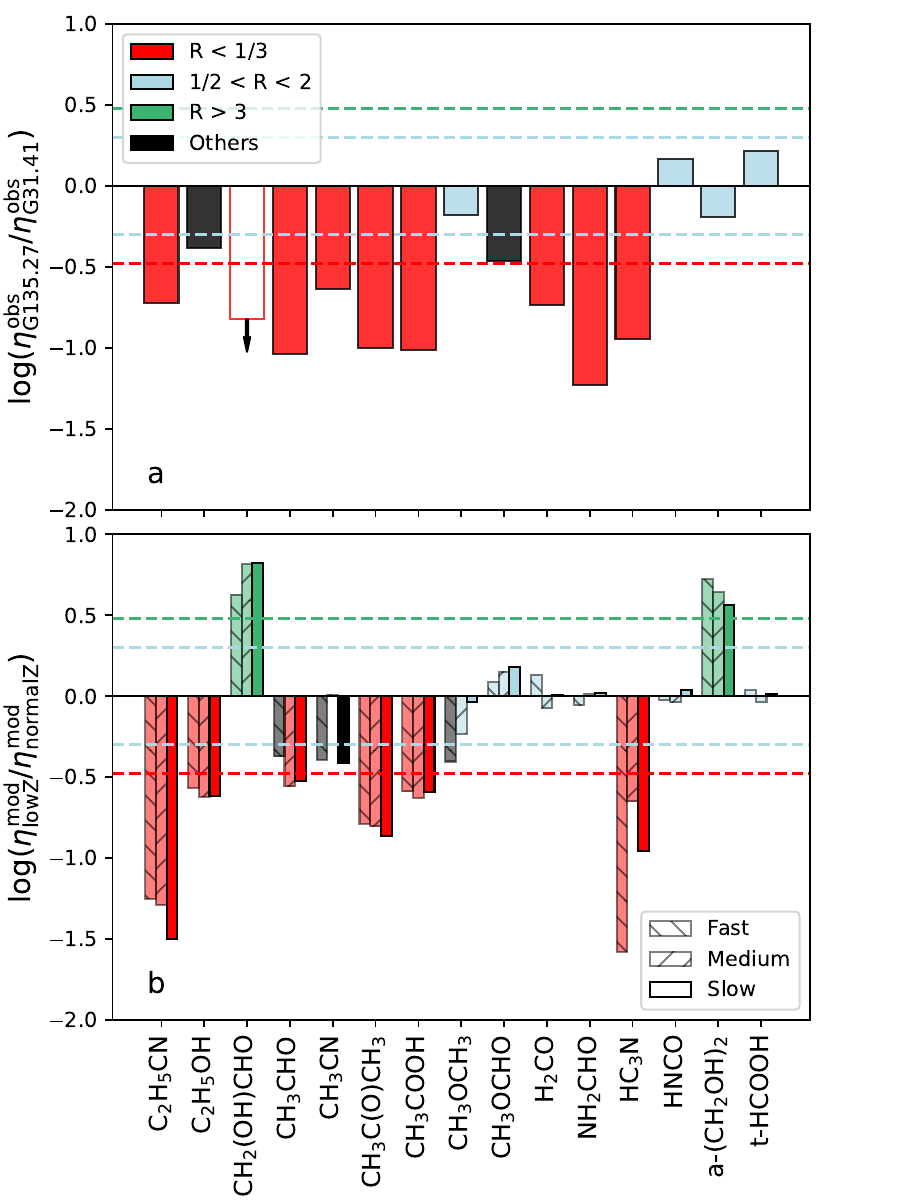}
    \caption{Panel a: Observed abundance ratios, $R$, between G135.27 MM1 and G31.41. Panel b: Abundance ratios, $R$, of models (\( A_{\rm{V,\,init}} = 3\,\rm{mag} \)) with low and normal metallicity. In both panels, the ratios correspond to $\eta$ = $X$/$X_{\rm CH_3OH}$. Molecules with $R$ lower than one-third (dashed red line) or higher than three (dashed green line) are highlighted in red or green, respectively, while those with $R$ within a factor of two of unity (1/2 < $R$ < 2; dashed blue lines) are highlighted in blue. The other molecules are shown in black. The unfilled bar with arrow in panel a represents an observational upper limit for G135.27 MM1. In panel b, the solid bars correspond to the best-fit model of G135.27 MM1 ($t_{\rm{wu}} = 1\times10^6$ yr), while the hatched bars represent the ratios for the other two timescales. }
    \label{fig_comp_ratio}
\end{figure}

We compared in Fig.~\ref{fig_Rob_mod} the outcomes of the chemical model under low and normal metallicities. In order to test if the behavior of the models with respect to metallicity is corroborated by the observations, we need to compare the chemical composition of G135.27 to the composition of a hot core in the inner Galaxy. We chose the hot core G31.41 because its detailed chemical composition has been derived \citep[][]{2020A&A...644A..84M,2023A&A...677A..15M,2021A&A...653A.129C,2022A&A...658A.150G,2024MNRAS.529.3244L} and it has a similar luminosity and envelope mass \citep[4.4$\times$10$^4$ L$_{\odot}$ and 70 M$_{\odot}$, respectively;][]{2020A&A...644A..84M} as G135.27. Figure~\ref{fig_comp_ratio}a shows the observational ratios between G135.27 and G31.41 for the molecules detected in both sources, and Fig.~\ref{fig_comp_ratio}b displays the corresponding modeled abundance ratios between reduced (appropriate for a Galactocentric distance of 13 kpc) and normal metallicity.

For most molecules, the observational ratios follow the qualitative trends predicted by the models, largely independent of the adopted model timescale. Nevertheless, quantitative differences of up to one to two orders of magnitude are found for several species. Overall, these results point to a general consistency with metallicity-dependent trends predicted by the models, while leaving room for significant deviations at the level of individual molecules. However, several species deviate from this general behavior. In particular, CH$_2$(OH)CHO, CH$_3$OCHO, and a-(CH$_2$OH)$_2$ exhibit the opposite trend to the models, showing lower relative abundances in the low-Z environment than in the normal-Z case, while the models predict the opposite for all modeled timescales. NH$_2$CHO also deviates strongly from the model predictions: while the model predicts a similar sensitivity to metallicity as methanol, the molecule is more than one order of magnitude less abundant (relative to methanol) in G135.27 than in G31.41. These four species contain, or are strongly related to, grain-surface chemistry involving the HCO radical. The reactions involving this species are strongly influenced by the atomic-H abundance on the grains, which is sensitive to the dust temperature. The use of only a single-point model, with an equally low minimum value of $T_{\rm dust}$ for both the local and outer-galaxy setups, could lead to an excess of HCO related species. A more dedicated evaluation of $T_{\rm dust}$ for outer-galaxy sources, based on lower dust content (and thus lower $A_{\rm V,init}$) would be warranted, along with the use of multi-dimensional models of hot-core chemo-dynamical evolution \citetext{see \citealp{2021A&A...651A..43B}; Bonfand et al. in prep.}. It is also useful to compare the behavior of these HCO group-bearing COMs with the behavior of HCO itself, as observed by \cite{2025A&A...704A.171G} using data from the CHEMOUT survey. Those authors found that HCO gas-phase abundances in cold cores scale with measured C/H ratios in the 8.34--24~kpc range of galactocentric radii. Methanol also shows a similar C/H dependence in those observations, indicating that HCO abundances with respect to methanol should be stable as a function of galactocentric radius (within the expressed range). Assuming that a meaningful comparison can be made between HCO-bearing COMs in hot gas and HCO in cold cores, our modeled abundances of methyl formate (CH$_3$OCHO) and formamide (NH$_2$CHO) appear to be more or less consistent with the CHEMOUT results. The other two species, glycolaldehyde (CH$_2$(OH)CHO) and ethylene glycol (a-(CH$_2$OH)$_2$), are still overproduced with respect to those data. However, for the reasons noted above, regarding dust temperature, it is unclear whether the abundance of cold HCO in the gas will provide an accurate reflection of COM production efficiencies on grains, or their gas-phase abundances at later times. CH$_3$CN and C$_2$H$_5$OH are assigned different colors in Fig.~\ref{fig_comp_ratio}a and b, but the agreement between the models and the observations is in fact relatively good with a ratio close to 1/3. 

These results suggest that metallicity plays a role in shaping the abundances of certain COMs. However, other physical parameters—such as density structure, temperature evolution, and dynamical history—may also influence the chemical composition. Additionally, it should be noted that the metallicity of G31.41 is likely higher than solar, approximately 1.5 $Z_{\odot}$ if we take the Galactic metallicity gradient of \citet{2011AJ....142..136L}. This implies that the current model, which assumes solar metallicity, may not be a good reference for the comparison shown in Fig~\ref{fig_comp_ratio}, and a revised model adjusted to the expected metallicity of G31.41 should be considered to provide more accurate predictions. 

Overall, molecules that are predicted by our best-fit model in Fig.~\ref{fig_comp_ratio}b to be more sensitive to metallicity relative to CH$_3$OH are generally confirmed to behave in this manner by the observational results, while only about half of the molecules that exhibit similar metallicity sensitivity to CH$_3$OH in the models do behave in this manner in the observations. One possibility is that our predictions focus primarily on the impact of metallicity, without accounting for other processes such as shocks, enhanced cosmic ray ionization, or local UV radiation, all of which can significantly alter molecular abundances. Another possibility is that the comparison is limited to a single source, making it difficult to draw statistically robust conclusions. Expanding the analysis to a larger sample of hot cores in the outer Galaxy will be essential to better constrain the role of metallicity relative to other physical and chemical effects.

\section{Conclusions} \label{Conclusions}

We have carried out an imaging spectral line survey with NOEMA to probe the chemical composition of G135.27+2.79, a hot core candidate in the outer Galaxy ($D\rm _{gc}$ = 13.1 kpc). At the angular resolution ($\sim$5000 au) of the survey, G135.27 consists of three continuum cores, which we called MM1, MM2, and MM3. We derived the molecular composition of these cores assuming LTE. The main results of our analysis are the following:

\begin{enumerate}
    \item We detected 28 molecules toward MM1 and several of their less abundant isotopologs, and we tentatively detected three additional molecules. Among these detected molecules, 13 (including a tentative detection of CH$_2$(OH)CHO) are COMs, most of which trace a hot and compact region. This confirms G135.27 MM1 as a bona fide hot core, the third one identified in the outer Galaxy.
    \item MM1 hosts a large-scale bipolar outflow traced by CO and $^{13}$CO. Interestingly, COMs such as CH$_3$OH and CH$_3$OCHO trace a velocity gradient along the outflow axis at a scale of 600 au but with a direction opposite to that of the large-scale CO outflow. This feature may be linked to a wide-angle disk wind.
    \item Only simple molecules trace MM2. Its CO spectrum shows an inverse P-Cygni profile but no sign of an outflow. MM2 may thus be a collapsing prestellar core. MM3 shows emission of simple molecules and extended CH$_3$CCH emission accompanied by a bipolar outflow traced by high-velocity SiO, suggesting that it is a young protostar.
\end{enumerate}

To assess the impact of a reduced metallicity and dust-to-gas ratio in the outer Galaxy on hot-core chemistry, we compared the composition of MM1 with that of two other hot cores: G31.41 in the inner Galaxy and WB89-789 further out in the outer Galaxy. The COM abundances relative to CH$_3$OH in MM1 show a good correlation with those measured in G31.41 and WB89-789, but with values intermediate between those of G31.41 and WB89-789. This may reflect the influence of metallicity on the formation of COMs.

We ran the astrochemical model \textsc{MAGICKAL} to simulate the abundances of $\sim$100 molecules under low-metallicity ($D_{\rm gc}$=13 kpc) and normal-metallicity ($D_{\rm gc}$=8.2 kpc) conditions. Slower adsorption onto grains due to the lower dust-to-gas ratio at low metallicity produces a more complete gas-phase conversion of the limited carbon available into CO, rather than it being adsorbed onto the grains to form methane or a related species. This behavior is further enhanced by the C:O ratio falling in the low-Z case. Larger hydrocarbon species, including radicals, are also depressed both in the gas phase and on the grains due to the reduced budget of free carbon. Thus, while other species may suffer, CO -- and those species that may form from it via grain-surface chemistry -- is less strongly affected by the decline in elemental C or O. Furthermore, the slow uptake of material onto the grains leads to lower coverages of CO, leading to more effective conversion to COMs of increasing degrees of saturation. COMs with production routes related to atomic C and/or pure hydrocarbon chemistry are therefore more likely to suffer diminished abundances at low-Z, including larger N-bearing species. The abundances of O-bearing COMs are often less diminished or even enhanced (with respect to methanol) under low-Z conditions.

The low-metallicity model that best fits the COM abundances of G135.27 MM1 relative to methanol corresponds to a slow warm-up timescale of \( t_{\mathrm{wu}} = 1 \times 10^6~\mathrm{yr} \) and an initial visual extinction of \( A_{V,\,\mathrm{init}} = 3~\mathrm{mag} \). The abundances of most common hot-core species are reproduced within one order of magnitude, but CH$_2$(OH)CHO, CH$_3$COOH, and NH$_2$CHO deviate more significantly.

In order to more carefully evaluate the impact of metallicity on the chemical composition of hot cores, we compared the abundance ratios of G135.27 MM1 and G31.41 to the abundance ratios of the low-Z and normal-Z models. Overall, the model reproduces well the behavior of most investigated molecules, demonstrating that it catches the influence of metallicity (and gas-to-dust ratio) on the chemistry relatively well. However, the model predicts opposite trends for CH$_2$(OH)CHO, a-(CH$_2$OH)$_2$, CH$_3$OCHO, and NH$_2$CHO compared to the observations, suggesting that their chemical networks require more attention or that the treatment of extinction and dust temperature need to be more tailored to specific local conditions.

Overall, these comparisons indicate that metallicity has a significant impact on the formation of COMs. Expanding the source sample will be essential to assessing the robustness of the observed deviations, while incorporating additional processes (e.g., shocks, cosmic rays, UV radiation) into the models may further improve the agreement between models and observations.

\section*{Data availability}

Appendices~E–K are available at Zenodo (DOI: \href{https://zenodo.org/records/17779973}{10.5281/zenodo.17779972}).

\begin{acknowledgements}
    We gratefully thank Karl M. Menten, whose wisdom, guidance, and encouragement inspired this research. Although he passed away before the completion of this article, his influence is deeply felt throughout this work. This paper is dedicated to his memory. 
    We thank the IRAM staff for carrying out the NOEMA observations of project S23AD and Michael Bremer for his support regarding the calibration of the data. IRAM is supported by INSU/CNRS (France), MPG (Germany) and IGN (Spain). 
    We thank L. Coudert and S. Spezzano for providing spectroscopic predictions for CH$_2$DOH. 
    Part of this work has been carried out within the Collaborative Research Center 1601 (SFB 1601 sub-project A1) funded by the Deutsche Forschungsgemeinschaft (DFG, German Research Foundation) – project ID 500700252. 
    YXW received financial support for this research from the International Max Planck Research School (IMPRS) for Astronomy and Astrophysics at the Universities of Bonn and Cologne.
    RTG thanks the National Science Foundation for funding through the Astronomy $\&$ Astrophysics program (grant number 2206516). RTG also thanks the Alexander von Humboldt Foundation for funding a renewed research stay in Germany in Spring 2025. 
\end{acknowledgements}

\bibliographystyle{bibtex/aa} 
\bibliography{AAA_hot} 

@ARTICLE{2023A&A...677A..15M,
       author = {{Mininni}, C. and {Beltr{\'a}n}, M.~T. and {Colzi}, L. and {Rivilla}, V.~M. and {Fontani}, F. and {Lorenzani}, A. and {L{\'o}pez-Gallifa}, {\'A}. and {Viti}, S. and {S{\'a}nchez-Monge}, {\'A}. and {Schilke}, P. and {Testi}, L.},
        title = "{The GUAPOS project. III. Characterization of the O- and N-bearing complex organic molecules content and search for chemical differentiation}",
      journal = {\aap},
         year = 2023,
        month = sep,
       volume = {677},
          eid = {A15},
        pages = {A15},
          doi = {10.1051/0004-6361/202245277},
archivePrefix = {arXiv},
       eprint = {2306.13563},
 primaryClass = {astro-ph.GA},
       adsurl = {https://ui.adsabs.harvard.edu/abs/2023A&A...677A..15M}
}

@ARTICLE{2009ApJ...696.1981H,
       author = {{Hachisuka}, K. and {Brunthaler}, A. and {Menten}, K.~M. and {Reid}, M.~J. and {Hagiwara}, Y. and {Mochizuki}, N.},
        title = "{The Distance to a Star-Forming Region in the Outer Arm of the Galaxy}",
      journal = {\apj},
         year = 2009,
        month = may,
       volume = {696},
       number = {2},
        pages = {1981-1986},
          doi = {10.1088/0004-637X/696/2/1981},
archivePrefix = {arXiv},
       eprint = {0901.2481},
 primaryClass = {astro-ph.GA},
       adsurl = {https://ui.adsabs.harvard.edu/abs/2009ApJ...696.1981H}
}

@ARTICLE{2009ARA&A..47..427H,
       author = {{Herbst}, Eric and {van Dishoeck}, Ewine F.},
        title = "{Complex Organic Interstellar Molecules}",
      journal = {\araa},
         year = 2009,
        month = sep,
       volume = {47},
       number = {1},
        pages = {427-480},
          doi = {10.1146/annurev-astro-082708-101654},
       adsurl = {https://ui.adsabs.harvard.edu/abs/2009ARA&A..47..427H}
}

@ARTICLE{2019NatAs...3..314L,
       author = {{Lee}, Jeong-Eun and {Lee}, Seokho and {Baek}, Giseon and {Aikawa}, Yuri and {Cieza}, Lucas and {Yoon}, Sung-Yong and {Herczeg}, Gregory and {Johnstone}, Doug and {Casassus}, Simon},
        title = "{The ice composition in the disk around V883 Ori revealed by its stellar outburst}",
      journal = {Nature Astronomy},
         year = 2019,
        month = feb,
       volume = {3},
        pages = {314-319},
          doi = {10.1038/s41550-018-0680-0},
archivePrefix = {arXiv},
       eprint = {1809.00353},
 primaryClass = {astro-ph.SR},
       adsurl = {https://ui.adsabs.harvard.edu/abs/2019NatAs...3..314L}
}

@INPROCEEDINGS{2013prpl.conf2S032W,
       author = {{Walsh}, Catherine and {Millar}, Tom and {Nomura}, Hideko and {Herbst}, Eric and {Widicus Weaver}, Susanna and {Aikawa}, Yuri and {Laas}, Jake and {Vasyunin}, Anton},
        title = "{Complex Organic Molecules in Protoplanetary Disks}",
    booktitle = {Protostars and Planets VI Posters},
         year = 2013,
        month = jul,
       adsurl = {https://ui.adsabs.harvard.edu/abs/2013prpl.conf2S032W}
}

@ARTICLE{2016A&A...595A.117J,
       author = {{J{\o}rgensen}, J.~K. and {van der Wiel}, M.~H.~D. and {Coutens}, A. and {Lykke}, J.~M. and {M{\"u}ller}, H.~S.~P. and {van Dishoeck}, E.~F. and {Calcutt}, H. and {Bjerkeli}, P. and {Bourke}, T.~L. and {Drozdovskaya}, M.~N. and {Favre}, C. and {Fayolle}, E.~C. and {Garrod}, R.~T. and {Jacobsen}, S.~K. and {{\"O}berg}, K.~I. and {Persson}, M.~V. and {Wampfler}, S.~F.},
        title = "{The ALMA Protostellar Interferometric Line Survey (PILS). First results from an unbiased submillimeter wavelength line survey of the Class 0 protostellar binary IRAS 16293-2422 with ALMA}",
      journal = {\aap},
         year = 2016,
        month = nov,
       volume = {595},
          eid = {A117},
        pages = {A117},
          doi = {10.1051/0004-6361/201628648},
archivePrefix = {arXiv},
       eprint = {1607.08733},
 primaryClass = {astro-ph.SR},
       adsurl = {https://ui.adsabs.harvard.edu/abs/2016A&A...595A.117J}
}

@ARTICLE{2020A&A...635A.198B,
       author = {{Belloche}, A. and {Maury}, A.~J. and {Maret}, S. and {Anderl}, S. and {Bacmann}, A. and {Andr{\'e}}, Ph. and {Bontemps}, S. and {Cabrit}, S. and {Codella}, C. and {Gaudel}, M. and {Gueth}, F. and {Lef{\`e}vre}, C. and {Lefloch}, B. and {Podio}, L. and {Testi}, L.},
        title = "{Questioning the spatial origin of complex organic molecules in young protostars with the CALYPSO survey}",
      journal = {\aap},
         year = 2020,
        month = mar,
       volume = {635},
          eid = {A198},
        pages = {A198},
          doi = {10.1051/0004-6361/201937352},
archivePrefix = {arXiv},
       eprint = {2002.00592},
 primaryClass = {astro-ph.GA},
       adsurl = {https://ui.adsabs.harvard.edu/abs/2020A&A...635A.198B}
}

@ARTICLE{2023A&A...678A.137C,
       author = {{Chen}, Y. and {van Gelder}, M.~L. and {Nazari}, P. and {Brogan}, C.~L. and {van Dishoeck}, E.~F. and {Linnartz}, H. and {J{\o}rgensen}, J.~K. and {Hunter}, T.~R. and {Wilkins}, O.~H. and {Blake}, G.~A. and {Caselli}, P. and {Chuang}, K. -J. and {Codella}, C. and {Cooke}, I. and {Drozdovskaya}, M.~N. and {Garrod}, R.~T. and {Ioppolo}, S. and {Jin}, M. and {Kulterer}, B.~M. and {Ligterink}, N.~F.~W. and {Lipnicky}, A. and {Loomis}, R. and {Rachid}, M.~G. and {Spezzano}, S. and {McGuire}, B.~A.},
        title = "{CoCCoA: Complex Chemistry in hot Cores with ALMA. Selected oxygen-bearing species}",
      journal = {\aap},
         year = 2023,
        month = oct,
       volume = {678},
          eid = {A137},
        pages = {A137},
          doi = {10.1051/0004-6361/202346491},
archivePrefix = {arXiv},
       eprint = {2308.02688},
 primaryClass = {astro-ph.GA},
       adsurl = {https://ui.adsabs.harvard.edu/abs/2023A&A...678A.137C}
}

@ARTICLE{2024A&A...688A...3H,
       author = {{Hamedani Golshan}, Roya and {S{\'a}nchez-Monge}, {\'A}lvaro and {Schilke}, Peter and {Sewi{\l}o}, Marta and {M{\"o}ller}, Thomas and {Veena}, Vadamattom S. and {Fuller}, Gary A.},
        title = "{High-mass star formation across the Large Magellanic Cloud. I. Chemical properties and hot molecular cores observed with ALMA at 1.2 mm}",
      journal = {\aap},
         year = 2024,
        month = aug,
       volume = {688},
          eid = {A3},
        pages = {A3},
          doi = {10.1051/0004-6361/202349077},
archivePrefix = {arXiv},
       eprint = {2405.01710},
 primaryClass = {astro-ph.GA},
       adsurl = {https://ui.adsabs.harvard.edu/abs/2024A&A...688A...3H}
}

@ARTICLE{2023ApJ...946L..41S,
       author = {{Shimonishi}, Takashi and {Tanaka}, Kei E.~I. and {Zhang}, Yichen and {Furuya}, Kenji},
        title = "{The Detection of Hot Molecular Cores in the Small Magellanic Cloud}",
      journal = {\apjl},
         year = 2023,
        month = apr,
       volume = {946},
       number = {2},
          eid = {L41},
        pages = {L41},
          doi = {10.3847/2041-8213/acc031},
archivePrefix = {arXiv},
       eprint = {2303.05630},
 primaryClass = {astro-ph.GA},
       adsurl = {https://ui.adsabs.harvard.edu/abs/2023ApJ...946L..41S}
}

@ARTICLE{1996ApJ...458..653R,
       author = {{Rudolph}, Alexander L. and {Brand}, Jan and {de Geus}, Eugene J. and {Wouterloot}, Jan G.~A.},
        title = "{Far Outer Galaxy H II Regions}",
      journal = {\apj},
         year = 1996,
        month = feb,
       volume = {458},
        pages = {653},
          doi = {10.1086/176847},
       adsurl = {https://ui.adsabs.harvard.edu/abs/1996ApJ...458..653R}
}

@ARTICLE{2017A&A...604A..60B,
       author = {{Bonfand}, M. and {Belloche}, A. and {Menten}, K.~M. and {Garrod}, R.~T. and {M{\"u}ller}, H.~S.~P.},
        title = "{Exploring molecular complexity with ALMA (EMoCA): Detection of three new hot cores in Sagittarius B2(N)}",
      journal = {\aap},
         year = 2017,
        month = aug,
       volume = {604},
          eid = {A60},
        pages = {A60},
          doi = {10.1051/0004-6361/201730648},
archivePrefix = {arXiv},
       eprint = {1703.09544},
 primaryClass = {astro-ph.GA},
       adsurl = {https://ui.adsabs.harvard.edu/abs/2017A&A...604A..60B}
}

@ARTICLE{2021ApJ...922..206S,
       author = {{Shimonishi}, Takashi and {Izumi}, Natsuko and {Furuya}, Kenji and {Yasui}, Chikako},
        title = "{The Detection of a Hot Molecular Core in the Extreme Outer Galaxy}",
      journal = {\apj},
         year = 2021,
        month = dec,
       volume = {922},
       number = {2},
          eid = {206},
        pages = {206},
          doi = {10.3847/1538-4357/ac289b},
archivePrefix = {arXiv},
       eprint = {2109.11123},
 primaryClass = {astro-ph.GA},
       adsurl = {https://ui.adsabs.harvard.edu/abs/2021ApJ...922..206S}
}

@ARTICLE{1994A&A...291..943O,
       author = {{Ossenkopf}, V. and {Henning}, Th.},
        title = "{Dust opacities for protostellar cores.}",
      journal = {\aap},
         year = 1994,
        month = nov,
       volume = {291},
        pages = {943-959},
       adsurl = {https://ui.adsabs.harvard.edu/abs/1994A&A...291..943O}
}

@ARTICLE{2017A&A...606L..12G,
       author = {{Giannetti}, A. and {Leurini}, S. and {K{\"o}nig}, C. and {Urquhart}, J.~S. and {Pillai}, T. and {Brand}, J. and {Kauffmann}, J. and {Wyrowski}, F. and {Menten}, K.~M.},
        title = "{Galactocentric variation of the gas-to-dust ratio and its relation with metallicity}",
      journal = {\aap},
         year = 2017,
        month = oct,
       volume = {606},
          eid = {L12},
        pages = {L12},
          doi = {10.1051/0004-6361/201731728},
archivePrefix = {arXiv},
       eprint = {1710.05721},
 primaryClass = {astro-ph.GA},
       adsurl = {https://ui.adsabs.harvard.edu/abs/2017A&A...606L..12G}
}

@ARTICLE{2019A&A...632A.101T,
       author = {{Tychoniec}, {\L}ukasz and {Hull}, Charles L.~H. and {Kristensen}, Lars E. and {Tobin}, John J. and {Le Gouellec}, Valentin J.~M. and {van Dishoeck}, Ewine F.},
        title = "{Chemical and kinematic structure of extremely high-velocity molecular jets in the Serpens Main star-forming region}",
      journal = {\aap},
         year = 2019,
        month = dec,
       volume = {632},
          eid = {A101},
        pages = {A101},
          doi = {10.1051/0004-6361/201935409},
archivePrefix = {arXiv},
       eprint = {1910.07857},
 primaryClass = {astro-ph.SR},
       adsurl = {https://ui.adsabs.harvard.edu/abs/2019A&A...632A.101T}
}

@ARTICLE{1996ARA&A..34..111B,
       author = {{Bachiller}, Rafael},
        title = "{Bipolar Molecular Outflows from Young Stars and Protostars}",
      journal = {\araa},
         year = 1996,
        month = jan,
       volume = {34},
        pages = {111-154},
          doi = {10.1146/annurev.astro.34.1.111},
       adsurl = {https://ui.adsabs.harvard.edu/abs/1996ARA&A..34..111B}
}

@ARTICLE{2010A&A...522A..91T,
       author = {{Tafalla}, M. and {Santiago-Garc{\'\i}a}, J. and {Hacar}, A. and {Bachiller}, R.},
        title = "{A molecular survey of outflow gas: velocity-dependent shock chemistry and the peculiar composition of the EHV gas}",
      journal = {\aap},
         year = 2010,
        month = nov,
       volume = {522},
          eid = {A91},
        pages = {A91},
          doi = {10.1051/0004-6361/201015158},
archivePrefix = {arXiv},
       eprint = {1007.4549},
 primaryClass = {astro-ph.GA},
       adsurl = {https://ui.adsabs.harvard.edu/abs/2010A&A...522A..91T}
}

@ARTICLE{2011A&A...526A..47M,
       author = {{Maret}, S. and {Hily-Blant}, P. and {Pety}, J. and {Bardeau}, S. and {Reynier}, E.},
        title = "{Weeds: a CLASS extension for the analysis of millimeter and sub-millimeter spectral surveys}",
      journal = {\aap},
         year = 2011,
        month = feb,
       volume = {526},
          eid = {A47},
        pages = {A47},
          doi = {10.1051/0004-6361/201015487},
archivePrefix = {arXiv},
       eprint = {1012.1747},
 primaryClass = {astro-ph.IM},
       adsurl = {https://ui.adsabs.harvard.edu/abs/2011A&A...526A..47M}
}

@ARTICLE{2016JMoSp.327...95E,
       author = {{Endres}, Christian P. and {Schlemmer}, Stephan and {Schilke}, Peter and {Stutzki}, J{\"u}rgen and {M{\"u}ller}, Holger S.~P.},
        title = "{The Cologne Database for Molecular Spectroscopy, CDMS, in the Virtual Atomic and Molecular Data Centre, VAMDC}",
      journal = {Journal of Molecular Spectroscopy},
         year = 2016,
        month = sep,
       volume = {327},
        pages = {95-104},
          doi = {10.1016/j.jms.2016.03.005},
       adsurl = {https://ui.adsabs.harvard.edu/abs/2016JMoSp.327...95E}
}

@ARTICLE{2001A&A...370L..49M,
       author = {{M{\"u}ller}, H.~S.~P. and {Thorwirth}, S. and {Roth}, D.~A. and {Winnewisser}, G.},
        title = "{The Cologne Database for Molecular Spectroscopy, CDMS}",
      journal = {\aap},
         year = 2001,
        month = apr,
       volume = {370},
        pages = {L49-L52},
          doi = {10.1051/0004-6361:20010367},
       adsurl = {https://ui.adsabs.harvard.edu/abs/2001A&A...370L..49M}
}

@ARTICLE{2015A&A...576A..45F,
       author = {{Fayolle}, Edith C. and {{\"O}berg}, Karin I. and {Garrod}, Robin T. and {van Dishoeck}, Ewine F. and {Bisschop}, Suzanne E.},
        title = "{Complex organic molecules in organic-poor massive young stellar objects}",
      journal = {\aap},
         year = 2015,
        month = apr,
       volume = {576},
          eid = {A45},
        pages = {A45},
          doi = {10.1051/0004-6361/201323114},
archivePrefix = {arXiv},
       eprint = {1501.03168},
 primaryClass = {astro-ph.SR},
       adsurl = {https://ui.adsabs.harvard.edu/abs/2015A&A...576A..45F}
}

@ARTICLE{2022A&A...665A..96B,
       author = {{Busch}, Laura A. and {Belloche}, Arnaud and {Garrod}, Robin T. and {M{\"u}ller}, Holger S.~P. and {Menten}, Karl M.},
        title = "{Resolving desorption of complex organic molecules in a hot core. Transition from non-thermal to thermal desorption or two-step thermal desorption?}",
      journal = {\aap},
         year = 2022,
        month = sep,
       volume = {665},
          eid = {A96},
        pages = {A96},
          doi = {10.1051/0004-6361/202243383},
archivePrefix = {arXiv},
       eprint = {2206.11174},
 primaryClass = {astro-ph.GA},
       adsurl = {https://ui.adsabs.harvard.edu/abs/2022A&A...665A..96B}
}

@ARTICLE{1999ApJ...517..209G,
       author = {{Goldsmith}, Paul F. and {Langer}, William D.},
        title = "{Population Diagram Analysis of Molecular Line Emission}",
      journal = {\apj},
         year = 1999,
        month = may,
       volume = {517},
       number = {1},
        pages = {209-225},
          doi = {10.1086/307195},
       adsurl = {https://ui.adsabs.harvard.edu/abs/1999ApJ...517..209G}
}

@ARTICLE{1996AAS..119..439W,
       author = {{Wouterloot}, J.~G.~A. and {Brand}, J.},
        title = "{IRAS sources beyond the solar circle. VII. The \^12\^C/\^13\^C ratio in the far outer Galaxy.}",
      journal = {\aaps},
         year = 1996,
        month = nov,
       volume = {119},
        pages = {439-457},
       adsurl = {https://ui.adsabs.harvard.edu/abs/1996A&AS..119..439W}
}

@ARTICLE{2008ApJ...682..283G,
       author = {{Garrod}, Robin T. and {Widicus Weaver}, Susanna L. and {Herbst}, Eric},
        title = "{Complex Chemistry in Star-forming Regions: An Expanded Gas-Grain Warm-up Chemical Model}",
      journal = {\apj},
         year = 2008,
        month = jul,
       volume = {682},
       number = {1},
        pages = {283-302},
          doi = {10.1086/588035},
archivePrefix = {arXiv},
       eprint = {0803.1214},
 primaryClass = {astro-ph},
       adsurl = {https://ui.adsabs.harvard.edu/abs/2008ApJ...682..283G}
}

@ARTICLE{1992ApJS...82..167H,
       author = {{Hasegawa}, Tatsuhiko I. and {Herbst}, Eric and {Leung}, Chun M.},
        title = "{Models of Gas-Grain Chemistry in Dense Interstellar Clouds with Complex Organic Molecules}",
      journal = {\apjs},
         year = 1992,
        month = sep,
       volume = {82},
        pages = {167},
          doi = {10.1086/191713},
       adsurl = {https://ui.adsabs.harvard.edu/abs/1992ApJS...82..167H}
}

@ARTICLE{1998ARA&A..36..317V,
       author = {{van Dishoeck}, Ewine F. and {Blake}, Geoffrey A.},
        title = "{Chemical Evolution of Star-Forming Regions}",
      journal = {\araa},
         year = 1998,
        month = jan,
       volume = {36},
        pages = {317-368},
          doi = {10.1146/annurev.astro.36.1.317},
       adsurl = {https://ui.adsabs.harvard.edu/abs/1998ARA&A..36..317V}
}

@ARTICLE{2014ApJ...795L...2V,
       author = {{Vastel}, C. and {Ceccarelli}, C. and {Lefloch}, B. and {Bachiller}, R.},
        title = "{The Origin of Complex Organic Molecules in Prestellar Cores}",
      journal = {\apjl},
         year = 2014,
        month = nov,
       volume = {795},
       number = {1},
          eid = {L2},
        pages = {L2},
          doi = {10.1088/2041-8205/795/1/L2},
archivePrefix = {arXiv},
       eprint = {1409.6565},
 primaryClass = {astro-ph.SR},
       adsurl = {https://ui.adsabs.harvard.edu/abs/2014ApJ...795L...2V}
}

@ARTICLE{2012A&A...541L..12B,
       author = {{Bacmann}, A. and {Taquet}, V. and {Faure}, A. and {Kahane}, C. and {Ceccarelli}, C.},
        title = "{Detection of complex organic molecules in a prestellar core: a new challenge for astrochemical models}",
      journal = {\aap},
         year = 2012,
        month = may,
       volume = {541},
          eid = {L12},
        pages = {L12},
          doi = {10.1051/0004-6361/201219207},
       adsurl = {https://ui.adsabs.harvard.edu/abs/2012A&A...541L..12B}
}

@ARTICLE{2024A&A...686A.252B,
       author = {{Bouscasse}, L. and {Csengeri}, T. and {Wyrowski}, F. and {Menten}, K.~M. and {Bontemps}, S.},
        title = "{Complex organic molecules uncover deeply embedded precursors of hot cores. An APEX unbiased spectral survey of infrared quiet massive clumps}",
      journal = {\aap},
         year = 2024,
        month = jun,
       volume = {686},
          eid = {A252},
        pages = {A252},
          doi = {10.1051/0004-6361/202348803},
archivePrefix = {arXiv},
       eprint = {2403.05237},
 primaryClass = {astro-ph.GA},
       adsurl = {https://ui.adsabs.harvard.edu/abs/2024A&A...686A.252B}
}

@ARTICLE{2022ApJ...941L..13Y,
       author = {{Yang}, Yao-Lun and {Green}, Joel D. and {Pontoppidan}, Klaus M. and {Bergner}, Jennifer B. and {Cleeves}, L. Ilsedore and {Evans}, II, Neal J. and {Garrod}, Robin T. and {Jin}, Miwha and {Kim}, Chul Hwan and {Kim}, Jaeyeong and {Lee}, Jeong-Eun and {Sakai}, Nami and {Shingledecker}, Christopher N. and {Shope}, Brielle and {Tobin}, John J. and {van Dishoeck}, Ewine F.},
        title = "{CORINOS. I. JWST/MIRI Spectroscopy and Imaging of a Class 0 Protostar IRAS 15398{\textendash}3359}",
      journal = {\apjl},
         year = 2022,
        month = dec,
       volume = {941},
       number = {1},
          eid = {L13},
        pages = {L13},
          doi = {10.3847/2041-8213/aca289},
archivePrefix = {arXiv},
       eprint = {2208.10673},
 primaryClass = {astro-ph.SR},
       adsurl = {https://ui.adsabs.harvard.edu/abs/2022ApJ...941L..13Y}
}

@ARTICLE{2024A&A...683A.124R,
       author = {{Rocha}, W.~R.~M. and {van Dishoeck}, E.~F. and {Ressler}, M.~E. and {van Gelder}, M.~L. and {Slavicinska}, K. and {Brunken}, N.~G.~C. and {Linnartz}, H. and {Ray}, T.~P. and {Beuther}, H. and {Caratti o Garatti}, A. and {Geers}, V. and {Kavanagh}, P.~J. and {Klaassen}, P.~D. and {Justtanont}, K. and {Chen}, Y. and {Francis}, L. and {Gieser}, C. and {Perotti}, G. and {Tychoniec}, {\L}. and {Barsony}, M. and {Majumdar}, L. and {le Gouellec}, V.~J.~M. and {Chu}, L.~E.~U. and {Lew}, B.~W.~P. and {Henning}, Th. and {Wright}, G.},
        title = "{JWST Observations of Young protoStars (JOYS+): Detecting icy complex organic molecules and ions. I. CH$_{4}$, SO$_{2}$, HCOO$^{{\ensuremath{-}}}$, OCN$^{{\ensuremath{-}}}$, H$_{2}$CO, HCOOH, CH$_{3}$CH$_{2}$OH, CH$_{3}$CHO, CH$_{3}$OCHO, and CH$_{3}$COOH}",
      journal = {\aap},
         year = 2024,
        month = mar,
       volume = {683},
          eid = {A124},
        pages = {A124},
          doi = {10.1051/0004-6361/202348427},
archivePrefix = {arXiv},
       eprint = {2312.06834},
 primaryClass = {astro-ph.SR},
       adsurl = {https://ui.adsabs.harvard.edu/abs/2024A&A...683A.124R}
}

@ARTICLE{2022A&A...665A..89R,
       author = {{Rachid}, M.~G. and {Rocha}, W.~R.~M. and {Linnartz}, H.},
        title = "{Infrared spectra of complex organic molecules in astronomically relevant ice mixtures. V. Methyl cyanide (acetonitrile)}",
      journal = {\aap},
         year = 2022,
        month = sep,
       volume = {665},
          eid = {A89},
        pages = {A89},
          doi = {10.1051/0004-6361/202243417},
archivePrefix = {arXiv},
       eprint = {2207.12502},
 primaryClass = {astro-ph.GA},
       adsurl = {https://ui.adsabs.harvard.edu/abs/2022A&A...665A..89R}
}

@ARTICLE{2011AJ....142..136L,
       author = {{Luck}, R. Earle and {Lambert}, David L.},
        title = "{The Distribution of the Elements in the Galactic Disk. III. A Reconsideration of Cepheids from l = 30{\textdegree} to 250{\textdegree}}",
      journal = {\aj},
         year = 2011,
        month = oct,
       volume = {142},
       number = {4},
          eid = {136},
        pages = {136},
          doi = {10.1088/0004-6256/142/4/136},
archivePrefix = {arXiv},
       eprint = {1108.1947},
 primaryClass = {astro-ph.GA},
       adsurl = {https://ui.adsabs.harvard.edu/abs/2011AJ....142..136L}
}

@ARTICLE{2006A&A...457..927G,
       author = {{Garrod}, R.~T. and {Herbst}, E.},
        title = "{Formation of methyl formate and other organic species in the warm-up phase of hot molecular cores}",
      journal = {\aap},
         year = 2006,
        month = oct,
       volume = {457},
       number = {3},
        pages = {927-936},
          doi = {10.1051/0004-6361:20065560},
archivePrefix = {arXiv},
       eprint = {astro-ph/0607560},
 primaryClass = {astro-ph},
       adsurl = {https://ui.adsabs.harvard.edu/abs/2006A&A...457..927G}
}

@INPROCEEDINGS{2000prpl.conf..299K,
       author = {{Kurtz}, S. and {Cesaroni}, R. and {Churchwell}, E. and {Hofner}, P. and {Walmsley}, C.~M.},
        title = "{Hot Molecular Cores and the Earliest Phases of High-Mass Star Formation}",
    booktitle = {Protostars and Planets IV},
         year = 2000,
       editor = {{Mannings}, V. and {Boss}, A.~P. and {Russell}, S.~S.},
        month = may,
        pages = {299-326},
       adsurl = {https://ui.adsabs.harvard.edu/abs/2000prpl.conf..299K}
}

@INPROCEEDINGS{2005IAUS..227...59C,
       author = {{Cesaroni}, Riccardo},
        title = "{Hot molecular cores}",
    booktitle = {Massive Star Birth: A Crossroads of Astrophysics},
         year = 2005,
       editor = {{Cesaroni}, R. and {Felli}, M. and {Churchwell}, E. and {Walmsley}, M.},
       series = {IAU Symposium},
       volume = {227},
        month = jan,
        pages = {59-69},
          doi = {10.1017/S1743921305004369},
       adsurl = {https://ui.adsabs.harvard.edu/abs/2005IAUS..227...59C}
}

@INPROCEEDINGS{2004ASPC..323..195C,
       author = {{Ceccarelli}, C.},
        title = "{The Hot Corinos of Solar Type Protostars}",
    booktitle = {Star Formation in the Interstellar Medium: In Honor of David Hollenbach},
         year = 2004,
       editor = {{Johnstone}, D. and {Adams}, F.~C. and {Lin}, D.~N.~C. and {Neufeeld}, D.~A. and {Ostriker}, E.~C.},
       series = {Astronomical Society of the Pacific Conference Series},
       volume = {323},
        month = dec,
        pages = {195},
       adsurl = {https://ui.adsabs.harvard.edu/abs/2004ASPC..323..195C}
}

@ARTICLE{2020ARA&A..58..727J,
       author = {{J{\o}rgensen}, Jes K. and {Belloche}, Arnaud and {Garrod}, Robin T.},
        title = "{Astrochemistry During the Formation of Stars}",
      journal = {\araa},
         year = 2020,
        month = aug,
       volume = {58},
        pages = {727-778},
          doi = {10.1146/annurev-astro-032620-021927},
archivePrefix = {arXiv},
       eprint = {2006.07071},
 primaryClass = {astro-ph.SR},
       adsurl = {https://ui.adsabs.harvard.edu/abs/2020ARA&A..58..727J}
}

@ARTICLE{2016A&A...587A..91B,
       author = {{Belloche}, A. and {M{\"u}ller}, H.~S.~P. and {Garrod}, R.~T. and {Menten}, K.~M.},
        title = "{Exploring molecular complexity with ALMA (EMoCA): Deuterated complex organic molecules in Sagittarius B2(N2)}",
      journal = {\aap},
         year = 2016,
        month = mar,
       volume = {587},
          eid = {A91},
        pages = {A91},
          doi = {10.1051/0004-6361/201527268},
archivePrefix = {arXiv},
       eprint = {1511.05721},
 primaryClass = {astro-ph.GA},
       adsurl = {https://ui.adsabs.harvard.edu/abs/2016A&A...587A..91B}
}

@ARTICLE{2019A&A...628A..10B,
       author = {{Belloche}, A. and {Garrod}, R.~T. and {M{\"u}ller}, H.~S.~P. and {Menten}, K.~M. and {Medvedev}, I. and {Thomas}, J. and {Kisiel}, Z.},
        title = "{Re-exploring Molecular Complexity with ALMA (ReMoCA): interstellar detection of urea}",
      journal = {\aap},
         year = 2019,
        month = aug,
       volume = {628},
          eid = {A10},
        pages = {A10},
          doi = {10.1051/0004-6361/201935428},
archivePrefix = {arXiv},
       eprint = {1906.04614},
 primaryClass = {astro-ph.GA},
       adsurl = {https://ui.adsabs.harvard.edu/abs/2019A&A...628A..10B}
}

@ARTICLE{2020A&A...644A..84M,
       author = {{Mininni}, C. and {Beltr{\'a}n}, M.~T. and {Rivilla}, V.~M. and {S{\'a}nchez-Monge}, A. and {Fontani}, F. and {M{\"o}ller}, T. and {Cesaroni}, R. and {Schilke}, P. and {Viti}, S. and {Jim{\'e}nez-Serra}, I. and {Colzi}, L. and {Lorenzani}, A. and {Testi}, L.},
        title = "{The GUAPOS project: G31.41+0.31 Unbiased ALMA sPectral Observational Survey. I. Isomers of C$_{2}$H$_{4}$O$_{2}$}",
      journal = {\aap},
         year = 2020,
        month = dec,
       volume = {644},
          eid = {A84},
        pages = {A84},
          doi = {10.1051/0004-6361/202038966},
archivePrefix = {arXiv},
       eprint = {2009.13297},
 primaryClass = {astro-ph.SR},
       adsurl = {https://ui.adsabs.harvard.edu/abs/2020A&A...644A..84M}
}

@ARTICLE{2018A&A...616A..90C,
       author = {{Calcutt}, H. and {J{\o}rgensen}, J.~K. and {M{\"u}ller}, H.~S.~P. and {Kristensen}, L.~E. and {Coutens}, A. and {Bourke}, T.~L. and {Garrod}, R.~T. and {Persson}, M.~V. and {van der Wiel}, M.~H.~D. and {van Dishoeck}, E.~F. and {Wampfler}, S.~F.},
        title = "{The ALMA-PILS survey: complex nitriles towards IRAS 16293-2422}",
      journal = {\aap},
         year = 2018,
        month = aug,
       volume = {616},
          eid = {A90},
        pages = {A90},
          doi = {10.1051/0004-6361/201732289},
archivePrefix = {arXiv},
       eprint = {1804.09210},
 primaryClass = {astro-ph.GA},
       adsurl = {https://ui.adsabs.harvard.edu/abs/2018A&A...616A..90C}
}

@ARTICLE{2018A&A...620A.170J,
       author = {{J{\o}rgensen}, J.~K. and {M{\"u}ller}, H.~S.~P. and {Calcutt}, H. and {Coutens}, A. and {Drozdovskaya}, M.~N. and {{\"O}berg}, K.~I. and {Persson}, M.~V. and {Taquet}, V. and {van Dishoeck}, E.~F. and {Wampfler}, S.~F.},
        title = "{The ALMA-PILS survey: isotopic composition of oxygen-containing complex organic molecules toward IRAS 16293-2422B}",
      journal = {\aap},
         year = 2018,
        month = dec,
       volume = {620},
          eid = {A170},
        pages = {A170},
          doi = {10.1051/0004-6361/201731667},
archivePrefix = {arXiv},
       eprint = {1808.08753},
 primaryClass = {astro-ph.SR},
       adsurl = {https://ui.adsabs.harvard.edu/abs/2018A&A...620A.170J}
}

@ARTICLE{2024A&A...681A.104B,
       author = {{Busch}, Laura A. and {Belloche}, Arnaud and {Garrod}, Robin T. and {M{\"u}ller}, Holger S.~P. and {Menten}, Karl M.},
        title = "{Shocking Sgr B2 (N1) with its own outflow. A new perspective on segregation between O- and N-bearing molecules}",
      journal = {\aap},
         year = 2024,
        month = jan,
       volume = {681},
          eid = {A104},
        pages = {A104},
          doi = {10.1051/0004-6361/202347256},
archivePrefix = {arXiv},
       eprint = {2310.11339},
 primaryClass = {astro-ph.GA},
       adsurl = {https://ui.adsabs.harvard.edu/abs/2024A&A...681A.104B}
}

@ARTICLE{2024arXiv241104451S,
       author = {{Shimonishi}, Takashi},
        title = "{The Role of Metallicity in the Chemical Evolution of Star-Forming Regions}",
      journal = {arXiv e-prints},
         year = 2024,
        month = nov,
          eid = {arXiv:2411.04451},
        pages = {arXiv:2411.04451},
          doi = {10.48550/arXiv.2411.04451},
archivePrefix = {arXiv},
       eprint = {2411.04451},
 primaryClass = {astro-ph.GA},
       adsurl = {https://ui.adsabs.harvard.edu/abs/2024arXiv241104451S}
}

@ARTICLE{2016ApJ...827...72S,
       author = {{Shimonishi}, Takashi and {Onaka}, Takashi and {Kawamura}, Akiko and {Aikawa}, Yuri},
        title = "{The Detection of a Hot Molecular Core in the Large Magellanic Cloud with ALMA}",
      journal = {\apj},
         year = 2016,
        month = aug,
       volume = {827},
       number = {1},
          eid = {72},
        pages = {72},
          doi = {10.3847/0004-637X/827/1/72},
archivePrefix = {arXiv},
       eprint = {1606.02823},
 primaryClass = {astro-ph.GA},
       adsurl = {https://ui.adsabs.harvard.edu/abs/2016ApJ...827...72S}
}

@ARTICLE{2020ApJ...891..164S,
       author = {{Shimonishi}, Takashi and {Das}, Ankan and {Sakai}, Nami and {Tanaka}, Kei E.~I. and {Aikawa}, Yuri and {Onaka}, Takashi and {Watanabe}, Yoshimasa and {Nishimura}, Yuri},
        title = "{Chemistry and Physics of a Low-metallicity Hot Core in the Large Magellanic Cloud}",
      journal = {\apj},
         year = 2020,
        month = mar,
       volume = {891},
       number = {2},
          eid = {164},
        pages = {164},
          doi = {10.3847/1538-4357/ab6e6b},
archivePrefix = {arXiv},
       eprint = {2001.06982},
 primaryClass = {astro-ph.GA},
       adsurl = {https://ui.adsabs.harvard.edu/abs/2020ApJ...891..164S}
}

@ARTICLE{2018ApJ...853L..19S,
       author = {{Sewi{\l}o}, Marta and {Indebetouw}, Remy and {Charnley}, Steven B. and {Zahorecz}, Sarolta and {Oliveira}, Joana M. and {van Loon}, Jacco Th. and {Ward}, Jacob L. and {Chen}, C. -H. Rosie and {Wiseman}, Jennifer and {Fukui}, Yasuo and {Kawamura}, Akiko and {Meixner}, Margaret and {Onishi}, Toshikazu and {Schilke}, Peter},
        title = "{The Detection of Hot Cores and Complex Organic Molecules in the Large Magellanic Cloud}",
      journal = {\apjl},
         year = 2018,
        month = feb,
       volume = {853},
       number = {2},
          eid = {L19},
        pages = {L19},
          doi = {10.3847/2041-8213/aaa079},
archivePrefix = {arXiv},
       eprint = {1801.10275},
 primaryClass = {astro-ph.GA},
       adsurl = {https://ui.adsabs.harvard.edu/abs/2018ApJ...853L..19S}
}

@ARTICLE{2022ApJ...931..102S,
       author = {{Sewi{\l}o}, Marta and {Cordiner}, Martin and {Charnley}, Steven B. and {Oliveira}, Joana M. and {Garcia-Berrios}, Emmanuel and {Schilke}, Peter and {Ward}, Jacob L. and {Wiseman}, Jennifer and {Indebetouw}, Remy and {Tokuda}, Kazuki and {van Loon}, Jacco Th. and {S{\'a}nchez-Monge}, {\'A}lvaro and {Allen}, Veronica and {Chen}, C. -H. Rosie and {Hamedani Golshan}, Roya and {Karska}, Agata and {Kristensen}, Lars E. and {Kurtz}, Stan E. and {M{\"o}ller}, Thomas and {Onishi}, Toshikazu and {Zahorecz}, Sarolta},
        title = "{ALMA Observations of Molecular Complexity in the Large Magellanic Cloud: The N 105 Star-forming Region}",
      journal = {\apj},
         year = 2022,
        month = jun,
       volume = {931},
       number = {2},
          eid = {102},
        pages = {102},
          doi = {10.3847/1538-4357/ac4e8f},
archivePrefix = {arXiv},
       eprint = {2201.09945},
 primaryClass = {astro-ph.GA},
       adsurl = {https://ui.adsabs.harvard.edu/abs/2022ApJ...931..102S}
}

@ARTICLE{2021A&A...653A.129C,
       author = {{Colzi}, L. and {Rivilla}, V.~M. and {Beltr{\'a}n}, M.~T. and {Jim{\'e}nez-Serra}, I. and {Mininni}, C. and {Melosso}, M. and {Cesaroni}, R. and {Fontani}, F. and {Lorenzani}, A. and {S{\'a}nchez-Monge}, A. and {Viti}, S. and {Schilke}, P. and {Testi}, L. and {Alonso}, E.~R. and {Kolesnikov{\'a}}, L.},
        title = "{The GUAPOS project. II. A comprehensive study of peptide-like bond molecules}",
      journal = {\aap},
         year = 2021,
        month = sep,
       volume = {653},
          eid = {A129},
        pages = {A129},
          doi = {10.1051/0004-6361/202141573},
archivePrefix = {arXiv},
       eprint = {2107.11258},
 primaryClass = {astro-ph.GA},
       adsurl = {https://ui.adsabs.harvard.edu/abs/2021A&A...653A.129C}
}

@ARTICLE{1988A&A...191..323W,
       author = {{Wouterloot}, J.~G.~A. and {Brand}, J. and {Henkel}, C.},
        title = "{Star formation in the outer Galaxy.}",
      journal = {\aap},
         year = 1988,
        month = feb,
       volume = {191},
        pages = {323-340},
       adsurl = {https://ui.adsabs.harvard.edu/abs/1988A&A...191..323W}
}

@ARTICLE{1993A&AS...98..589W,
       author = {{Wouterloot}, J.~G.~A. and {Brand}, J. and {Fiegle}, K.},
        title = "{IRAS sources beyond the solar circle. III. Observations of H2O, OH, CH3OH and CO.}",
      journal = {\aaps},
         year = 1993,
        month = may,
       volume = {98},
        pages = {589-636},
       adsurl = {https://ui.adsabs.harvard.edu/abs/1993A&AS...98..589W}
}

@ARTICLE{2017MNRAS.469.1383G,
       author = {{Green}, J.~A. and {Breen}, S.~L. and {Fuller}, G.~A. and {McClure-Griffiths}, N.~M. and {Ellingsen}, S.~P. and {Voronkov}, M.~A. and {Avison}, A. and {Brooks}, K. and {Burton}, M.~G. and {Chrysostomou}, A. and {Cox}, J. and {Diamond}, P.~J. and {Gray}, M.~D. and {Hoare}, M.~G. and {Masheder}, M.~R.~W. and {Pestalozzi}, M. and {Phillips}, C. and {Quinn}, L.~J. and {Richards}, A.~M.~S. and {Thompson}, M.~A. and {Walsh}, A.~J. and {Ward-Thompson}, D. and {Wong-McSweeney}, D. and {Yates}, J.~A.},
        title = "{The 6-GHz multibeam maser survey - II. Statistical analysis and Galactic distribution of 6668-MHz methanol masers}",
      journal = {\mnras},
         year = 2017,
        month = aug,
       volume = {469},
       number = {2},
        pages = {1383-1402},
          doi = {10.1093/mnras/stx887},
       adsurl = {https://ui.adsabs.harvard.edu/abs/2017MNRAS.469.1383G}
}

@ARTICLE{2005A&A...432..737P,
       author = {{Pestalozzi}, M.~R. and {Minier}, V. and {Booth}, R.~S.},
        title = "{A general catalogue of 6.7-GHz methanol masers. I. Data.}",
      journal = {\aap},
         year = 2005,
        month = mar,
       volume = {432},
       number = {2},
        pages = {737-742},
          doi = {10.1051/0004-6361:20035855},
archivePrefix = {arXiv},
       eprint = {astro-ph/0411564},
 primaryClass = {astro-ph},
       adsurl = {https://ui.adsabs.harvard.edu/abs/2005A&A...432..737P}
}

@ARTICLE{2019ApJ...885..131R,
       author = {{Reid}, M.~J. and {Menten}, K.~M. and {Brunthaler}, A. and {Zheng}, X.~W. and {Dame}, T.~M. and {Xu}, Y. and {Li}, J. and {Sakai}, N. and {Wu}, Y. and {Immer}, K. and {Zhang}, B. and {Sanna}, A. and {Moscadelli}, L. and {Rygl}, K.~L.~J. and {Bartkiewicz}, A. and {Hu}, B. and {Quiroga-Nu{\~n}ez}, L.~H. and {van Langevelde}, H.~J.},
        title = "{Trigonometric Parallaxes of High-mass Star-forming Regions: Our View of the Milky Way}",
      journal = {\apj},
         year = 2019,
        month = nov,
       volume = {885},
       number = {2},
          eid = {131},
        pages = {131},
          doi = {10.3847/1538-4357/ab4a11},
archivePrefix = {arXiv},
       eprint = {1910.03357},
 primaryClass = {astro-ph.GA},
       adsurl = {https://ui.adsabs.harvard.edu/abs/2019ApJ...885..131R}
}

@ARTICLE{1989A&AS...80..149W,
       author = {{Wouterloot}, J.~G.~A. and {Brand}, J.},
        title = "{IRAS sources beyond the solar circle. I. CO observations.}",
      journal = {\aaps},
         year = 1989,
        month = oct,
       volume = {80},
        pages = {149-187},
       adsurl = {https://ui.adsabs.harvard.edu/abs/1989A&AS...80..149W}
}

@ARTICLE{2022ApJS..259....1G,
       author = {{Garrod}, Robin T. and {Jin}, Miwha and {Matis}, Kayla A. and {Jones}, Dylan and {Willis}, Eric R. and {Herbst}, Eric},
        title = "{Formation of Complex Organic Molecules in Hot Molecular Cores through Nondiffusive Grain-surface and Ice-mantle Chemistry}",
      journal = {\apjs},
         year = 2022,
        month = mar,
       volume = {259},
       number = {1},
          eid = {1},
        pages = {1},
          doi = {10.3847/1538-4365/ac3131},
archivePrefix = {arXiv},
       eprint = {2110.09743},
 primaryClass = {astro-ph.GA},
       adsurl = {https://ui.adsabs.harvard.edu/abs/2022ApJS..259....1G}
}

@ARTICLE{2024ApJ...972..146S,
       author = {{Shope}, Brielle M. and {El-Abd}, Samer J. and {Brogan}, Crystal L. and {Hunter}, Todd R. and {Willis}, Eric R. and {McGuire}, Brett A. and {Garrod}, Robin T.},
        title = "{Interstellar Glycolaldehyde, Methyl Formate, and Acetic Acid. II. Chemical Modeling of the Bimodal Abundance Pattern in NGC 6334I}",
      journal = {\apj},
         year = 2024,
        month = sep,
       volume = {972},
       number = {2},
          eid = {146},
        pages = {146},
          doi = {10.3847/1538-4357/ad5d5f},
archivePrefix = {arXiv},
       eprint = {2409.13673},
 primaryClass = {astro-ph.GA},
       adsurl = {https://ui.adsabs.harvard.edu/abs/2024ApJ...972..146S}
}

@ARTICLE{2001A&A...377..644B,
       author = {{Brand}, J. and {Wouterloot}, J.~G.~A. and {Rudolph}, A.~L. and {de Geus}, E.~J.},
        title = "{Clumpy outer Galaxy molecular clouds and the steepening of the IMF}",
      journal = {\aap},
         year = 2001,
        month = oct,
       volume = {377},
        pages = {644-671},
          doi = {10.1051/0004-6361:20011057},
archivePrefix = {arXiv},
       eprint = {astro-ph/0107494},
 primaryClass = {astro-ph},
       adsurl = {https://ui.adsabs.harvard.edu/abs/2001A&A...377..644B}
}

@ARTICLE{1973AJ.....78..929P,
       author = {{Panagia}, Nino},
        title = "{Some Physical parameters of early-type stars}",
      journal = {\aj},
         year = 1973,
        month = nov,
       volume = {78},
        pages = {929-934},
          doi = {10.1086/111498},
       adsurl = {https://ui.adsabs.harvard.edu/abs/1973AJ.....78..929P}
}

@ARTICLE{2003A&A...403.1095M,
       author = {{Minier}, V. and {Ellingsen}, S.~P. and {Norris}, R.~P. and {Booth}, R.~S.},
        title = "{The protostellar mass limit for 6.7 GHz methanol masers.  I. A low-mass YSO survey}",
      journal = {\aap},
         year = 2003,
        month = jun,
       volume = {403},
        pages = {1095-1100},
          doi = {10.1051/0004-6361:20030465},
       adsurl = {https://ui.adsabs.harvard.edu/abs/2003A&A...403.1095M}
}

@ARTICLE{2006ApJ...638..241E,
       author = {{Ellingsen}, S.~P.},
        title = "{Methanol Masers: Reliable Tracers of the Early Stages of High-Mass Star Formation}",
      journal = {\apj},
         year = 2006,
        month = feb,
       volume = {638},
       number = {1},
        pages = {241-261},
          doi = {10.1086/498673},
archivePrefix = {arXiv},
       eprint = {astro-ph/0510218},
 primaryClass = {astro-ph},
       adsurl = {https://ui.adsabs.harvard.edu/abs/2006ApJ...638..241E}
}

@ARTICLE{2012ApJ...748...16T,
       author = {{Tobin}, John J. and {Hartmann}, Lee and {Bergin}, Edwin and {Chiang}, Hsin-Fang and {Looney}, Leslie W. and {Chandler}, Claire J. and {Maret}, S{\'e}bastien and {Heitsch}, Fabian},
        title = "{Complex Structure in Class 0 Protostellar Envelopes. III. Velocity Gradients in Non-axisymmetric Envelopes, Infall, or Rotation?}",
      journal = {\apj},
         year = 2012,
        month = mar,
       volume = {748},
       number = {1},
          eid = {16},
        pages = {16},
          doi = {10.1088/0004-637X/748/1/16},
archivePrefix = {arXiv},
       eprint = {1201.2174},
 primaryClass = {astro-ph.GA},
       adsurl = {https://ui.adsabs.harvard.edu/abs/2012ApJ...748...16T}
}

@ARTICLE{2024A&A...686L..13D,
       author = {{De Simone}, M. and {Podio}, L. and {Chahine}, L. and {Codella}, C. and {Chandler}, C.~J. and {Ceccarelli}, C. and {L{\'o}pez-Sepulcre}, A. and {Loinard}, L. and {Svoboda}, B. and {Sakai}, N. and {Johnstone}, D. and {M{\'e}nard}, F. and {Aikawa}, Y. and {Bouvier}, M. and {Sabatini}, G. and {Miotello}, A. and {Vastel}, C. and {Cuello}, N. and {Bianchi}, E. and {Caselli}, P. and {Caux}, E. and {Hanawa}, T. and {Herbst}, E. and {Segura-Cox}, D. and {Zhang}, Z. and {Yamamoto}, S.},
        title = "{FAUST. XV. A disc wind mapped by CH$_{3}$OH and SiO in the inner 300 au of the NGC 1333 IRAS 4A2 protostar}",
      journal = {\aap},
         year = 2024,
        month = jun,
       volume = {686},
          eid = {L13},
        pages = {L13},
          doi = {10.1051/0004-6361/202449949},
archivePrefix = {arXiv},
       eprint = {2404.19690},
 primaryClass = {astro-ph.SR},
       adsurl = {https://ui.adsabs.harvard.edu/abs/2024A&A...686L..13D}
}

@ARTICLE{2008AsBio...8...59B,
       author = {{Blair}, Samantha K. and {Magnani}, Loris and {Brand}, Jan and {Wouterloot}, Jan G.~A.},
        title = "{Formaldehyde in the Far Outer Galaxy: Constraining the Outer Boundary of the Galactic Habitable Zone}",
      journal = {Astrobiology},
         year = 2008,
        month = feb,
       volume = {8},
       number = {1},
        pages = {59-73},
          doi = {10.1089/ast.2007.0171},
       adsurl = {https://ui.adsabs.harvard.edu/abs/2008AsBio...8...59B}
}

@INPROCEEDINGS{1995RMxAC...1..137W,
       author = {{Walmsley}, M.},
        title = "{Dense Cores in Molecular Clouds}",
    booktitle = {Revista Mexicana de Astronomia y Astrofisica Conference Series},
         year = 1995,
       editor = {{Lizano}, S. and {Torrelles}, J.~M.},
       series = {Revista Mexicana de Astronomia y Astrofisica Conference Series},
       volume = {1},
        month = apr,
        pages = {137},
       adsurl = {https://ui.adsabs.harvard.edu/abs/1995RMxAC...1..137W}
}

@ARTICLE{2025A&A...698A.143B,
       author = {{Belloche}, A. and {Garrod}, R.~T. and {M{\"u}ller}, H.~S.~P. and {Morin}, N.~J. and {Willis}, S.~A. and {Menten}, K.~M.},
        title = "{Re-exploring Molecular Complexity with ALMA: Insights into chemical differentiation from the molecular composition of hot cores in Sgr B2(N2)}",
      journal = {\aap},
         year = 2025,
        month = jun,
       volume = {698},
          eid = {A143},
        pages = {A143},
          doi = {10.1051/0004-6361/202554411},
archivePrefix = {arXiv},
       eprint = {2505.03262},
 primaryClass = {astro-ph.GA},
       adsurl = {https://ui.adsabs.harvard.edu/abs/2025A&A...698A.143B}
}

@manual{cortes2025alma,
  author       = {Cortés, P. C. and Carpenter, J. and Kameno, S. and Loomis, R. and Vila-Vilaró, B. and Immer, K. and Vlahakis, C. and Law, J. and Stoehr, F. and Saini, K. and Hales, A.},
  title        = {ALMA Technical Handbook},
  year         = {2025},
  note         = {{ALMA} Doc. 12.3, ver. 1.5},
  isbn         = {978-3-923524-66-2},
  doi          = {10.5281/zenodo.14933753},
  url          = {https://doi.org/10.5281/zenodo.14933753}
}

@ARTICLE{2008A&A...487..993K,
       author = {{Kauffmann}, J. and {Bertoldi}, F. and {Bourke}, T.~L. and {Evans}, II, N.~J. and {Lee}, C.~W.},
        title = "{MAMBO mapping of Spitzer c2d small clouds and cores}",
      journal = {\aap},
         year = 2008,
        month = sep,
       volume = {487},
       number = {3},
        pages = {993-1017},
          doi = {10.1051/0004-6361:200809481},
archivePrefix = {arXiv},
       eprint = {0805.4205},
 primaryClass = {astro-ph},
       adsurl = {https://ui.adsabs.harvard.edu/abs/2008A&A...487..993K}
}

@ARTICLE{1994ARA&A..32..191W,
       author = {{Wilson}, T.~L. and {Rood}, R.},
        title = "{Abundances in the Interstellar Medium}",
      journal = {\araa},
         year = 1994,
        month = jan,
       volume = {32},
        pages = {191-226},
          doi = {10.1146/annurev.aa.32.090194.001203},
       adsurl = {https://ui.adsabs.harvard.edu/abs/1994ARA&A..32..191W}
}

@ARTICLE{2002A&A...383..892B,
       author = {{Beuther}, H. and {Schilke}, P. and {Sridharan}, T.~K. and {Menten}, K.~M. and {Walmsley}, C.~M. and {Wyrowski}, F.},
        title = "{Massive molecular outflows}",
      journal = {\aap},
         year = 2002,
        month = mar,
       volume = {383},
        pages = {892-904},
          doi = {10.1051/0004-6361:20011808},
archivePrefix = {arXiv},
       eprint = {astro-ph/0110372},
 primaryClass = {astro-ph},
       adsurl = {https://ui.adsabs.harvard.edu/abs/2002A&A...383..892B}
}

@ARTICLE{2015MNRAS.453..645M,
       author = {{Maud}, L.~T. and {Moore}, T.~J.~T. and {Lumsden}, S.~L. and {Mottram}, J.~C. and {Urquhart}, J.~S. and {Hoare}, M.~G.},
        title = "{A distance-limited sample of massive molecular outflows}",
      journal = {\mnras},
         year = 2015,
        month = oct,
       volume = {453},
       number = {1},
        pages = {645-665},
          doi = {10.1093/mnras/stv1635},
archivePrefix = {arXiv},
       eprint = {1509.00199},
 primaryClass = {astro-ph.SR},
       adsurl = {https://ui.adsabs.harvard.edu/abs/2015MNRAS.453..645M}
}

@ARTICLE{2016ARA&A..54..491B,
       author = {{Bally}, John},
        title = "{Protostellar Outflows}",
      journal = {\araa},
         year = 2016,
        month = sep,
       volume = {54},
        pages = {491-528},
          doi = {10.1146/annurev-astro-081915-023341},
       adsurl = {https://ui.adsabs.harvard.edu/abs/2016ARA&A..54..491B}
}

@ARTICLE{1997ApJ...487L..93B,
       author = {{Bachiller}, R. and {P{\'e}rez Guti{\'e}rrez}, M.},
        title = "{Shock Chemistry in the Young Bipolar Outflow L1157}",
      journal = {\apjl},
         year = 1997,
        month = sep,
       volume = {487},
       number = {1},
        pages = {L93-L96},
          doi = {10.1086/310877},
       adsurl = {https://ui.adsabs.harvard.edu/abs/1997ApJ...487L..93B}
}

@ARTICLE{2025ApJ...979...17L,
       author = {{Liu}, Chun-Fan and {Shang}, Hsien and {Johnstone}, Doug and {Ai}, Tsung-Han and {Lee}, Tsz Ming and {Krasnopolsky}, Ruben and {Hirano}, Naomi and {Dutta}, Somnath and {Hsu}, Shih-Ying and {L{\'o}pez-V{\'a}zquez}, Jes{\'u}s Alejandro and {Liu}, Sheng-Yuan and {Liu}, Tie and {Tatematsu}, Ken'ichi and {Zhang}, Qizhou and {Rawlings}, Mark G. and {Eden}, David and {Ren}, Zhiyuan and {Sanhueza}, Patricio and {Kwon}, Woojin and {Lee}, Chang Won and {Kuan}, Yi-Jehng and {Bandopadhyay}, Somdeb and {V{\"a}is{\"a}l{\"a}}, Miikka S. and {Lee}, Chin-Fei and {Das}, Indrani},
        title = "{ALMA Survey of Orion Planck Galactic Cold Clumps (ALMASOP): Nested Morphological and Kinematic Structures of Outflows Revealed in SiO and CO Emission}",
      journal = {\apj},
         year = 2025,
        month = jan,
       volume = {979},
       number = {1},
          eid = {17},
        pages = {17},
          doi = {10.3847/1538-4357/ad9275},
archivePrefix = {arXiv},
       eprint = {2411.08827},
 primaryClass = {astro-ph.SR},
       adsurl = {https://ui.adsabs.harvard.edu/abs/2025ApJ...979...17L}
}

@ARTICLE{2001ApJ...562..770D,
       author = {{Di Francesco}, James and {Myers}, Philip C. and {Wilner}, David J. and {Ohashi}, Nagayoshi and {Mardones}, Diego},
        title = "{Infall, Outflow, Rotation, and Turbulent Motions of Dense Gas within NGC 1333 IRAS 4}",
      journal = {\apj},
         year = 2001,
        month = dec,
       volume = {562},
       number = {2},
        pages = {770-789},
          doi = {10.1086/323854},
archivePrefix = {arXiv},
       eprint = {astro-ph/0108022},
 primaryClass = {astro-ph},
       adsurl = {https://ui.adsabs.harvard.edu/abs/2001ApJ...562..770D}
}

@ARTICLE{2013A&A...558A.126M,
       author = {{Mottram}, J.~C. and {van Dishoeck}, E.~F. and {Schmalzl}, M. and {Kristensen}, L.~E. and {Visser}, R. and {Hogerheijde}, M.~R. and {Bruderer}, S.},
        title = "{Waterfalls around protostars. Infall motions towards Class 0/I envelopes as probed by water}",
      journal = {\aap},
         year = 2013,
        month = oct,
       volume = {558},
          eid = {A126},
        pages = {A126},
          doi = {10.1051/0004-6361/201321828},
archivePrefix = {arXiv},
       eprint = {1308.5119},
 primaryClass = {astro-ph.GA},
       adsurl = {https://ui.adsabs.harvard.edu/abs/2013A&A...558A.126M}
}

@ARTICLE{2015ApJ...814...22E,
       author = {{Evans}, II, Neal J. and {Di Francesco}, James and {Lee}, Jeong-Eun and {J{\o}rgensen}, Jes K. and {Choi}, Minho and {Myers}, Philip C. and {Mardones}, Diego},
        title = "{Detection of Infall in the Protostar B335 with ALMA}",
      journal = {\apj},
         year = 2015,
        month = nov,
       volume = {814},
       number = {1},
          eid = {22},
        pages = {22},
          doi = {10.1088/0004-637X/814/1/22},
archivePrefix = {arXiv},
       eprint = {1510.03476},
 primaryClass = {astro-ph.GA},
       adsurl = {https://ui.adsabs.harvard.edu/abs/2015ApJ...814...22E}
}

@ARTICLE{2021A&A...656A..46M,
       author = {{Mart{\'\i}n}, S. and {Mangum}, J.~G. and {Harada}, N. and {Costagliola}, F. and {Sakamoto}, K. and {Muller}, S. and {Aladro}, R. and {Tanaka}, K. and {Yoshimura}, Y. and {Nakanishi}, K. and {Herrero-Illana}, R. and {M{\"u}hle}, S. and {Aalto}, S. and {Behrens}, E. and {Colzi}, L. and {Emig}, K.~L. and {Fuller}, G.~A. and {Garc{\'\i}a-Burillo}, S. and {Greve}, T.~R. and {Henkel}, C. and {Holdship}, J. and {Humire}, P. and {Hunt}, L. and {Izumi}, T. and {Kohno}, K. and {K{\"o}nig}, S. and {Meier}, D.~S. and {Nakajima}, T. and {Nishimura}, Y. and {Padovani}, M. and {Rivilla}, V.~M. and {Takano}, S. and {van der Werf}, P.~P. and {Viti}, S. and {Yan}, Y.~T.},
        title = "{ALCHEMI, an ALMA Comprehensive High-resolution Extragalactic Molecular Inventory. Survey presentation and first results from the ACA array}",
      journal = {\aap},
         year = 2021,
        month = dec,
       volume = {656},
          eid = {A46},
        pages = {A46},
          doi = {10.1051/0004-6361/202141567},
archivePrefix = {arXiv},
       eprint = {2109.08638},
 primaryClass = {astro-ph.GA},
       adsurl = {https://ui.adsabs.harvard.edu/abs/2021A&A...656A..46M}
}

@ARTICLE{2025A&A...698A.261B,
       author = {{Bouvier}, M. and {Viti}, S. and {Mangum}, J.~G. and {Eibensteiner}, C. and {Behrens}, E. and {Rivilla}, V.~M. and {L{\'o}pez-Gallifa}, {\'A}. and {Mart{\'\i}n}, S. and {Harada}, N. and {Muller}, S. and {Colzi}, L. and {Sakamoto}, K.},
        title = "{Complex organic molecules towards the central molecular zone of NGC 253}",
      journal = {\aap},
         year = 2025,
        month = jun,
       volume = {698},
          eid = {A261},
        pages = {A261},
          doi = {10.1051/0004-6361/202554420},
archivePrefix = {arXiv},
       eprint = {2504.19631},
 primaryClass = {astro-ph.GA},
       adsurl = {https://ui.adsabs.harvard.edu/abs/2025A&A...698A.261B}
}

@ARTICLE{2008A&A...482..549J,
       author = {{Jim{\'e}nez-Serra}, I. and {Caselli}, P. and {Mart{\'\i}n-Pintado}, J. and {Hartquist}, T.~W.},
        title = "{Parametrization of C-shocks. Evolution of the sputtering of grains}",
      journal = {\aap},
         year = 2008,
        month = may,
       volume = {482},
       number = {2},
        pages = {549-559},
          doi = {10.1051/0004-6361:20078054},
archivePrefix = {arXiv},
       eprint = {0802.0594},
 primaryClass = {astro-ph},
       adsurl = {https://ui.adsabs.harvard.edu/abs/2008A&A...482..549J}
}

@ARTICLE{2021A&A...655A..65T,
       author = {{Tychoniec}, {\L}ukasz and {van Dishoeck}, Ewine F. and {van't Hoff}, Merel L.~R. and {van Gelder}, Martijn L. and {Tabone}, Beno{\^\i}t and {Chen}, Yuan and {Harsono}, Daniel and {Hull}, Charles L.~H. and {Hogerheijde}, Michiel R. and {Murillo}, Nadia M. and {Tobin}, John J.},
        title = "{Which molecule traces what: Chemical diagnostics of protostellar sources}",
      journal = {\aap},
         year = 2021,
        month = nov,
       volume = {655},
          eid = {A65},
        pages = {A65},
          doi = {10.1051/0004-6361/202140692},
archivePrefix = {arXiv},
       eprint = {2107.03696},
 primaryClass = {astro-ph.SR},
       adsurl = {https://ui.adsabs.harvard.edu/abs/2021A&A...655A..65T}
}

@ARTICLE{2022A&A...667A..20A,
       author = {{Artur de la Villarmois}, E. and {Guzm{\'a}n}, V.~V. and {J{\o}rgensen}, J.~K. and {Kristensen}, L.~E. and {Bergin}, E.~A. and {Harsono}, D. and {Sakai}, N. and {van Dishoeck}, E.~F. and {Yamamoto}, S.},
        title = "{Physical properties of accretion shocks toward the Class I protostellar system Oph-IRS 44}",
      journal = {\aap},
         year = 2022,
        month = nov,
       volume = {667},
          eid = {A20},
        pages = {A20},
          doi = {10.1051/0004-6361/202244312},
archivePrefix = {arXiv},
       eprint = {2209.02827},
 primaryClass = {astro-ph.SR},
       adsurl = {https://ui.adsabs.harvard.edu/abs/2022A&A...667A..20A}
}

@ARTICLE{2014MNRAS.440.1844S,
       author = {{Suutarinen}, A.~N. and {Kristensen}, L.~E. and {Mottram}, J.~C. and {Fraser}, H.~J. and {van Dishoeck}, E.~F.},
        title = "{Water and methanol in low-mass protostellar outflows: gas-phase synthesis, ice sputtering and destruction}",
      journal = {\mnras},
         year = 2014,
        month = may,
       volume = {440},
       number = {2},
        pages = {1844-1855},
          doi = {10.1093/mnras/stu406},
archivePrefix = {arXiv},
       eprint = {1402.7214},
 primaryClass = {astro-ph.GA},
       adsurl = {https://ui.adsabs.harvard.edu/abs/2014MNRAS.440.1844S}
}

@ARTICLE{2002ApJ...571L.173W,
       author = {{Watanabe}, Naoki and {Kouchi}, Akira},
        title = "{Efficient Formation of Formaldehyde and Methanol by the Addition of Hydrogen Atoms to CO in H$_{2}$O-CO Ice at 10 K}",
      journal = {\apjl},
         year = 2002,
        month = jun,
       volume = {571},
       number = {2},
        pages = {L173-L176},
          doi = {10.1086/341412},
       adsurl = {https://ui.adsabs.harvard.edu/abs/2002ApJ...571L.173W}
}

@ARTICLE{2010MNRAS.409...29F,
       author = {{Flower}, D.~R. and {Pineau des For{\^e}ts}, G. and {Rabli}, Djamal},
        title = "{Methanol line formation in outflow sources}",
      journal = {\mnras},
         year = 2010,
        month = nov,
       volume = {409},
       number = {1},
        pages = {29-37},
          doi = {10.1111/j.1365-2966.2010.17501.x},
       adsurl = {https://ui.adsabs.harvard.edu/abs/2010MNRAS.409...29F}
}

@ARTICLE{2020A&A...634A..52S,
       author = {{Simons}, M.~A.~J. and {Lamberts}, T. and {Cuppen}, H.~M.},
        title = "{Formation of COMs through CO hydrogenation on interstellar grains}",
      journal = {\aap},
         year = 2020,
        month = feb,
       volume = {634},
          eid = {A52},
        pages = {A52},
          doi = {10.1051/0004-6361/201936522},
archivePrefix = {arXiv},
       eprint = {2001.04895},
 primaryClass = {astro-ph.SR},
       adsurl = {https://ui.adsabs.harvard.edu/abs/2020A&A...634A..52S}
}

@ARTICLE{2022MNRAS.511.3463Q,
       author = {{Qin}, Sheng-Li and {Liu}, Tie and {Liu}, Xunchuan and {Goldsmith}, Paul F. and {Li}, Di and {Zhang}, Qizhou and {Liu}, Hong-Li and {Wu}, Yuefang and {Bronfman}, Leonardo and {Juvela}, Mika and {Lee}, Chang Won and {Garay}, Guido and {Zhang}, Yong and {He}, Jinhua and {Hsu}, Shih-Ying and {Shen}, Zhi-Qiang and {Lee}, Jeong-Eun and {Wang}, Ke and {Tang}, Ningyu and {Tang}, Mengyao and {Zhang}, Chao and {Yue}, Yinghua and {Xue}, Qiaowei and {Li}, Shanghuo and {Peng}, Yaping and {Dutta}, Somnath and {Ge}, Jixing and {Xu}, Fengwei and {Chen}, Long-Fei and {Baug}, Tapas and {Dewangan}, Lokesh and {Tej}, Anandmayee},
        title = "{ATOMS: ALMA Three-millimeter Observations of Massive Star-forming regions - VIII. A search for hot cores by using C$_{2}$H$_{5}$CN, CH$_{3}$OCHO, and CH$_{3}$OH lines}",
      journal = {\mnras},
         year = 2022,
        month = apr,
       volume = {511},
       number = {3},
        pages = {3463-3476},
          doi = {10.1093/mnras/stac219},
archivePrefix = {arXiv},
       eprint = {2201.10044},
 primaryClass = {astro-ph.GA},
       adsurl = {https://ui.adsabs.harvard.edu/abs/2022MNRAS.511.3463Q}
}

@ARTICLE{2008ApJ...672..352R,
       author = {{Requena-Torres}, M.~A. and {Mart{\'\i}n-Pintado}, J. and {Mart{\'\i}n}, S. and {Morris}, M.~R.},
        title = "{The Galactic Center: The Largest Oxygen-bearing Organic Molecule Repository}",
      journal = {\apj},
         year = 2008,
        month = jan,
       volume = {672},
       number = {1},
        pages = {352-360},
          doi = {10.1086/523627},
archivePrefix = {arXiv},
       eprint = {0709.0542},
 primaryClass = {astro-ph},
       adsurl = {https://ui.adsabs.harvard.edu/abs/2008ApJ...672..352R}
}

@ARTICLE{2018MNRAS.478.2962Z,
       author = {{Zeng}, S. and {Jim{\'e}nez-Serra}, I. and {Rivilla}, V.~M. and {Mart{\'\i}n}, S. and {Mart{\'\i}n-Pintado}, J. and {Requena-Torres}, M.~A. and {Armijos-Abenda{\~n}o}, J. and {Riquelme}, D. and {Aladro}, R.},
        title = "{Complex organic molecules in the Galactic Centre: the N-bearing family}",
      journal = {\mnras},
         year = 2018,
        month = aug,
       volume = {478},
       number = {3},
        pages = {2962-2975},
          doi = {10.1093/mnras/sty1174},
archivePrefix = {arXiv},
       eprint = {1804.11321},
 primaryClass = {astro-ph.GA},
       adsurl = {https://ui.adsabs.harvard.edu/abs/2018MNRAS.478.2962Z}
}

@ARTICLE{2019A&A...632A.123I,
       author = {{Immer}, K. and {Li}, J. and {Quiroga-Nu{\~n}ez}, L.~H. and {Reid}, M.~J. and {Zhang}, B. and {Moscadelli}, L. and {Rygl}, K.~L.~J.},
        title = "{Anomalous peculiar motions of high-mass young stars in the Scutum spiral arm}",
      journal = {\aap},
         year = 2019,
        month = dec,
       volume = {632},
          eid = {A123},
        pages = {A123},
          doi = {10.1051/0004-6361/201834208},
archivePrefix = {arXiv},
       eprint = {1911.06806},
 primaryClass = {astro-ph.GA},
       adsurl = {https://ui.adsabs.harvard.edu/abs/2019A&A...632A.123I}
}

@ARTICLE{2007A&A...464..909B,
       author = {{Brand}, J. and {Wouterloot}, J.~G.~A.},
        title = "{A star cluster at the edge of the Galaxy}",
      journal = {\aap},
         year = 2007,
        month = mar,
       volume = {464},
       number = {3},
        pages = {909-920},
          doi = {10.1051/0004-6361:20065437},
archivePrefix = {arXiv},
       eprint = {astro-ph/0702541},
 primaryClass = {astro-ph},
       adsurl = {https://ui.adsabs.harvard.edu/abs/2007A&A...464..909B}
}

@ARTICLE{2020ApJS..249...26J,
       author = {{Jin}, Mihwa and {Garrod}, Robin T.},
        title = "{Formation of Complex Organic Molecules in Cold Interstellar Environments through Nondiffusive Grain-surface and Ice-mantle Chemistry}",
      journal = {\apjs},
         year = 2020,
        month = aug,
       volume = {249},
       number = {2},
          eid = {26},
        pages = {26},
          doi = {10.3847/1538-4365/ab9ec8},
archivePrefix = {arXiv},
       eprint = {2006.11127},
 primaryClass = {astro-ph.GA},
       adsurl = {https://ui.adsabs.harvard.edu/abs/2020ApJS..249...26J}
}

@ARTICLE{2017A&A...604A..58H,
       author = {{Hocuk}, S. and {Sz{\H{u}}cs}, L. and {Caselli}, P. and {Cazaux}, S. and {Spaans}, M. and {Esplugues}, G.~B.},
        title = "{Parameterizing the interstellar dust temperature}",
      journal = {\aap},
         year = 2017,
        month = aug,
       volume = {604},
          eid = {A58},
        pages = {A58},
          doi = {10.1051/0004-6361/201629944},
archivePrefix = {arXiv},
       eprint = {1704.02763},
 primaryClass = {astro-ph.GA},
       adsurl = {https://ui.adsabs.harvard.edu/abs/2017A&A...604A..58H}
}

@ARTICLE{2022ApJ...935..133J,
       author = {{Jin}, Miwha and {Lam}, Ka Ho and {McClure}, Melissa K. and {van Scheltinga}, Jeroen Terwisscha and {Li}, Zhi-Yun and {Boogert}, Adwin and {Herbst}, Eric and {Davis}, Shane W. and {Garrod}, Robin T.},
        title = "{Ice Age: Chemodynamical Modeling of Cha-MMS1 to Predict New Solid-phase Species for Detection with JWST}",
      journal = {\apj},
         year = 2022,
        month = aug,
       volume = {935},
       number = {2},
          eid = {133},
        pages = {133},
          doi = {10.3847/1538-4357/ac8006},
archivePrefix = {arXiv},
       eprint = {2207.04269},
 primaryClass = {astro-ph.SR},
       adsurl = {https://ui.adsabs.harvard.edu/abs/2022ApJ...935..133J}
}

@ARTICLE{2016MNRAS.455.1855C,
       author = {{Choudhury}, Samyaday and {Subramaniam}, Annapurni and {Cole}, Andrew A.},
        title = "{Photometric metallicity map of the Large Magellanic Cloud}",
      journal = {\mnras},
         year = 2016,
        month = jan,
       volume = {455},
       number = {2},
        pages = {1855-1880},
          doi = {10.1093/mnras/stv2414},
archivePrefix = {arXiv},
       eprint = {1510.05769},
 primaryClass = {astro-ph.GA},
       adsurl = {https://ui.adsabs.harvard.edu/abs/2016MNRAS.455.1855C}
}

@ARTICLE{2018MNRAS.475.4279C,
       author = {{Choudhury}, S. and {Subramaniam}, A. and {Cole}, A.~A. and {Sohn}, Y. -J.},
        title = "{Photometric metallicity map of the Small Magellanic Cloud}",
      journal = {\mnras},
         year = 2018,
        month = apr,
       volume = {475},
       number = {4},
        pages = {4279-4297},
          doi = {10.1093/mnras/sty087},
archivePrefix = {arXiv},
       eprint = {1801.03403},
 primaryClass = {astro-ph.GA},
       adsurl = {https://ui.adsabs.harvard.edu/abs/2018MNRAS.475.4279C}
}

@ARTICLE{2021MNRAS.507.4752C,
       author = {{Choudhury}, Samyaday and {de Grijs}, Richard and {Bekki}, Kenji and {Cioni}, Maria-Rosa L. and {Ivanov}, Valentin D. and {van Loon}, Jacco Th and {Miller}, Amy E. and {Niederhofer}, Florian and {Oliveira}, Joana M. and {Ripepi}, Vincenzo and {Sun}, Ning-Chen and {Subramanian}, Smitha},
        title = "{The VMC survey - XLIV: mapping metallicity trends in the large magellanic cloud using near-infrared passbands}",
      journal = {\mnras},
         year = 2021,
        month = nov,
       volume = {507},
       number = {4},
        pages = {4752-4763},
          doi = {10.1093/mnras/stab2446},
archivePrefix = {arXiv},
       eprint = {2108.10529},
 primaryClass = {astro-ph.GA},
       adsurl = {https://ui.adsabs.harvard.edu/abs/2021MNRAS.507.4752C}
}

@ARTICLE{2016ApJ...823..125C,
       author = {{Chen}, Huei-Ru Vivien and {Keto}, Eric and {Zhang}, Qizhou and {Sridharan}, T.~K. and {Liu}, Sheng-Yuan and {Su}, Yu-Nung},
        title = "{A Hot and Massive Accretion Disk around the High-mass Protostar IRAS 20126+4104}",
      journal = {\apj},
         year = 2016,
        month = jun,
       volume = {823},
       number = {2},
          eid = {125},
        pages = {125},
          doi = {10.3847/0004-637X/823/2/125},
archivePrefix = {arXiv},
       eprint = {1604.00523},
 primaryClass = {astro-ph.GA},
       adsurl = {https://ui.adsabs.harvard.edu/abs/2016ApJ...823..125C}
}

@ARTICLE{2019ApJ...882...49L,
       author = {{Long}, Feng and {Herczeg}, Gregory J. and {Harsono}, Daniel and {Pinilla}, Paola and {Tazzari}, Marco and {Manara}, Carlo F. and {Pascucci}, Ilaria and {Cabrit}, Sylvie and {Nisini}, Brunella and {Johnstone}, Doug and {Edwards}, Suzan and {Salyk}, Colette and {Menard}, Francois and {Lodato}, Giuseppe and {Boehler}, Yann and {Mace}, Gregory N. and {Liu}, Yao and {Mulders}, Gijs D. and {Hendler}, Nathanial and {Ragusa}, Enrico and {Fischer}, William J. and {Banzatti}, Andrea and {Rigliaco}, Elisabetta and {van de Plas}, Gerrit and {Dipierro}, Giovanni and {Gully-Santiago}, Michael and {Lopez-Valdivia}, Ricardo},
        title = "{Compact Disks in a High-resolution ALMA Survey of Dust Structures in the Taurus Molecular Cloud}",
      journal = {\apj},
         year = 2019,
        month = sep,
       volume = {882},
       number = {1},
          eid = {49},
        pages = {49},
          doi = {10.3847/1538-4357/ab2d2d},
archivePrefix = {arXiv},
       eprint = {1906.10809},
 primaryClass = {astro-ph.SR},
       adsurl = {https://ui.adsabs.harvard.edu/abs/2019ApJ...882...49L}
}

@ARTICLE{2023A&A...670A..98Y,
       author = {{Yan}, Y.~T. and {Henkel}, C. and {Kobayashi}, C. and {Menten}, K.~M. and {Gong}, Y. and {Zhang}, J.~S. and {Yu}, H.~Z. and {Yang}, K. and {Xie}, J.~J. and {Wang}, Y.~X.},
        title = "{Direct measurements of carbon and sulfur isotope ratios in the Milky Way}",
      journal = {\aap},
         year = 2023,
        month = feb,
       volume = {670},
          eid = {A98},
        pages = {A98},
          doi = {10.1051/0004-6361/202244584},
archivePrefix = {arXiv},
       eprint = {2212.03252},
 primaryClass = {astro-ph.GA},
       adsurl = {https://ui.adsabs.harvard.edu/abs/2023A&A...670A..98Y}
}

@ARTICLE{2014JChPh.140f4307C,
       author = {{Coudert}, L.~H. and {Zemouli}, M. and {Motiyenko}, R.~A. and {Margul{\`e}s}, L. and {Klee}, S.},
        title = "{Analysis of the microwave, terahertz, and far infrared spectra of monodeuterated methanol CH$_{2}$DOH up to J = 26, K = 11, and o$_{1}$}",
      journal = {\jcp},
         year = 2014,
        month = feb,
       volume = {140},
       number = {6},
          eid = {064307},
        pages = {064307},
          doi = {10.1063/1.4864203},
       adsurl = {https://ui.adsabs.harvard.edu/abs/2014JChPh.140f4307C}
}

@ARTICLE{2018MNRAS.479.2007A,
       author = {{{\'A}lvarez-Barcia}, S. and {Russ}, P. and {K{\"a}stner}, J. and {Lamberts}, T.},
        title = "{Hydrogen transfer reactions of interstellar complex organic molecules}",
      journal = {\mnras},
         year = 2018,
        month = sep,
       volume = {479},
       number = {2},
        pages = {2007-2015},
          doi = {10.1093/mnras/sty1478},
archivePrefix = {arXiv},
       eprint = {1806.02062},
 primaryClass = {astro-ph.GA},
       adsurl = {https://ui.adsabs.harvard.edu/abs/2018MNRAS.479.2007A}
}

@ARTICLE{2005JMoSt.742..215M,
       author = {{M{\"u}ller}, Holger S.~P. and {Schl{\"o}der}, Frank and {Stutzki}, J{\"u}rgen and {Winnewisser}, Gisbert},
        title = "{The Cologne Database for Molecular Spectroscopy, CDMS: a useful tool for astronomers and spectroscopists}",
      journal = {Journal of Molecular Structure},
         year = 2005,
        month = may,
       volume = {742},
       number = {1-3},
        pages = {215-227},
          doi = {10.1016/j.molstruc.2005.01.027},
       adsurl = {https://ui.adsabs.harvard.edu/abs/2005JMoSt.742..215M}
}

@ARTICLE{2025A&A...699A.348M,
       author = {{Motiyenko}, Roman A. and {Margul{\`e}s}, Laurent},
        title = "{Lille spectroscopic database for astrophysically and atmospherically relevant molecules}",
      journal = {\aap},
         year = 2025,
        month = jul,
       volume = {699},
          eid = {A348},
        pages = {A348},
          doi = {10.1051/0004-6361/202554385},
       adsurl = {https://ui.adsabs.harvard.edu/abs/2025A&A...699A.348M}
}

@ARTICLE{1998JQSRT..60..883P,
       author = {{Pickett}, H.~M. and {Poynter}, R.~L. and {Cohen}, E.~A. and {Delitsky}, M.~L. and {Pearson}, J.~C. and {M{\"u}ller}, H.~S.~P.},
        title = "{Submillimeter, millimeter and microwave spectral line catalog.}",
      journal = {\jqsrt},
         year = 1998,
        month = nov,
       volume = {60},
       number = {5},
        pages = {883-890},
          doi = {10.1016/S0022-4073(98)00091-0},
       adsurl = {https://ui.adsabs.harvard.edu/abs/1998JQSRT..60..883P}
}

@ARTICLE{2024MNRAS.529.3244L,
       author = {{L{\'o}pez-Gallifa}, {\'A}. and {Rivilla}, V.~M. and {Beltr{\'a}n}, M.~T. and {Colzi}, L. and {Mininni}, C. and {S{\'a}nchez-Monge}, {\'A}. and {Fontani}, F. and {Viti}, S. and {Jim{\'e}nez-Serra}, I. and {Testi}, L. and {Cesaroni}, R. and {Lorenzani}, A.},
        title = "{The GUAPOS project - V: The chemical ingredients of a massive stellar protocluster in the making}",
      journal = {\mnras},
         year = 2024,
        month = apr,
       volume = {529},
       number = {4},
        pages = {3244-3283},
          doi = {10.1093/mnras/stae676},
archivePrefix = {arXiv},
       eprint = {2403.02191},
 primaryClass = {astro-ph.GA},
       adsurl = {https://ui.adsabs.harvard.edu/abs/2024MNRAS.529.3244L}
}

@ARTICLE{2022A&A...658A.150G,
       author = {{Garc{\'\i}a de la Concepci{\'o}n}, J. and {Colzi}, L. and {Jim{\'e}nez-Serra}, I. and {Molpeceres}, G. and {Corchado}, J.~C. and {Rivilla}, V.~M. and {Mart{\'\i}n-Pintado}, J. and {Beltr{\'a}n}, M.~T. and {Mininni}, C.},
        title = "{The trans/cis ratio of formic (HCOOH) and thioformic (HC(O)SH) acids in the interstellar medium}",
      journal = {\aap},
         year = 2022,
        month = feb,
       volume = {658},
          eid = {A150},
        pages = {A150},
          doi = {10.1051/0004-6361/202142287},
archivePrefix = {arXiv},
       eprint = {2111.10842},
 primaryClass = {astro-ph.GA},
       adsurl = {https://ui.adsabs.harvard.edu/abs/2022A&A...658A.150G}
}

@ARTICLE{2019A&A...621A..76M,
       author = {{Maury}, A.~J. and {Andr{\'e}}, Ph. and {Testi}, L. and {Maret}, S. and {Belloche}, A. and {Hennebelle}, P. and {Cabrit}, S. and {Codella}, C. and {Gueth}, F. and {Podio}, L. and {Anderl}, S. and {Bacmann}, A. and {Bontemps}, S. and {Gaudel}, M. and {Ladjelate}, B. and {Lef{\`e}vre}, C. and {Tabone}, B. and {Lefloch}, B.},
        title = "{Characterizing young protostellar disks with the CALYPSO IRAM-PdBI survey: large Class 0 disks are rare}",
      journal = {\aap},
         year = 2019,
        month = jan,
       volume = {621},
          eid = {A76},
        pages = {A76},
          doi = {10.1051/0004-6361/201833537},
archivePrefix = {arXiv},
       eprint = {1810.11221},
 primaryClass = {astro-ph.GA},
       adsurl = {https://ui.adsabs.harvard.edu/abs/2019A&A...621A..76M}
}

@ARTICLE{2011ApJ...735...15G,
       author = {{Garrod}, R.~T. and {Pauly}, T.},
        title = "{On the Formation of CO$_{2}$ and Other Interstellar Ices}",
      journal = {\apj},
         year = 2011,
        month = jul,
       volume = {735},
       number = {1},
          eid = {15},
        pages = {15},
          doi = {10.1088/0004-637X/735/1/15},
archivePrefix = {arXiv},
       eprint = {1106.0540},
 primaryClass = {astro-ph.GA},
       adsurl = {https://ui.adsabs.harvard.edu/abs/2011ApJ...735...15G}
}

@ARTICLE{2025AJ....169..307F,
       author = {{Fadul}, Abubakar M.~A. and {Schwarz}, Kamber R. and {van'T Hoff}, Merel L.~R. and {Huang}, Jane and {Bergner}, Jennifer B. and {Suhasaria}, Tushar and {Calahan}, Jenny K.},
        title = "{A Deep Search for Complex Organic Molecules toward the Protoplanetary Disk of V883 Ori}",
      journal = {\aj},
         year = 2025,
        month = jun,
       volume = {169},
       number = {6},
          eid = {307},
        pages = {307},
          doi = {10.3847/1538-3881/adc998},
archivePrefix = {arXiv},
       eprint = {2504.06005},
 primaryClass = {astro-ph.SR},
       adsurl = {https://ui.adsabs.harvard.edu/abs/2025AJ....169..307F}
}

@ARTICLE{2025ApJ...989...56H,
       author = {{Hsu}, Shih-Ying and {Lee}, Chin-Fei and {Johnstone}, Doug and {Liu}, Sheng-Yuan and {Liu}, Tie and {Bronfman}, Leonardo and {Chen}, Huei-Ru Vivien and {Dutta}, Somnath and {Eden}, David J. and {Hirano}, Naomi and {Juvela}, Mika and {Kim}, Kee-Tae and {Kuan}, Yi-Jehng and {Kwon}, Woojin and {Lee}, Chang Won and {Lee}, Jeong-Eun and {Li}, Shanghuo and {Lin}, Sheng-Jun and {Liu}, Chun-Fan and {Liu}, Xunchuan and {L{\'o}pez-V{\'a}zquez}, J.~A. and {Luo}, Qiuyi and {Rawlings}, Mark G. and {Sahu}, Dipen and {Sanhueza}, Patricio and {Shang}, Hsien and {Tatematsu}, Ken'ichi and {Yang}, Yao-Lun},
        title = "{ALMASOP. A Rotating Feature Rich in Complex Organic Molecules in a Protostellar Core}",
      journal = {\apj},
         year = 2025,
        month = aug,
       volume = {989},
       number = {1},
          eid = {56},
        pages = {56},
          doi = {10.3847/1538-4357/ade7fc},
archivePrefix = {arXiv},
       eprint = {2506.15140},
 primaryClass = {astro-ph.GA},
       adsurl = {https://ui.adsabs.harvard.edu/abs/2025ApJ...989...56H}
}

@ARTICLE{2025A&A...699A..70H,
       author = {{Huang}, K.-Y. and {Behrens}, E. and {Bouvier}, M. and {Viti}, S. and {Mangum}, J.~G. and {Eibensteiner}, C.},
        title = "{Investigating the chemical link between H$_{2}$CO and CH$_{3}$OH within the central molecular zone of NGC 253}",
      journal = {\aap},
         year = 2025,
        month = jul,
       volume = {699},
          eid = {A70},
        pages = {A70},
          doi = {10.1051/0004-6361/202554156},
archivePrefix = {arXiv},
       eprint = {2505.16255},
 primaryClass = {astro-ph.GA},
       adsurl = {https://ui.adsabs.harvard.edu/abs/2025A&A...699A..70H}
}

@ARTICLE{2025ApJ...988..111I,
       author = {{Ikeda}, Toki and {Shimonishi}, Takashi and {Izumi}, Natsuko and {Kaneko}, Hiroyuki and {Takahashi}, Satoko and {Tanaka}, Kei E.~I. and {Furuya}, Kenji and {Yasui}, Chikako},
        title = "{The Detection of Spatially Resolved Protostellar Outflows and Episodic Jets in the Outer Galaxy}",
      journal = {\apj},
         year = 2025,
        month = jul,
       volume = {988},
       number = {1},
          eid = {111},
        pages = {111},
          doi = {10.3847/1538-4357/ade235},
archivePrefix = {arXiv},
       eprint = {2506.08601},
 primaryClass = {astro-ph.GA},
       adsurl = {https://ui.adsabs.harvard.edu/abs/2025ApJ...988..111I}
}

@ARTICLE{2021A&A...651A..43B,
       author = {{Barger}, C.~J. and {Lam}, K.~H. and {Li}, Z.-Y. and {Davis}, S.~W. and {Herbst}, E. and {Garrod}, R.~T.},
        title = "{Combined hydrodynamic and gas-grain chemical modeling of hot cores. I. One-dimensional simulations}",
      journal = {\aap},
         year = 2021,
        month = jul,
       volume = {651},
          eid = {A43},
        pages = {A43},
          doi = {10.1051/0004-6361/202039226},
       adsurl = {https://ui.adsabs.harvard.edu/abs/2021A&A...651A..43B}
}

@ARTICLE{2022A&A...660A..76F,
       author = {{Fontani}, F. and {Colzi}, L. and {Bizzocchi}, L. and {Rivilla}, V.~M. and {Elia}, D. and {Beltr{\'a}n}, M.~T. and {Caselli}, P. and {Magrini}, L. and {S{\'a}nchez-Monge}, A. and {Testi}, L. and {Romano}, D.},
        title = "{CHEMOUT: CHEMical complexity in star-forming regions of the OUTer Galaxy. I. Organic molecules and tracers of star-formation activity}",
      journal = {\aap},
         year = 2022,
        month = apr,
       volume = {660},
          eid = {A76},
        pages = {A76},
          doi = {10.1051/0004-6361/202142923},
archivePrefix = {arXiv},
       eprint = {2203.00719},
 primaryClass = {astro-ph.GA},
       adsurl = {https://ui.adsabs.harvard.edu/abs/2022A&A...660A..76F}
}

@ARTICLE{2022A&A...664A.154F,
       author = {{Fontani}, F. and {Schmiedeke}, A. and {S{\'a}nchez-Monge}, A. and {Colzi}, L. and {Elia}, D. and {Rivilla}, V.~M. and {Beltr{\'a}n}, M.~T. and {Bizzocchi}, L. and {Caselli}, P. and {Magrini}, L. and {Romano}, D.},
        title = "{CHEMOUT: CHEMical complexity in star-forming regions of the OUTer Galaxy. II. Methanol formation at low metallicity}",
      journal = {\aap},
         year = 2022,
        month = aug,
       volume = {664},
          eid = {A154},
        pages = {A154},
          doi = {10.1051/0004-6361/202243532},
archivePrefix = {arXiv},
       eprint = {2205.09136},
 primaryClass = {astro-ph.GA},
       adsurl = {https://ui.adsabs.harvard.edu/abs/2022A&A...664A.154F}
}

@ARTICLE{2025A&A...704A.171G,
       author = {{Gigli}, D. and {Fontani}, F. and {Colzi}, L. and {Vermari{\"e}n}, G. and {Viti}, S. and {Rivilla}, V.~M. and {S{\'a}nchez-Monge}, A.},
        title = "{CHEMOUT: CHEMical complexity in star-forming regions of the OUTer Galaxy: V. Chemical composition gradients as a function of the galactocentric radius}",
      journal = {\aap},
         year = 2025,
        month = dec,
       volume = {704},
          eid = {A171},
        pages = {A171},
          doi = {10.1051/0004-6361/202555956},
archivePrefix = {arXiv},
       eprint = {2509.26556},
 primaryClass = {astro-ph.GA},
       adsurl = {https://ui.adsabs.harvard.edu/abs/2025A&A...704A.171G}
}

@ARTICLE{2021ApJ...922..106B,
       author = {{Bernal}, J.~J. and {Sephus}, C.~D. and {Ziurys}, L.~M.},
        title = "{Methanol at the Edge of the Galaxy: New Observations to Constrain the Galactic Habitable Zone}",
      journal = {\apj},
         year = 2021,
        month = dec,
       volume = {922},
       number = {2},
          eid = {106},
        pages = {106},
          doi = {10.3847/1538-4357/ac27a6},
       adsurl = {https://ui.adsabs.harvard.edu/abs/2021ApJ...922..106B}
}

@ARTICLE{2025ApJ...986..122K,
       author = {{Koelemay}, L.~A. and {Ziurys}, L.~M.},
        title = "{NO and c-C$_{3}$H$_{2}$ in the Outer Galaxy: High Molecular Abundances at R$_{GC}$ > 13 kpc}",
      journal = {\apj},
         year = 2025,
        month = jun,
       volume = {986},
       number = {2},
          eid = {122},
        pages = {122},
          doi = {10.3847/1538-4357/add32d},
       adsurl = {https://ui.adsabs.harvard.edu/abs/2025ApJ...986..122K}
}

@ARTICLE{2024A&A...691A.180F,
       author = {{Fontani}, F. and {Vermari{\"e}n}, G. and {Viti}, S. and {Gigli}, D. and {Colzi}, L. and {Beltr{\'a}n}, M.~T. and {Caselli}, P. and {Rivilla}, V.~M. and {S{\'a}nchez-Monge}, A.},
        title = "{CHEMOUT: CHEMical complexity in star-forming regions of the OUTer Galaxy: IV. ALMA observations of organic species at a galactocentric radius of  23 kpc}",
      journal = {\aap},
         year = 2024,
        month = nov,
       volume = {691},
          eid = {A180},
        pages = {A180},
          doi = {10.1051/0004-6361/202451500},
archivePrefix = {arXiv},
       eprint = {2409.07243},
 primaryClass = {astro-ph.GA},
       adsurl = {https://ui.adsabs.harvard.edu/abs/2024A&A...691A.180F}
}

@ARTICLE{2023NatAs...7..431M,
       author = {{McClure}, M.~K. and {Rocha}, W.~R.~M. and {Pontoppidan}, K.~M. and {Crouzet}, N. and {Chu}, L.~E.~U. and {Dartois}, E. and {Lamberts}, T. and {Noble}, J.~A. and {Pendleton}, Y.~J. and {Perotti}, G. and {Qasim}, D. and {Rachid}, M.~G. and {Smith}, Z.~L. and {Sun}, Fengwu and {Beck}, Tracy L. and {Boogert}, A.~C.~A. and {Brown}, W.~A. and {Caselli}, P. and {Charnley}, S.~B. and {Cuppen}, Herma M. and {Dickinson}, H. and {Drozdovskaya}, M.~N. and {Egami}, E. and {Erkal}, J. and {Fraser}, H. and {Garrod}, R.~T. and {Harsono}, D. and {Ioppolo}, S. and {Jim{\'e}nez-Serra}, I. and {Jin}, M. and {J{\o}rgensen}, J.~K. and {Kristensen}, L.~E. and {Lis}, D.~C. and {McCoustra}, M.~R.~S. and {McGuire}, Brett A. and {Melnick}, G.~J. and {{\"O}berg}, Karin I. and {Palumbo}, M.~E. and {Shimonishi}, T. and {Sturm}, J.~A. and {van Dishoeck}, E.~F. and {Linnartz}, H.},
        title = "{An Ice Age JWST inventory of dense molecular cloud ices}",
      journal = {Nature Astronomy},
         year = 2023,
        month = apr,
       volume = {7},
        pages = {431-443},
          doi = {10.1038/s41550-022-01875-w},
archivePrefix = {arXiv},
       eprint = {2301.09140},
 primaryClass = {astro-ph.GA},
       adsurl = {https://ui.adsabs.harvard.edu/abs/2023NatAs...7..431M}
}

@ARTICLE{2025A&A...696A.198C,
       author = {{Chen}, Y. and {Garrod}, R.~T. and {Rachid}, M. and {van Dishoeck}, E.~F. and {Brogan}, C.~L. and {Loomis}, R. and {Lipnicky}, A. and {McGuire}, B.~A.},
        title = "{CoCCoA: Complex Chemistry in hot Cores with ALMA: The chemical evolution of acetone from ice to gas}",
      journal = {\aap},
         year = 2025,
        month = apr,
       volume = {696},
          eid = {A198},
        pages = {A198},
          doi = {10.1051/0004-6361/202453389},
archivePrefix = {arXiv},
       eprint = {2503.13770},
 primaryClass = {astro-ph.GA},
       adsurl = {https://ui.adsabs.harvard.edu/abs/2025A&A...696A.198C}
}

@ARTICLE{2020MNRAS.496.1051A,
       author = {{Arellano-C{\'o}rdova}, K.~Z. and {Esteban}, C. and {Garc{\'\i}a-Rojas}, J. and {M{\'e}ndez-Delgado}, J.~E.},
        title = "{The Galactic radial abundance gradients of C, N, O, Ne, S, Cl, and Ar from deep spectra of H II regions}",
      journal = {\mnras},
         year = 2020,
        month = aug,
       volume = {496},
       number = {2},
        pages = {1051-1076},
          doi = {10.1093/mnras/staa1523},
archivePrefix = {arXiv},
       eprint = {2005.11372},
 primaryClass = {astro-ph.GA},
       adsurl = {https://ui.adsabs.harvard.edu/abs/2020MNRAS.496.1051A}
}

@ARTICLE{2022MNRAS.510.4436M,
       author = {{M{\'e}ndez-Delgado}, J.~E. and {Amayo}, A. and {Arellano-C{\'o}rdova}, K.~Z. and {Esteban}, C. and {Garc{\'\i}a-Rojas}, J. and {Carigi}, L. and {Delgado-Inglada}, G.},
        title = "{Gradients of chemical abundances in the Milky Way from H II regions: distances derived from Gaia EDR3 parallaxes and temperature inhomogeneities}",
      journal = {\mnras},
         year = 2022,
        month = mar,
       volume = {510},
       number = {3},
        pages = {4436-4455},
          doi = {10.1093/mnras/stab3782},
archivePrefix = {arXiv},
       eprint = {2112.12600},
 primaryClass = {astro-ph.GA},
       adsurl = {https://ui.adsabs.harvard.edu/abs/2022MNRAS.510.4436M}
}

@ARTICLE{2026ApJ...999..194I,
       author = {{Ikeda}, Toki and {Shimonishi}, Takashi and {Kaneko}, Hiroyuki and {Furuya}, Kenji and {Tanaka}, Kei E.~I. and {Izumi}, Natsuko},
        title = "{Digging into the Chemical Complexity in the Outer Galaxy: A Hot Molecular Core in Sh 2-283}",
      journal = {\apj},
         year = 2026,
        month = mar,
       volume = {999},
       number = {2},
          eid = {194},
        pages = {194},
          doi = {10.3847/1538-4357/ae40b8},
archivePrefix = {arXiv},
       eprint = {2602.03122},
 primaryClass = {astro-ph.GA},
       adsurl = {https://ui.adsabs.harvard.edu/abs/2026ApJ...999..194I}
}

@ARTICLE{2025A&A...704A.288L,
       author = {{L{\'o}pez-Gallifa}, {\'A}. and {Rivilla}, V.~M. and {Beltr{\'a}n}, M.~T. and {Colzi}, L. and {Fontani}, F. and {S{\'a}nchez-Monge}, {\'A}. and {Mininni}, C. and {Cesaroni}, R. and {Jim{\'e}nez-Serra}, I. and {Viti}, S. and {Lorenzani}, A.},
        title = "{The GUAPOS project: VI. The chemical inventory of shocked gas}",
      journal = {\aap},
         year = 2025,
        month = dec,
       volume = {704},
          eid = {A288},
        pages = {A288},
          doi = {10.1051/0004-6361/202556837},
archivePrefix = {arXiv},
       eprint = {2509.16094},
 primaryClass = {astro-ph.GA},
       adsurl = {https://ui.adsabs.harvard.edu/abs/2025A&A...704A.288L}
}

\onecolumn
\appendix

\section{Observational parameters}

Tables~\ref{setup} and \ref{noise} summarize the observational setups, along with the synthesized beam sizes and rms noise levels in the final data cubes.

\begin{table*}[!htbp]
    \centering
    \caption{Observational setups of the spectral line survey.}
    \label{setup}
    \begin{tabular}{c c c c c c c c c c c}
    \hline\hline  %
    \noalign{\vskip 2pt}
     Setup & \multicolumn{2}{c}{Frequency range} & Date of        &  Baseline & t\textsubscript{int}$^{(a)}$ & \multicolumn{4}{c}{Calibrators$^{(b)}$} \\
     \cmidrule(r){2-3} \cmidrule(r){7-10} 
        & LSB             & USB             & observation &  range    &            &  P            & A  & B & F  \\
        & GHz             & GHz             & yyyy.mm.dd  &  m        & h        &              &  &  &   &  \\
    \midrule 
     S1    & 215.1$-$223.2 & 230.5$-$238.7 & 2023.12.03  & 23.1$-$328.0  & 2.6  & 1,2  & 1,2  & 3  & 4 \\
         &  &  & 2023.12.05  & 19.5$-$313.6  & 3.4  & 1,2  & 1,2  & 3  & 4 \\
     S2    & 222.8$-$230.9 & 238.3$-$246.4 & 2023.10.05  & 17.7$-$368.0  & 5.2  & 1,2  & 1,2  & 3  & 4,5 \\
    \bottomrule  %
    \end{tabular}
    \tablefoot{\textsuperscript{(a)} On-source integration time. \textsuperscript{(b)} Phase (P), amplitude (A), bandpass (B), and absolute flux (F) calibrators. The calibrators are: 1: J0244+6228, 2: 0224+671, 3: 3C84, 4: LKHA101, 5: MWC349.}
\end{table*}

\begin{table*}[!htbp]
    \centering
    \caption{Beam sizes and noise levels.}
    \label{noise}
    \begin{tabular}{c c c c c c c c c}
    \hline\hline  %
    \noalign{\vskip 2pt}
    Setup & SPW$^{(a)}$ & mode$^{(b)}$ & Frequency & \multicolumn{2}{c}{Synthesized beam} & Linear size & \multicolumn{2}{c}{rms} \\
    \cmidrule(r){5-6} \cmidrule(r){8-9}
          &   &  & range     & HPBW & PA & & line & continuum \\
          &   &  & MHz       & "$\times$" & $^{\circ}$ & au & mJy beam$^{-1}$ & mJy beam$^{-1}$ \\
    \hline
    \noalign{\vskip 4pt}
    S1 & LO & SURVEY & 215065$-$219123 & 0.89 $\times$ 0.78 & 137 & 5340 $\times$ 4680 & 6.27 & 0.12 \\
       & LI & SURVEY & 219130$-$223188 & 0.86 $\times$ 0.76 & 136 & 5160 $\times$ 4560 & 5.56 & 0.14 \\
       & UI$+$CO & SPLIT & 230549$-$230772 & 0.82 $\times$ 0.72 & 135 & 4920 $\times$ 4320 & 6.76 & 0.39 \\
       & UI$-$CO & SURVEY & 230772$-$234605 & 0.82 $\times$ 0.72 & 135 & 4920 $\times$ 4320 & 5.95 & 0.13 \\
       & UO & SURVEY & 234614$-$238673 & 0.81 $\times$ 0.70 & 137 & 4860 $\times$ 4200 & 6.91 & 0.15 \\
    S2 & LO & SURVEY & 222807$-$226863 & 0.90 $\times$ 0.70 & 170 & 5400 $\times$ 4200 & 5.77 & 0.13 \\
       & LI$-$CO & SURVEY & 226872$-$230368 & 0.87 $\times$ 0.70 & 0 & 5220 $\times$ 4200 & 5.54 & 0.12 \\
       & LI$+$CO & SPLIT & 230368$-$230930 & 0.87 $\times$ 0.70 & 0 & 5220 $\times$ 4200 & 6.07 & 0.25 \\
       & UI & SURVEY & 238290$-$242346 & 0.84 $\times$ 0.66 & 169 & 5040 $\times$ 3960 & 6.16 & 0.19 \\
       & UO & SURVEY & 242355$-$246413 & 0.83 $\times$ 0.65 & 0 & 4980 $\times$ 3900 & 6.77 & 0.17 \\ 
    \hline
    \end{tabular}
    \tablefoot{$^{(a)}$ Spectral window. S1-UI and S2-LI are each divided into two subwindows: +CO indicates the subwindow that contains the CO 2$-$1 line and $-$CO the subwindow that does not. $^{(b)}$ Pipeline mode used in IMAGER for imaging.}
\end{table*}

\section{LTE model parameters}
Tables~\ref{tab_best_fit}-\ref{tab_best_fit_MM3} list the input parameters of our best-fit LTE models of MM1, MM2, and MM3. 

\begin{longtable}{p{3.5cm}ccrrrrrrrrr}  
\captionsetup{justification=raggedright,singlelinecheck=false}
\caption{Parameters of our best-fit LTE model of MM1.} \label{tab_best_fit} \\
\hline
Molecule & Tag$^{a}$ & $N_{\rm det}$$^{b}$ & size$^{c}$ & $T_{\rm rot}$$^{d}$ & $N$$^{e}$ & $T_{\rm rot,pd}$$^{f}$ & $N_{\rm pd}$$^{g}$ & $C_{\rm vib}$$^{h}$ & $C_{\rm conf}$$^{i}$ & $V_{\rm off}$$^{j}$ & $\Delta V$$^{k}$ \\
  & &  & " & (K) & (cm$^{-2}$) & (K) & (cm$^{-2}$) & & &  \multicolumn{2}{c}{(km~s$^{-1}$)} \\
\hline
\endfirsthead

\multicolumn{10}{c}%
{{\tablename\ \thetable{.} Continued.}} \\
\hline
Molecule & Tag$^{a}$ & $N_{\rm det}$$^{b}$ & size$^{c}$ & $T_{\rm rot}$$^{d}$ & $N$$^{e}$ & $T_{\rm rot,pd}$$^{f}$ & $N_{\rm pd}$$^{g}$ & $C_{\rm vib}$$^{h}$ & $C_{\rm conf}$$^{i}$ & $V_{\rm off}$$^{j}$ & $\Delta V$$^{k}$ \\
  & &  & " & (K) & (cm$^{-2}$) & (K) & (cm$^{-2}$) & & &  \multicolumn{2}{c}{(km~s$^{-1}$)} \\
\hline
\endhead

\hline 
\endfoot

\hline
\endlastfoot
\setlength{\tabcolsep}{3pt}  
HDO & 19002 & 2 & 0.2 & 260 & 2.5(17) & 260$\pm$0 & 3.0$\pm$0.1(17) & 1.00 & 1.00 & 0.4 & 5.0 \\
 &  &  & 0.2 & 260 & 2.0(16) &  &  & 1.00 & 1.00 & $-$1.4 & 2.0 \\
\hline
CN & 26504 & 4 & 3.9 & 10 & 2.5(14) & ... & ... & 1.00 & 1.00 & 0.0 & 4.0 \\
\hline
CO & 28503 & 1 & 3 & 75 & 1.0(18) & ... & ... & 1.00 & 1.00 & $-$7.4 & 51.0 \\
 &  &  & 3 & 75 & 1.0(18) &  &  & 1.00 & 1.00 & 2.0 & 13.0 \\
 &  &  & 3 & 75 & 6.0(16) &  &  & 1.00 & 1.00 & $-$82.6 & 18.5 \\
$^{13}$CO & 29802 & 1 & 2 & 75 & 4.0(17) & ... & ... & 1.00 & 1.00 & 0.0 & 4.5 \\
C$^{18}$O & 30502 & 1 & 2.2 & 75 & 5.6(16) & ... & ... & 1.00 & 1.00 & 0.0 & 3.0 \\
C$^{17}$O & 29603 & 1 & 2.2 & 75 & 1.2(16) & ... & ... & 1.00 & 1.00 & 0.0 & 3.0 \\
\hline
DCN & 28802 & 1 & 1.9 & 75 & 5.6(13) & ... & ... & 1.00 & 1.00 & 0.1 & 4.3 \\
\hline
H$_2$CO & 30501 & 5 & 0.3 & 175 & 7.5(16) & 103$\pm$3 & 7.6$\pm$0.3(16) & 1.00 & 1.00 & 0.0 & 4.9 \\
 &  &  & 0.9 & 50 & 1.9(15) &  &  & 1.00 & 1.00 & 0.0 & 2.8 \\
H$_2$$^{13}$CO & 31503 & 1 & 0.3 & 175 & 1.5(15) & ... & ... & 1.00 & 1.00 & 0.0 & 4.9 \\
\hline
CH$_3$OH, $v=0$ & 32701 & 52 & 0.3 & 175 & 9.5(17) & 175$\pm$2 & 2.1$\pm$0.1(18) & 1.00 & 1.00 & 1.5 & 3.5 \\
 &  &  & 0.3 & 175 & 6.5(17) &  &  & 1.00 & 1.00 & $-$1.3 & 3.0 \\
CH$_3$OH, $v_{\rm t}=1$ & 32702 & 12 & 0.3 & 175 & 9.5(17) & 175$\pm$2 & 2.1$\pm$0.1(18) & 1.00 & 1.00 & 1.5 & 3.5 \\
 &  &  & 0.3 & 175 & 6.5(17) &  &  & 1.00 & 1.00 & $-$1.3 & 3.0 \\
CH$_3$OH, $v_{\rm t}=2$ & 32703 & 4 & 0.3 & 175 & 9.5(17) & 175$\pm$2 & 2.1$\pm$0.1(18) & 1.00 & 1.00 & 1.5 & 3.5 \\
 &  &  & 0.3 & 175 & 6.5(17) &  &  & 1.00 & 1.00 & $-$1.3 & 3.0 \\
CH$_2$DOH & 33902 & 4 & 0.3 & 80 & 9.0(15) & 76$\pm$19 & 1.1$\pm$0.5(16) & 1.00 & 1.00 & $-$1.0 & 2.9 \\
$^{13}$CH$_3$OH, $v=0$ & 33401 & 15 & 0.3 & 115 & 3.0(16) & 117$\pm$9 & 5.9$\pm$0.6(16) & 1.00 & 1.00 & 1.5 & 3.5 \\
 &  &  & 0.3 & 115 & 2.5(16) &  &  & 1.00 & 1.00 & $-$1.3 & 3.0 \\
$^{13}$CH$_3$OH, $v_{\rm t}=1$$^{(l)}$ & 33402 & 0 & 0.3 & 115 & 3.0(16) & ... & ... & 1.00 & 1.00 & 1.5 & 3.5 \\
 &  &  & 0.3 & 115 & 2.5(16) &  &  & 1.00 & 1.00 & $-$1.3 & 3.0 \\
\hline
H$_2$S & 34502 & 1 & 1.0 & 75 & 4.2(15) & ... & ... & 1.00 & 1.00 & 0.3 & 4.0 \\
\hline
CH$_3$CCH, $v=0$ & 40502 & 6 & 1.9 & 75 & 6.4(14) & 78$\pm$17 & 6.4$\pm$2.4(14) & 1.00 & 1.00 & 0.0 & 2.7 \\
\hline
CH$_3$CN, $v=0$ & 41601 & 17 & 0.3 & 190 & 6.0(15) & 193$\pm$6 & 1.2$\pm$0.1(16) & 1.00 & 1.00 & 1.5 & 3.5 \\
 &  &  & 0.3 & 190 & 4.0(15) &  &  & 1.00 & 1.00 & $-$1.3 & 3.0 \\
CH$_3$CN, $v_8=1$ & 41603 & 17 & 0.3 & 190 & 6.0(15) & 193$\pm$6 & 1.2$\pm$0.1(16) & 1.00 & 1.00 & 0.5 & 6.1 \\
 &  &  & 0.3 & 190 & 4.0(15) &  &  & 1.00 & 1.00 & $-$1.3 & 3.0 \\
\hline
CH$_2$CO, $v=0$ & 42501 & 4 & 0.3 & 110 & 2.0(15) & 103$\pm$19 & 3.6$\pm$1.0(15) & 1.00 & 1.00 & 1.2 & 3.4 \\
 &  &  & 0.3 & 110 & 1.5(15) &  &  & 1.00 & 1.00 & $-$1.8 & 2.6 \\
\hline
HNCO, $v=0$ & 43511 & 7 & 0.3 & 180 & 2.5(16) & 177$\pm$4 & 2.9$\pm$0.1(16) & 1.02 & 1.00 & 0.9 & 5.0 \\
 &  &  & 0.3 & 180 & 2.5(15) &  &  & 1.02 & 1.00 & $-$1.2 & 2.0 \\
\hline
CH$_3$CHO, $v=0$ & 44301 & 9 & 0.8 & 65 & 3.5(14) & 63$\pm$13 & 5.4$\pm$1.9(14) & 1.00 & 1.00 & 1.5 & 3.5 \\
 &  &  & 0.8 & 65 & 1.5(14) &  &  & 1.00 & 1.00 & $-$1.3 & 3.0 \\
\hline
CS & 44501 & 1 & 1.7 & 75 & 5.2(15) & ... & ... & 1.00 & 1.00 & 0.0 & 4.0 \\
$^{13}$CS & 45403 & 1 & 0.8 & 75 & 3.6(14) & ... & ... & 1.00 & 1.00 & 0.0 & 4.0 \\
C$^{34}$S & 46501 & 1 & 1.1 & 75 & 8.7(14) & ... & ... & 1.00 & 1.00 & 0.0 & 4.0 \\
C$^{33}$S & 45404 & 1 & 0.8 & 75 & 4.1(14) & ... & ... & 1.00 & 1.00 & 0.0 & 4.0 \\
$^{13}$C$^{34}$S$^{(l)}$ & 47501 & 1 & 0.8 & 75 & 1.8(13) & ... & ... & 1.00 & 1.00 & 0.0 & 4.0 \\
\hline
SiO & 44505 & 1 & 1.2 & 75 & 8.5(13) & ... & ... & 1.00 & 1.00 & 0.5 & 4.0 \\
\hline
NH$_2$CHO, $v=0$ & 45512 & 11 & 0.3 & 110 & 1.0(15) & 110$\pm$26 & 2.0$\pm$0.6(15) & 1.02 & 1.00 & 1.5 & 3.5 \\
 &  &  & 0.3 & 110 & 6.0(14) &  &  & 1.02 & 1.00 & $-$1.3 & 3.0 \\
\hline
t-HCOOH & 46506 & 14 & 0.3 & 190 & 3.7(16) & 191$\pm$22 & 3.9$\pm$0.5(16) & 1.02 & 1.00 & 0.7 & 5.0 \\
\hline
H$_2$CS & 46509 & 7 & 1. & 105 & 2.6(15) & 103$\pm$7 & 2.8$\pm$0.3(15) & 1.00 & 1.00 & 0.2 & 4.9 \\
\hline
CH$_3$OCH$_3$, $v=0$ & 46514 & 11 & 0.3 & 120 & 5.1(16) & 117$\pm$5 & 9.0$\pm$0.8(16) & 1.00 & 1.00 & $-$1.2 & 3.2 \\
 &  &  & 0.3 & 120 & 3.5(16) &  &  & 1.00 & 1.00 & 1.7 & 2.4 \\
\hline
C$_2$H$_5$OH, $v=0$ & 46524 & 23 & 0.3 & 85 & 2.0(16) & 88$\pm$10 & 3.2$\pm$0.7(16) & 1.05 & 1.00 & 1.0 & 3.0 \\
 &  &  & 0.3 & 85 & 1.1(16) &  &  & 1.05 & 1.00 & $-$2.1 & 1.8 \\
\hline
SO & 48501 & 4 & 1.2 & 75 & 1.1(16) & ... & ... & 1.00 & 1.00 & 0.0 & 4.0 \\
$^{33}$SO & 49501 & 1 & 0.4 & 75 & 1.5(15) & ... & ... & 1.00 & 1.00 & 0.0 & 4.0 \\
$^{34}$SO & 50501 & 1 & 0.5 & 75 & 4.5(15) & ... & ... & 1.00 & 1.00 & 0.0 & 4.0 \\
S$^{18}$O & 50502 & 2 & 0.5 & 75 & 3.2(14) & ... & ... & 1.00 & 1.00 & 0.0 & 4.0 \\
\hline
HC$_3$N, $v=0$ & 51501 & 4 & 0.5 & 135 & 5.1(14) & 136$\pm$14 & 5.1$\pm$1.0(14) & 1.41 & 1.00 & 0.9 & 5.0 \\
HC$_3$N, $v_7=1$ & 51502 & 2 & 0.5 & 135 & 5.1(14) & ... & ... & 1.41 & 1.00 & 0.9 & 5.0 \\
\hline
C$_2$H$_5$CN, $v=0$ & 55913 & 20 & 0.3 & 190 & 1.7(15) & ... & ... & 1.73 & 1.00 & 1.5 & 3.5 \\
\hline
CH$_3$C(O)CH$_3$, $v=0$ & 58912 & 13 & 0.3 & 82 & 5.0(15) & 82$\pm$8 & 9.1$\pm$1.7(15) & 1.00 & 1.00 & 1.5 & 3.5 \\
 &  &  & 0.3 & 82 & 4.0(15) & &  & 1.00 & 1.00 & $-$1.3 & 3.0 \\
\hline
CH$_2$(OH)CHO, $v=0$$^{(l)}$ & 60501 & 3 & 0.3 & 135 & 1.2(15) & ... & ... & 1.21 & 1.00 & 0.5 & 2.7 \\
\hline
CH$_3$COOH, $v_{\rm t}=0$ & 60523 & 13 & 0.3 & 135 & 1.0(16) & ... & ... & 1.01 & 1.00 & 0.6 & 2.7 \\
\hline
CH$_3$OCHO, $v=0$ & 60103 & 71 & 0.3 & 135 & 5.7(16) & 136$\pm$4 & 1.2$\pm$0.1(17) & 1.14 & 1.00 & 1.7 & 2.7 \\
 &  &  & 0.3 & 135 & 5.7(16) &  &  & 1.14 & 1.00 & $-$1.7 & 3.0 \\
CH$_3$OCHO, $v_{\rm t}=1$ & 60104 & 44 & 0.3 & 135 & 5.7(16) & 136$\pm$4 & 1.2$\pm$0.1(17) & 1.14 & 1.00 & 1.7 & 2.7 \\
 &  &  & 0.3 & 135 & 5.7(16) &  &  & 1.14 & 1.00 & $-$1.7 & 3.0 \\
\hline
OCS,v=0 & 60503 & 3 & 0.4 & 115 & 5.0(16) & 111$\pm$7 & 5.7$\pm$0.5(16) & 1.00 & 1.00 & 0.0 & 5.0 \\
O$^{13}$CS & 61502 & 1 & 0.3 & 115 & 2.0(15) & ... & ... & 1.00 & 1.00 & 0.0 & 5.0 \\
OC$^{34}$S & 62505 & 2 & 0.3 & 115 & 6.8(15) & ... & ... & 1.00 & 1.00 & 0.0 & 5.0 \\
OC$^{33}$S$^{(l)}$ & 61503 & 3 & 0.4 & 115 & 8.0(14) & ... & ... & 1.00 & 1.00 & 0.0 & 5.0 \\
\hline
a-(CH$_2$OH)$_2$ & 62503 & 23 & 0.3 & 175 & 1.5(16) & ... & ... & 1.91 & 1.19 & 0.5 & 3.9 \\
\hline
SO$_2$, $v=0$ & 64502 & 19 & 0.3 & 142 & 3.0(17) & 149$\pm$4 & 3.6$\pm$0.3(17) & 1.00 & 1.00 & 0.0 & 4.5 \\
SO$_2$, $v_2=1$$^{(l)}$ & 64503 & 0 & 0.3 & 142 & 3.0(17) & ... & ... & 1.00 & 1.00 & 0.0 & 4.5 \\
$^{33}$SO$_2$, $v=0$ & 65501 & 3 & 0.3 & 142 & 3.8(15) & ... & ... & 1.00 & 1.00 & 0.5 & 4.5 \\
$^{34}$SO$_2$, $v=0$ & 66501 & 13 & 0.3 & 142 & 1.8(16) & ... & ... & 1.01 & 1.00 & 0.0 & 4.5 \\
\end{longtable}
\tablefoot{ 
$^{(a)}$ Entry number in our Weeds local database; see Sect.~\ref{Method}. $^{(b)}$ Number of detected lines. One line of a given species may mean a group of transitions of this species that are blended together. $^{(c)}$ Source diameter (FWHM). $^{(d)}$ Rotational temperature used for the Weeds model. $^{(e)}$ Column density used for the Weeds model. X(Y) means X$\times$10$^{Y}$. $^{(f)}$ Rotational temperature derived from the population diagram. $^{(g)}$ Column density derived from the population diagram. In cases where molecules have two velocity components, the value represents the total column density summed over both components. $^{(h)}$ Correction factor that was applied to the column density to account for the contribution of vibrationally or torsionally excited states, in the cases where this contribution was not included in the partition function of the spectroscopic predictions. In most cases, this factor was estimated in the harmonic approximation \citep[Eq. (1) in][]{2025A&A...698A.143B}. $^{(i)}$ Correction factor that was applied to the column density to account for the contribution of conformers, in the cases where this contribution was not included in the partition function of the spectroscopic predictions. $^{(j)}$ Velocity offset with respect to the systemic velocity of MM1 ($-$71.7 km s$^{-1}$). $^{(k)}$ Linewidth (FWHM). $^{(l)}$ Tentative detection.
}

\begin{table}[htbp]
\centering
\captionsetup{justification=raggedright,singlelinecheck=false}
\caption{Parameters of our best-fit LTE model of MM2.} \label{tab_best_fit_MM2}
\renewcommand{\arraystretch}{1.2}
\begin{tabular}{p{3.5cm}ccrrrrrrrrr}
\hline
Molecule & Tag$^{a}$ & $N_{\rm det}$$^{b}$ & size$^{c}$ & $T_{\rm rot}$$^{d}$ & $N$$^{e}$ & $T_{\rm rot,pd}$$^{f}$ & $N_{\rm pd}$$^{g}$ & $C_{\rm vib}$$^{h}$ & $C_{\rm conf}$$^{i}$ & $V_{\rm off}$$^{j}$ & $\Delta V$$^{k}$ \\
  & &  & " & (K) & (cm$^{-2}$) & (K) & (cm$^{-2}$) & & &  \multicolumn{2}{c}{(km~s$^{-1}$)} \\
\hline
CN & 26504 & 1 & 4.5 & 7 & 3.0(14) & ... & ... & 1.00 & 1.00 & $-$0.7 & 1.5 \\ 
\hline
CO & 28503 & 1 & 4.5 & 12 & 1.0(17) & ... & ... & 1.00 & 1.00 & 2.0 & 13 \\ 
$^{13}$CO & 29802 & 1 & 4.5 & 12 & 2.0(15) & ... & ... & 1.00 & 1.00 & 0.0 & 4.5 \\ 
\hline
H$_2$CO & 30501 & 4 & 4.5 & 41 & 1.3(14) & 41$\pm$5 & 1.3$\pm$0.2(14) & 1.00 & 1.00 & $-$1.0 & 1.5 \\ 
\hline
CS & 44501 & 1 & 4.5 & 12 & 6.2(14) & ... & ... & 1.00 & 1.00 & $-$1.5 & 1.5 \\ 
C$^{34}$S & 46501 & 1 & 4.5 & 12 & 3.1(13) & ... & ... & 1.00 & 1.00 & $-$1.0 & 1.5 \\ 
\hline
CH$_3$OH, $v=0$ & 32701 & 3 & 4.5 & 60 & 2.0(14) & 49$\pm$3 & 5.7$\pm$0.6(14) & 1.00 & 1.00 & $-$0.5 & 2 \\ 
\hline
H$_2$CS & 46509 & 4 & 4.5 & 12 & 1.3(14) & 12$\pm$5 & 1.4$\pm$2.8(14) & 1.00 & 1.00 & $-$1.0 & 1.5 \\ 
\hline
\end{tabular}
\tablefoot{Same as Table~\ref{tab_best_fit}, but a fixed source size of 4.5" was adopted, for a beam filling factor of about unity, as the source size could not be fitted and the molecular lines trace extended emission.
}
\end{table}

\begin{longtable}{p{3.5cm}ccrrrrrrrrr}  
\captionsetup{justification=raggedright,singlelinecheck=false}
\caption{Parameters of our best-fit LTE model of MM3.} \label{tab_best_fit_MM3} \\
\hline
Molecule & Tag$^{a}$ & $N_{\rm det}$$^{b}$ & size$^{c}$ & $T_{\rm rot}$$^{d}$ & $N$$^{e}$ & $T_{\rm rot,pd}$$^{f}$ & $N_{\rm pd}$$^{g}$ & $C_{\rm vib}$$^{h}$ & $C_{\rm conf}$$^{i}$ & $V_{\rm off}$$^{j}$ & $\Delta V$$^{k}$ \\
  & &  & " & (K) & (cm$^{-2}$) & (K) & (cm$^{-2}$) & & &  \multicolumn{2}{c}{(km~s$^{-1}$)} \\
\hline
\endfirsthead

\multicolumn{10}{c}%
{{\tablename\ \thetable{.} Continued.}} \\
\hline
Molecule & Tag$^{a}$ & $N_{\rm det}$$^{b}$ & size$^{c}$ & $T_{\rm rot}$$^{d}$ & $N$$^{e}$ & $T_{\rm rot,pd}$$^{f}$ & $N_{\rm pd}$$^{g}$ & $C_{\rm vib}$$^{h}$ & $C_{\rm conf}$$^{i}$ & $V_{\rm off}$$^{j}$ & $\Delta V$$^{k}$ \\
  & &  & " & (K) & (cm$^{-2}$) & (K) & (cm$^{-2}$) & & &  \multicolumn{2}{c}{(km~s$^{-1}$)} \\
\hline
\endhead

\hline 
\endfoot

\hline
\endlastfoot
\setlength{\tabcolsep}{3pt} 
CN & 26504 & 1 & 4.5 & 10 & 4.5(14) & ... & ... & 1.00 & 1.00 & 1.0 & 2.0 \\
\hline
CO & 28503 & 1 & 4.5 & 43 & 6.0(17) & ... & ... & 1.00 & 1.00 & 5.0 & 4.0 \\
 &  &  & 4.5 & 43 & 4.0(16) &  &  & 1.00 & 1.00 & 0.0 & 2.5 \\
 &  &  & 4.5 & 43 & 3.0(16) &  &  & 1.00 & 1.00 & $-$3.5 & 2.0 \\
 &  &  & 4.5 & 43 & 6.0(16) &  &  & 1.00 & 1.00 & $-$18.0 & 12.0 \\
$^{13}$CO & 29802 & 1 & 4.5 & 43 & 2.5(16) & ... & ... & 1.00 & 1.00 & 1.0 & 2.0 \\
 &  &  & 4.5 & 43 & 2.5(16) &  &  & 1.00 & 1.00 & 4.0 & 3.0 \\
C$^{17}$O & 29603 & 1 & 4.5 & 43 & 1.6(15) & ... & ... & 1.00 & 1.00 & 1.0 & 2.0 \\
C$^{18}$O & 30502 & 1 & 4.5 & 43 & 5.5(15) & ... & ... & 1.00 & 1.00 & 1.0 & 2.0 \\
 &  &  & 4.5 & 43 & 3.0(15) &  &  & 1.00 & 1.00 & 4.0 & 3.0 \\
\hline
DNC & 28508 & 1 & 4.5 & 43 & 2.0(12) & ... & ... &  &  & 1.0 & 1.5 \\
\hline
DCN & 28802 & 1 & 4.5 & 43 & 1.1(13) & ... & ... & 1.00 & 1.00 & 1.0 & 1.5 \\
\hline
H$_2$CO & 30501 & 4 & 4.5 & 87 & 2.8(14) & 89$\pm$8 & 3.5$\pm$0.4(14) & 1.00 & 1.00 & 1.0 & 1.5 \\
\hline
CH$_3$OH, $v=0$ & 32701 & 10 & 4.5 & 37 & 4.0(14) & 36$\pm$8 & 4.4$\pm$1.7(14) & 1.00 & 1.00 & 1.0 & 1.5 \\
\hline
H$_2$S & 34502 & 1 & 4.5 & 43 & 2.0(14) & ... & ... & 1.00 & 1.00 & 1.5 & 1.5 \\
\hline
CH$_3$CCH, $v=0$ & 40502 & 7 & 4.5 & 55 & 1.8(14) & 54$\pm$12 & 1.8$\pm$8.7(13) & 1.00 & 1.00 & 1.0 & 1.5 \\
\hline
CS & 44501 & 1 & 4.5 & 43 & 2.1(14) & ... & ... & 1.00 & 1.00 & 1.0 & 2.0 \\
 &  &  & 4.5 & 43 & 3.5(13) &  &  & 1.00 & 1.00 & 4.0 & 3.0 \\
$^{13}$CS & 45403 & 1 & 4.5 & 43 & 1.0(13) & ... & ... & 1.00 & 1.00 & 1.0 & 2.0 \\
C$^{33}$S & 45404 & 1 & 4.5 & 43 & 1.5(13) & ... & ... & 1.00 & 1.00 & 1.0 & 2.0 \\
C$^{34}$S & 46501 & 1 & 4.5 & 43 & 4.5(13) & ... & ... & 1.00 & 1.00 & 1.0 & 2.0 \\
\hline
SiO & 44505 & 1 & 4.5 & 43 & 1.5(12) & ... & ... & 1.00 & 1.00 & 1.0 & 1.5 \\
 &  &  & 4.5 & 43 & 1.6(12) &  &  & 1.00 & 1.00 & 4.5 & 2.0 \\
\hline
H$_2$CS & 46509 & 6 & 4.5 & 43 & 1.3(14) & 42$\pm$2 & 1.4$\pm$0.2(14) & 1.00 & 1.00 & 1.0 & 1.5 \\
\hline
SO & 48501 & 1 & 4.5 & 43 & 1.3(14) & ... & ... & 1.00 & 1.00 & 1.0 & 2.0 \\
 &  &  & 4.5 & 43 & 6.1(13) &  &  & 1.00 & 1.00 & 4.0 & 3.0 \\
\hline
HC$_3$N, $v=0$ & 51501 & 4 & 4.5 & 27 & 8.0(13) & 27$\pm$6 & 8.8$\pm$11.3(13) & 1.41 & 1.00 & 1.0 & 1.5 \\
\hline
SO$_2$, $v=0$ & 64502 & 1 & 4.5 & 87 & 1.0(14) & ... & ... & 1.00 & 1.00 & 1.0 & 2.0 \\
 &  &  & 4.5 & 87 & 8.0(13) &  &  & 1.00 & 1.00 & 4.0 & 3.0 \\
\hline
\end{longtable}
\tablefoot{Same as Table~\ref{tab_best_fit}, but a fixed source size of 4.5" was adopted, for a beam filling factor of about unity, as the source size could not be fitted and the molecular lines trace extended emission.
}

\section{Emission size}

\begin{figure*}[!htbp]
    \centering
    \includegraphics[width=\linewidth]{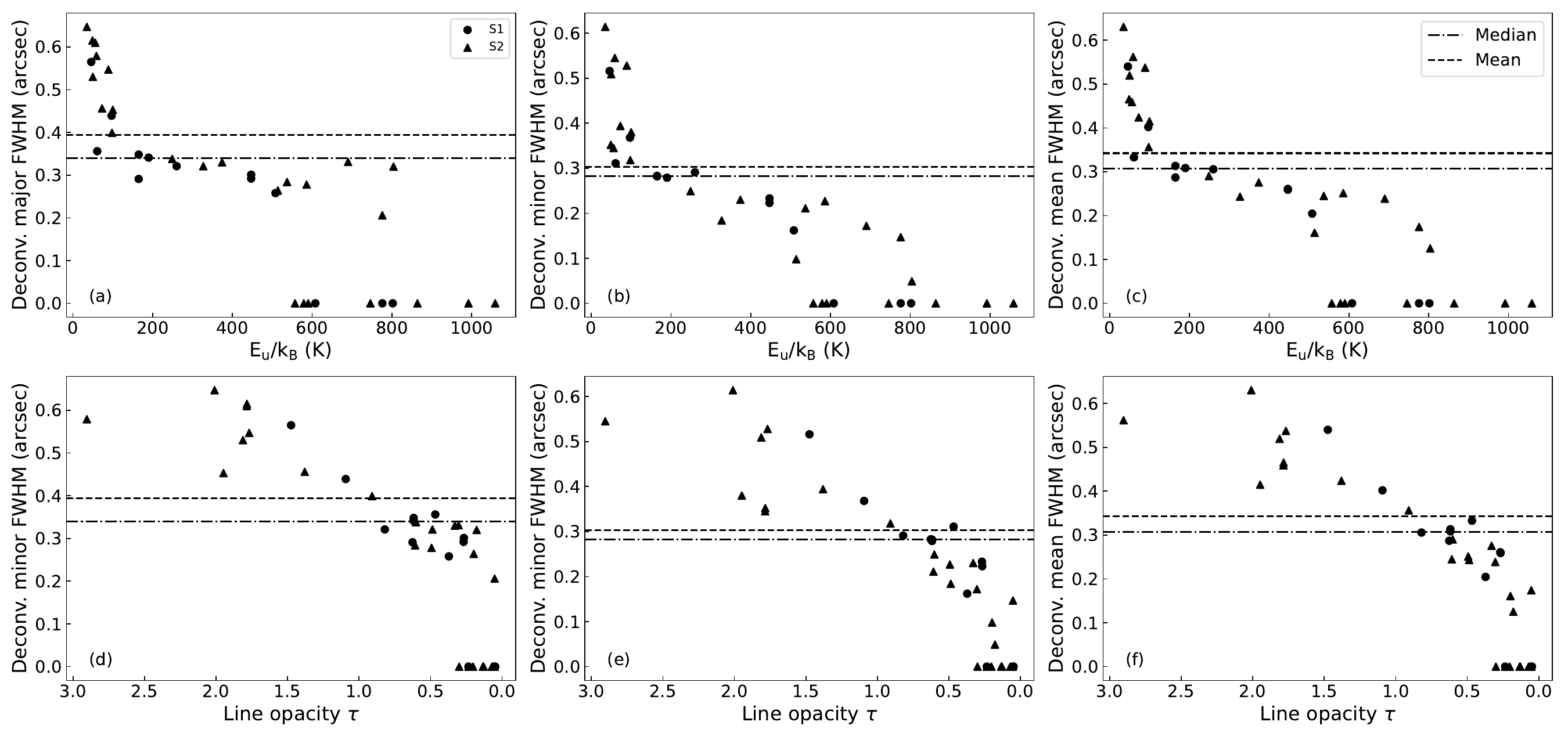}
    \caption{Deconvolved emission size of methanol transitions as a function of upper-level energy (top) and line opacity (bottom). The left (a and d), middle (b and e), and right (c and f) panels show the major, minor, and mean FWHM, respectively. The filled circle and triangle symbols indicate the spectral setups S1 and S2, respectively. The dashed line represents the mean value, while the dot-dashed line represents the median value.}
    \label{fig_size}
\end{figure*}

\begin{figure*}[!htbp]
    \centering
    \includegraphics[width=\linewidth]{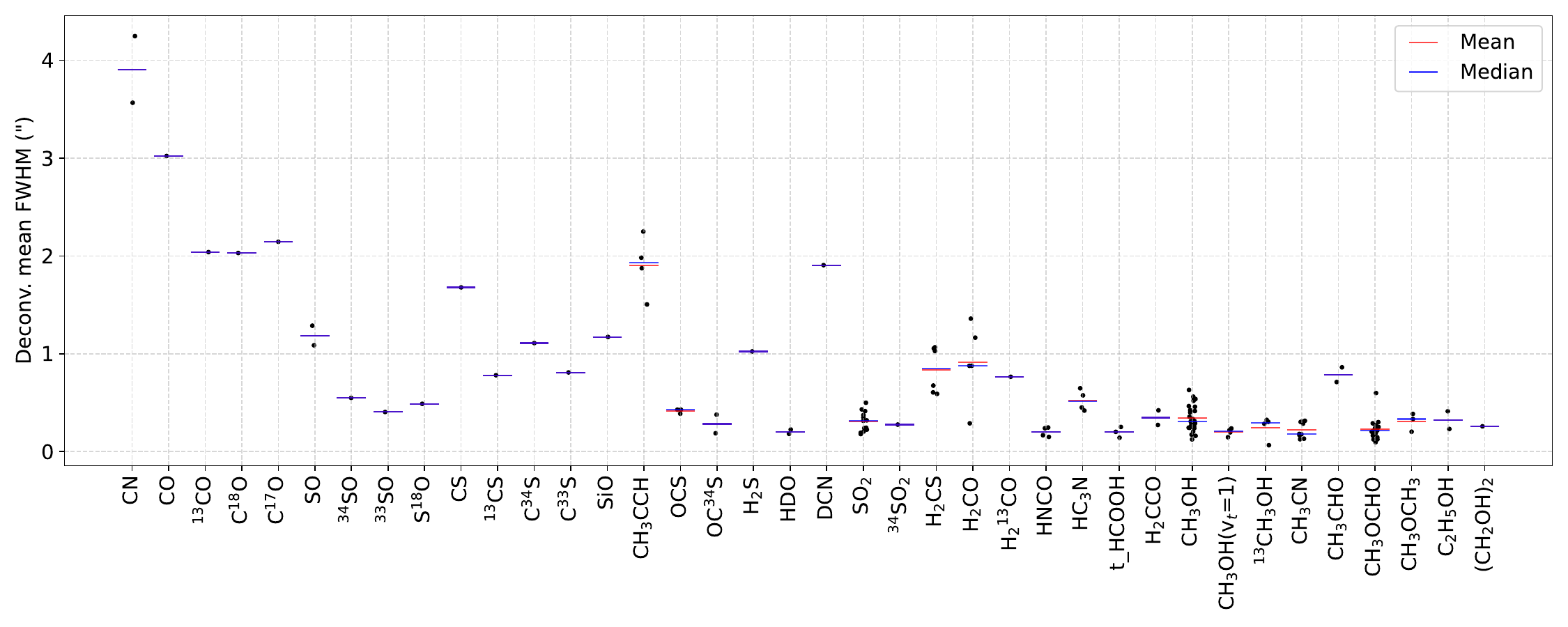}
    \caption{Deconvolved mean source size determined for all identified molecules through 2D Gaussian fitting. For each molecule, the deconvolved emission sizes of different transitions are shown as black dots, while the mean and median values are indicated by red and blue lines, respectively.} 
    \label{fig_all_size} 
\end{figure*}

Figure~\ref{fig_size} presents the deconvolved size of CH$_3$OH plotted against upper-level energy and line optical depth. Figure~\ref{fig_all_size} displays the deconvolved emission sizes for all detected molecules.

\section{Observed spectrum of MM1}

Figures~\ref{fig_all_spec} and \ref{fig_all_spec_unzoom} present the full spectral survey of G135.27 MM1. In Fig.~\ref{fig_all_spec}, the spectra are shown with a zoomed y-axis to highlight the weak COM lines, while Fig.~\ref{fig_all_spec_unzoom} shows the spectra un-zoomed. The LTE model parameters of the synthetic spectra overlaid on the observed
spectra are listed in Table~\ref{tab_best_fit}.

\begin{figure*}[!htbp]
    \centering
    \includegraphics[width=\linewidth]{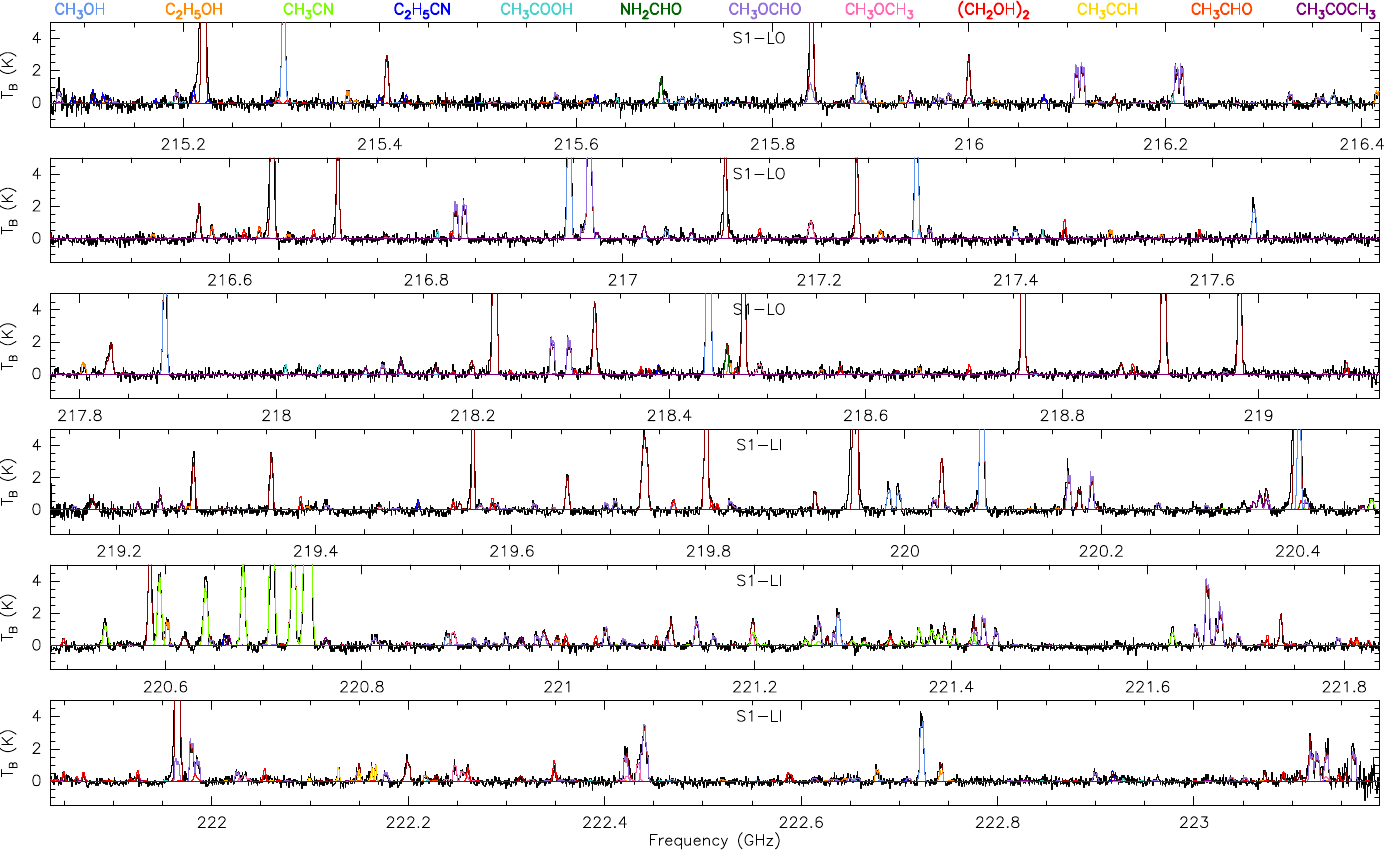} 
    \includegraphics[width=\linewidth]{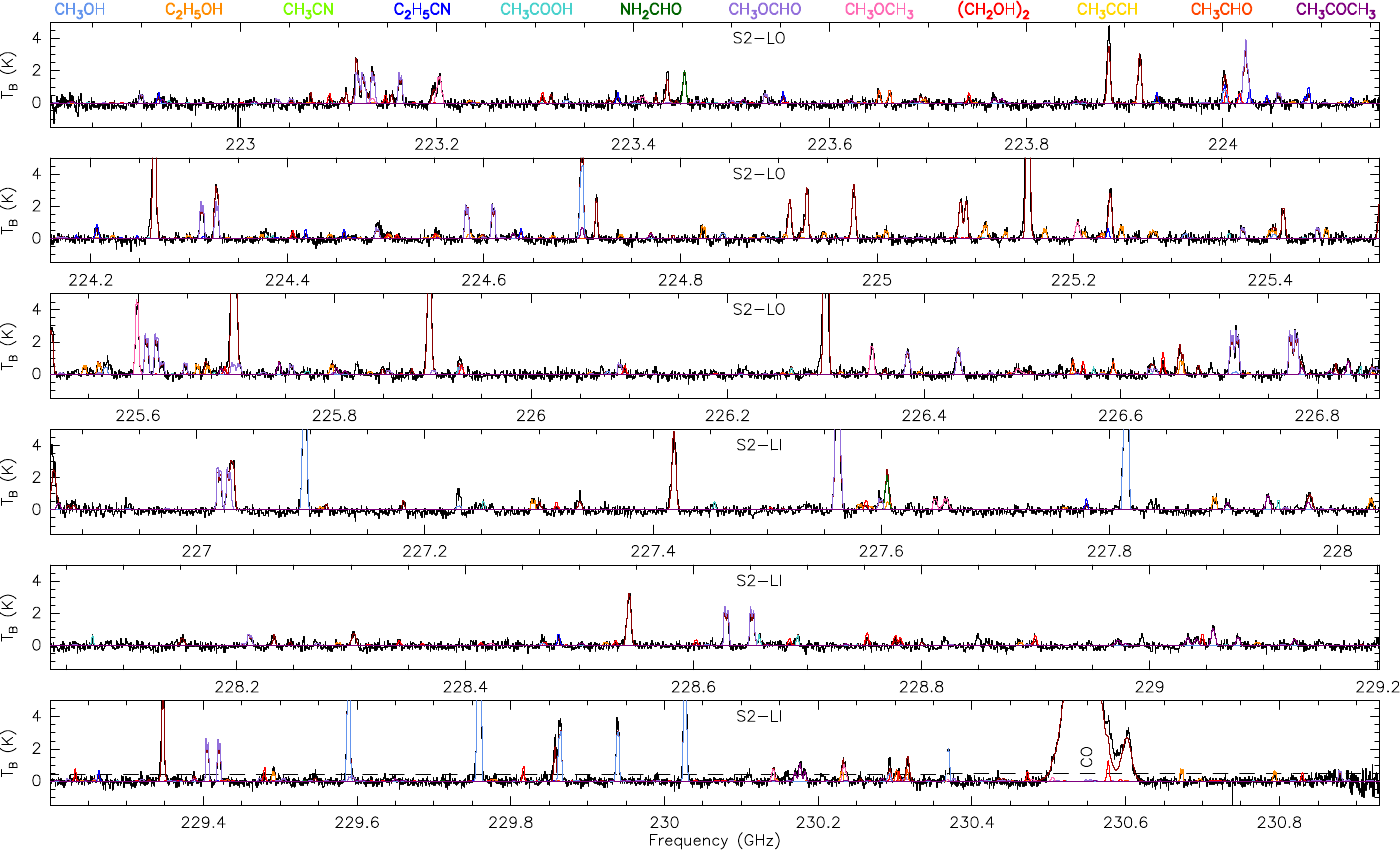} 
    \caption{NOEMA spectrum of the outer Galaxy hot core G135.27 MM1. The observed spectrum and the synthetic spectrum that contains the contribution of all identified molecules are shown in black and dark red, respectively. The synthetic spectra of selected COMs are overlaid in various colors, as labeled at the top. The horizontal dashed line marks the 3$\sigma$ noise level. The spectra have been truncated to emphasize the faint lines. The synthetic spectrum of S1-UI does not contain the CO contribution because the CO line profile is incomplete in this spectral window.}
    \label{fig_all_spec}
\end{figure*}

\addtocounter{figure}{-1}
\begin{figure*}[t!]
    \centering 
    \includegraphics[width=\linewidth]{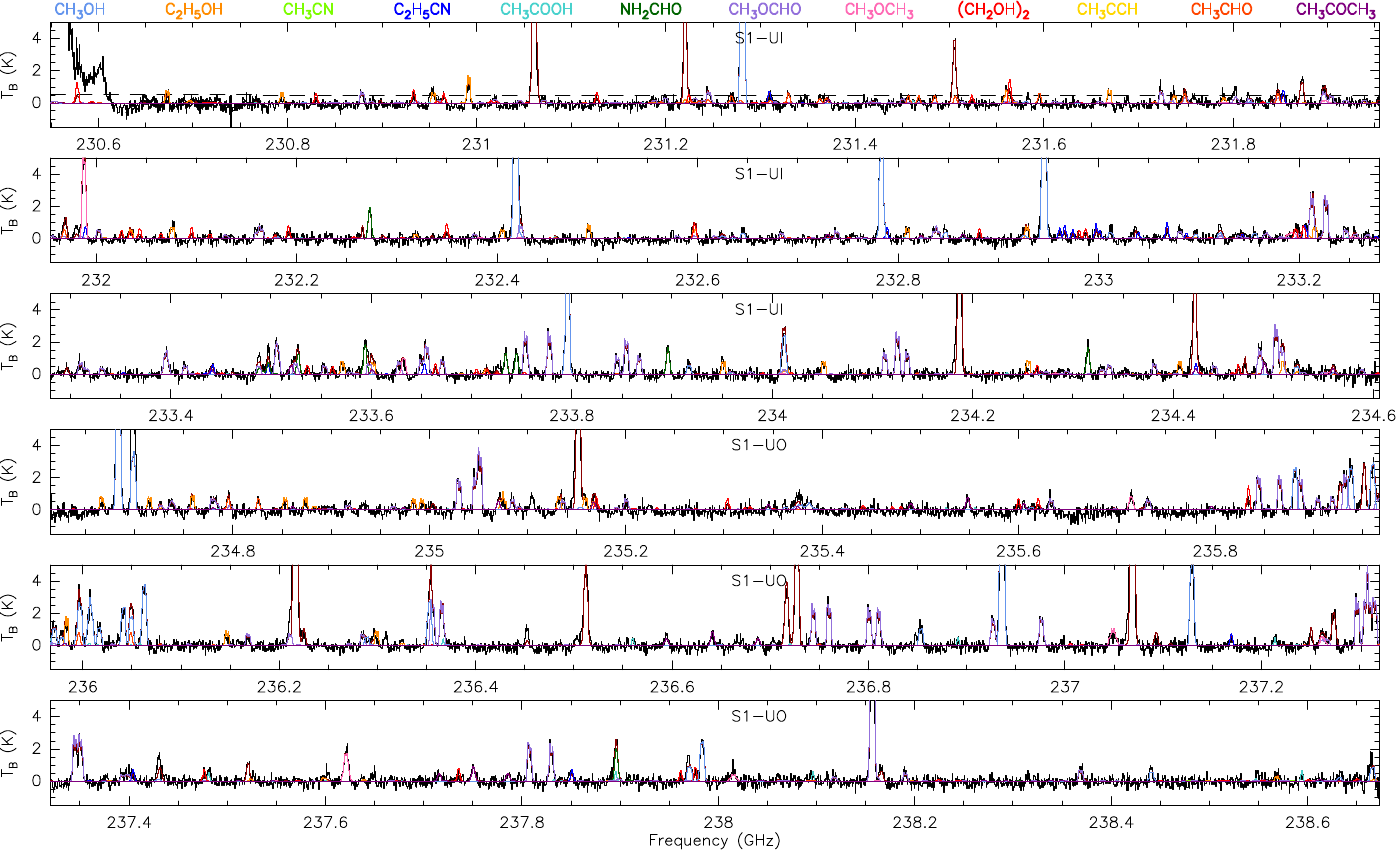} 
    \includegraphics[width=\linewidth]{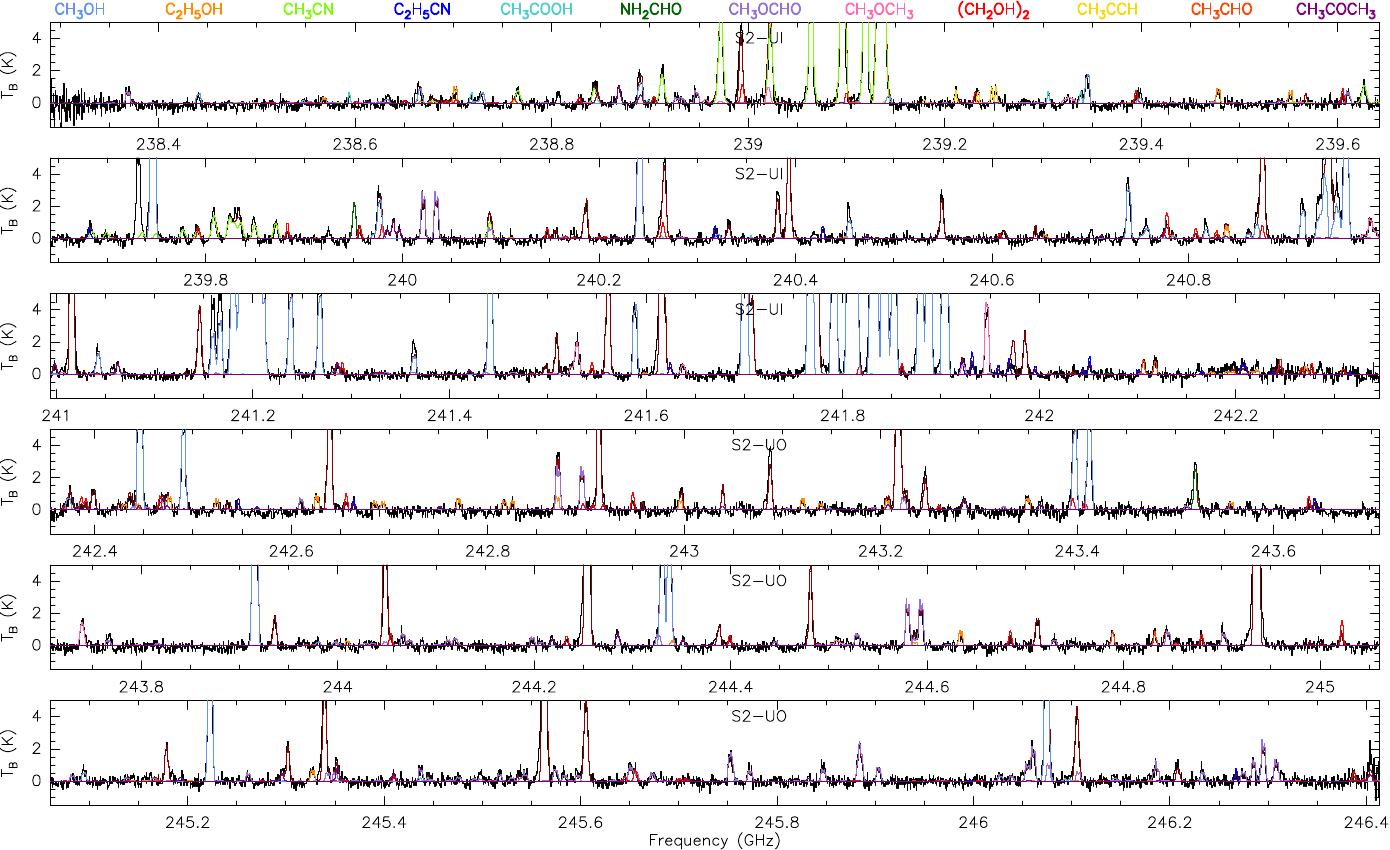}
    \caption{Continued.}
\end{figure*}

\begin{figure*}[!htbp]
    \centering
    \includegraphics[width=\linewidth]{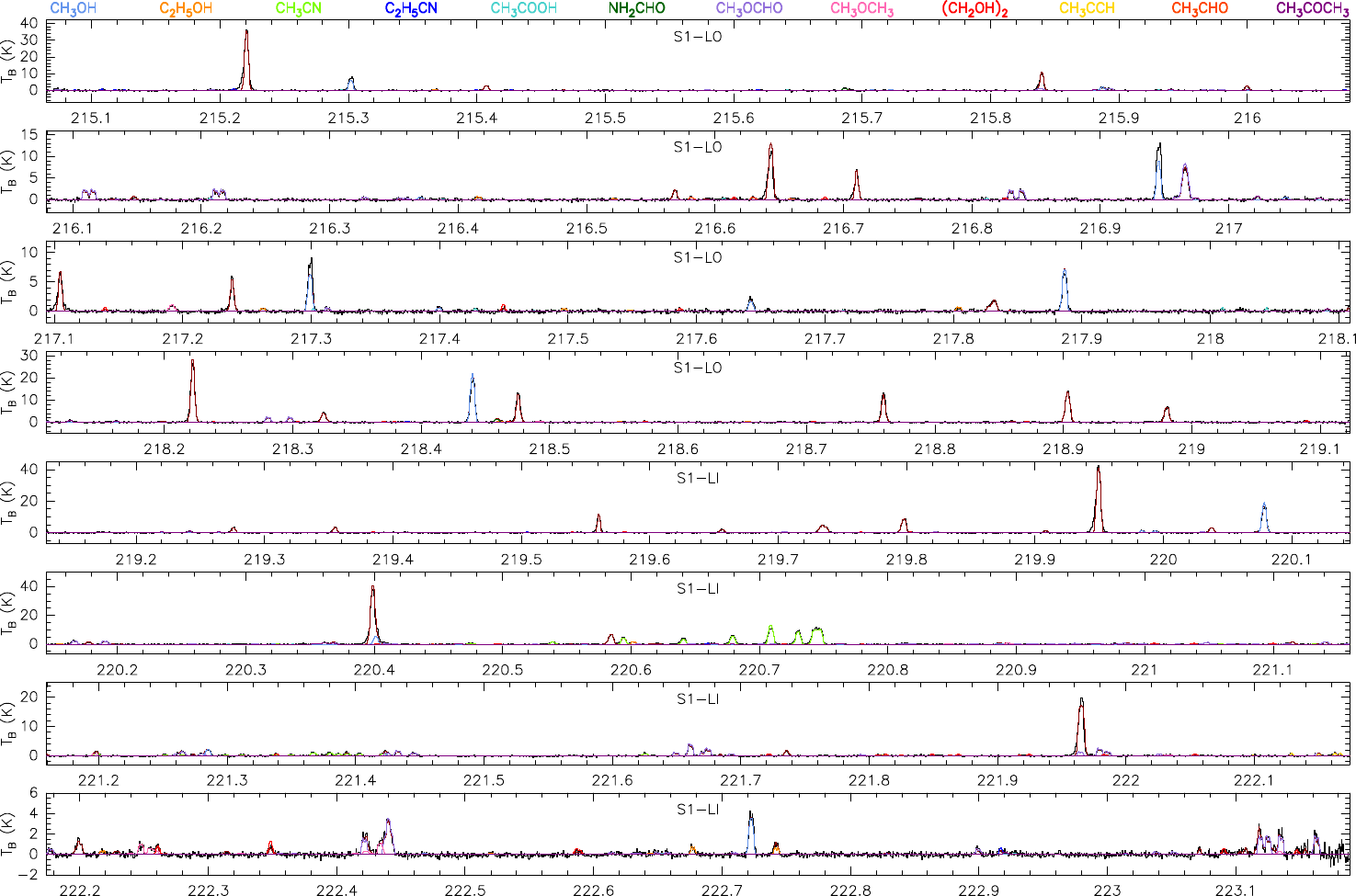} 
    \includegraphics[width=\linewidth]{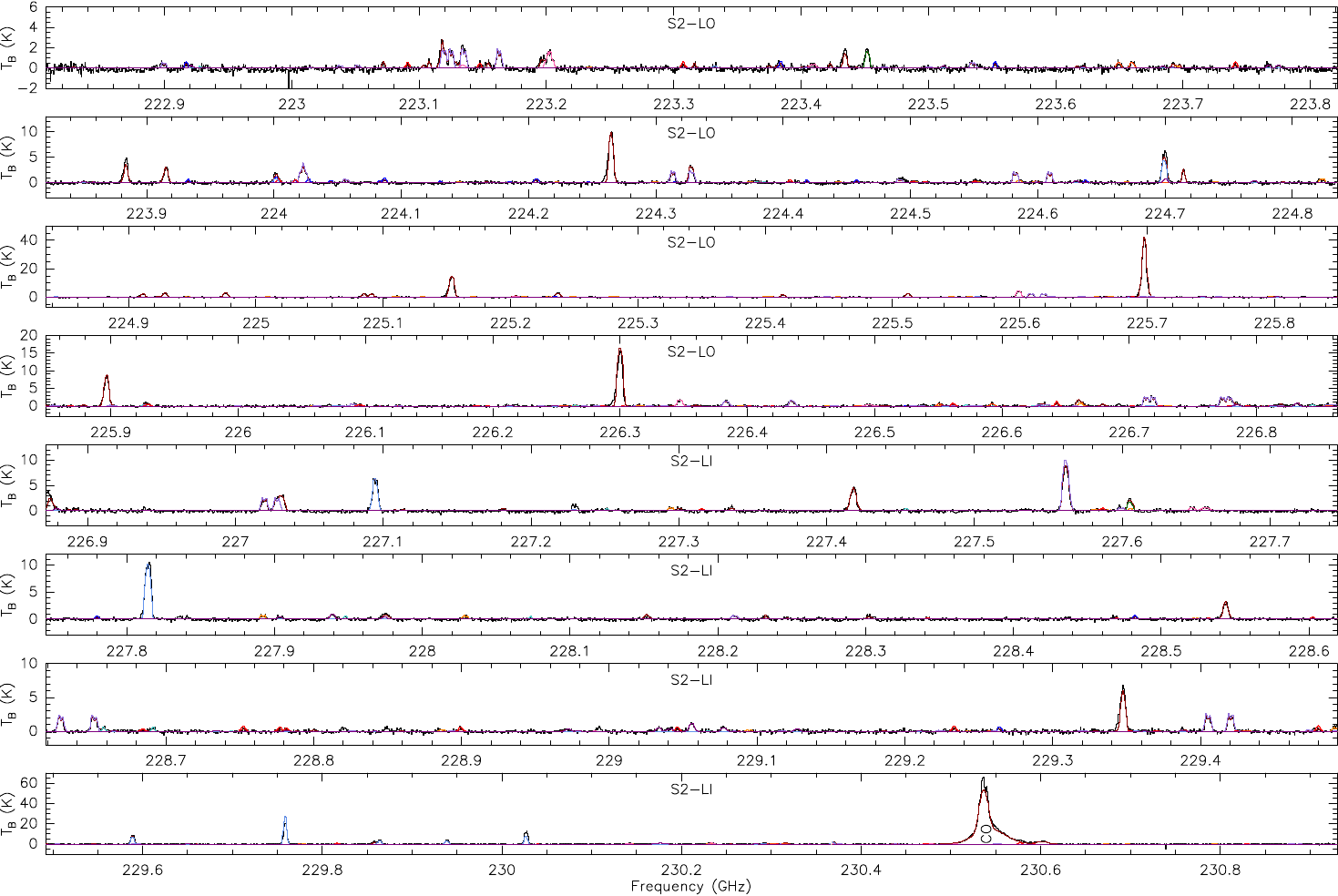} 
    \caption{Same as Fig.~\ref{fig_all_spec} but showing the un-zoomed spectrum.}
    \label{fig_all_spec_unzoom}
\end{figure*}

\addtocounter{figure}{-1}
\begin{figure*}[!htbp]
    \centering 
    \includegraphics[width=\linewidth]{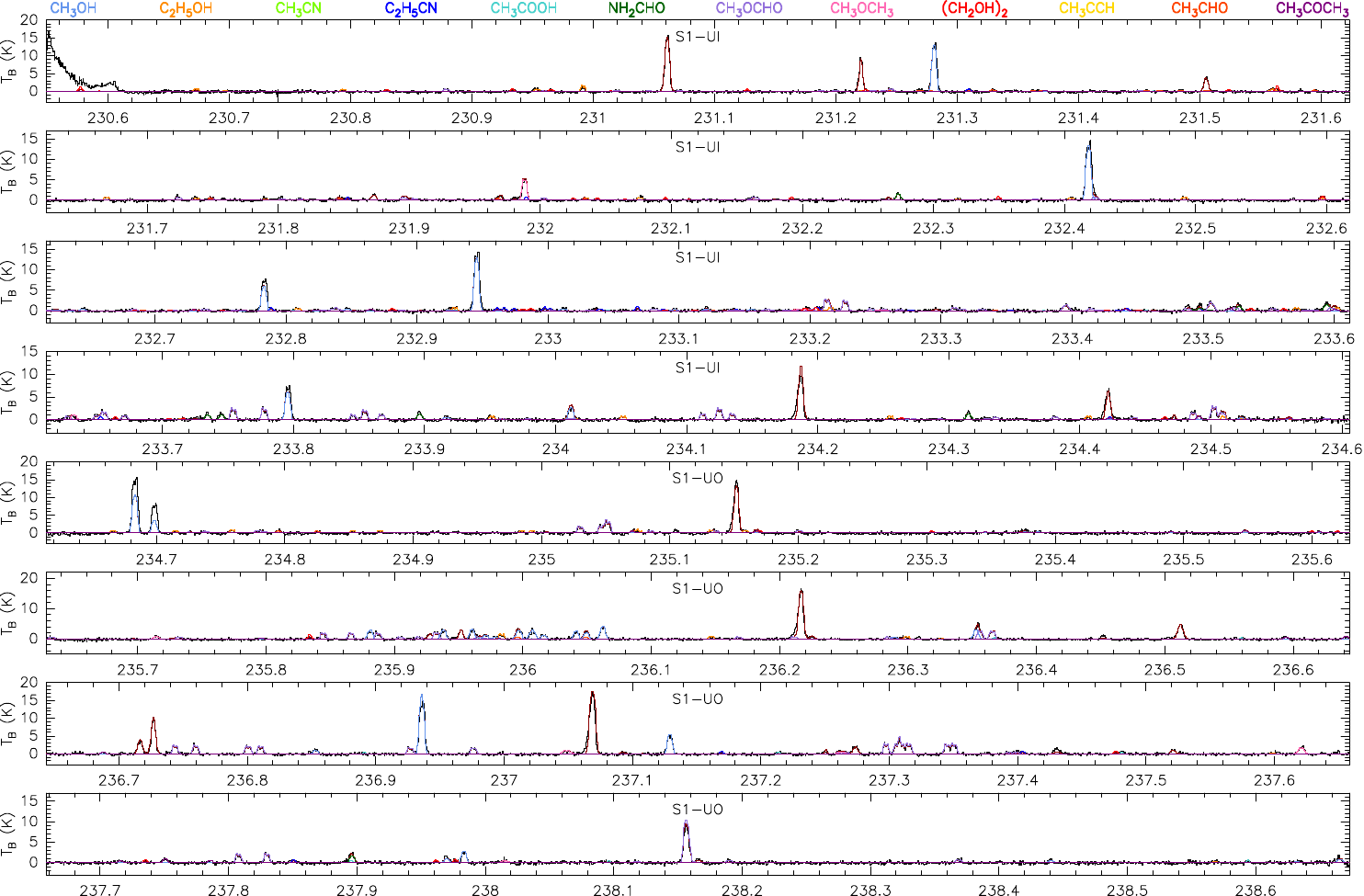} 
    \includegraphics[width=\linewidth]{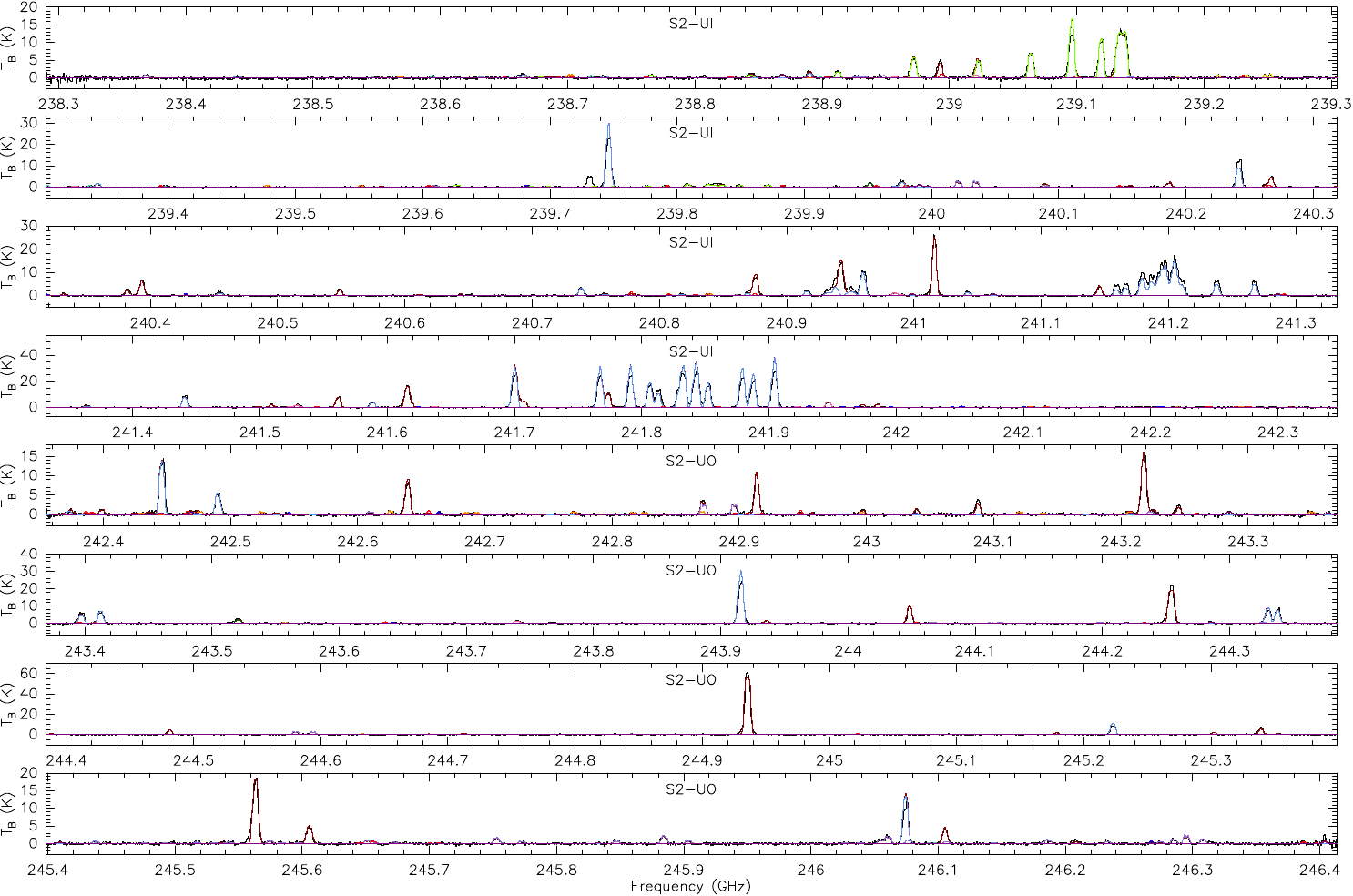}
    \caption{Continued.}
\end{figure*}

\end{document}